\documentclass[aps,prb,superscriptaddress,amsfonts,amsmath,amssymb,showpacs,floatfix,reprint,longbibliography,nofootinbib]{revtex4-2}

\usepackage{url}
\usepackage{bm}
\usepackage{graphicx}
\usepackage{amsmath}
\usepackage{amstext}
\usepackage{amssymb}
\usepackage{amsfonts}
\usepackage{amsbsy}
\usepackage{verbatim}
\usepackage{color}
\usepackage[colorlinks=true, urlcolor=blue, linkcolor=blue, citecolor=blue, pdftex]{hyperref}
\usepackage{floatrow}
\usepackage{float}
\usepackage{tikz}
\usepackage{gensymb}
\usepackage{textcomp}
\usepackage{enumitem}
\usepackage[version=3]{mhchem}
\usepackage{afterpage}
\usepackage{pbox}
\usepackage{dsfont}
\usepackage{siunitx}
\usepackage{fancyhdr}
\usepackage{stackengine,wasysym,scalerel}
\usepackage{latexsym}
\usepackage{cancel}
\usepackage{comment}
\usepackage{array, multirow, bigdelim, makecell, booktabs}
\usepackage[misc]{ifsym}
\usepackage[capitalise]{cleveref}

\begin{document}
\title{Projective symmetry group classification of Abrikosov fermion mean-field ans\"atze on the square-octagon lattice}
\author{Atanu Maity}
\affiliation{Department of Physics and Quantum Center for Diamond and Emergent Materials (QuCenDiEM), Indian Institute of Technology Madras, Chennai 600036, India}
\author{Francesco Ferrari}
\affiliation{Institut f\"ur Theoretische Physik, Goethe Universit\"at Frankfurt, Max-von-Laue-Straße 1, D-60438 Frankfurt am Main, Germany}
\affiliation{Department of Physics and Quantum Center for Diamond and Emergent Materials (QuCenDiEM), Indian Institute of Technology Madras, Chennai 600036, India}
\author{Ronny Thomale} 
\affiliation{Institut f\"{u}r Theoretische Physik und Astrophysik, 
  Julius-Maximilians-Universit\"at W\"{u}rzburg, Am Hubland, D-97074 W\"{u}rzburg, Germany}
\affiliation{Department of Physics and Quantum Center for Diamond and Emergent Materials (QuCenDiEM), Indian Institute of Technology Madras, Chennai 600036, India}
\author{Saptarshi Mandal}
\affiliation{Institute of Physics, Bhubaneswar 751005, Odisha, India}
\affiliation{Homi Bhabha National Institute, Mumbai 400 094, Maharashtra, India}
\author{Yasir Iqbal}
\email{yiqbal@physics.iitm.ac.in}
\affiliation{Department of Physics and Quantum Center for Diamond and Emergent Materials (QuCenDiEM), Indian Institute of Technology Madras, Chennai 600036, India}

\begin{abstract}
We perform a projective symmetry group (PSG) classification of symmetric quantum spin liquids with different gauge groups on the square-octagon lattice. Employing the Abrikosov fermion representation for spin-$1/2$, we obtain $32$ $SU(2)$, $1808$ $U(1)$ and $384$ $\mathds{Z}_{2}$ algebraic PSGs. Constraining ourselves to mean-field parton ansätze with short-range amplitudes, the classification reduces to a limited number, with 4 $SU(2)$, 24 $U(1)$ and 36 $\mathds{Z}_{2}$, distinct phases. We discuss their ground state properties and spinon dispersions within a self-consistent treatment of the Heisenberg Hamiltonian with frustrating couplings. 
\end{abstract}

\date{\today}

\maketitle
\section{Introduction}
In the context of magnetism, the umbrella term \textit{frustration} denotes the effects arising from competing interactions between magnetic moments which cannot be simultaneously satisfied. The archetypal picture of magnetic frustration is exemplified by the  arrangement of three antiferromagnetically coupled (Ising) spins on the vertices of a triangle, whose energy can be minimized by six different spin patterns~\cite{Wannier-1950,Vannimenus_1977,Toulouse-1980,balents2010}. This minimal example of geometric frustration carries on to extended lattice structures with antiferromagnetic Heisenberg-like interactions and triangular motifs (or loops with an odd number of spins, in general), which can possess an extensively degenerate ground state manifold of classical spin arrangements~\cite{moessner1998}. In this setting, when the temperature of the system is sufficiently low, the effects of quantum fluctuations, combined with the presence of competing ordering tendencies, can inhibit the onset of magnetic order instead favoring the formation of unconventional phases of matter such as quantum spin-liquid states~\cite{Savary-2016,Zhou-2017,Broholm-2020}. The quantum spin liquid paradigm encompasses those zero-temperature strongly correlated spin states which do not show spontaneous symmetry breaking and, thus, cannot be distinguished by any local order parameter~\cite{misguich2011}. In a seminal paper, Wen introduced the concept of quantum order to characterize different quantum spin liquid states based on the so-called projective symmetry group (PSG) classification~\cite{Wen-2002}. This approach is based on the construction of effective low-energy theories of quantum spin liquids for spin models by recasting the original spin degrees of freedom in terms of Abrikosov pseudofermion parton operators ~\cite{Abrikosov-1965}\textemdash charge neutral quasiparticles carrying spin $S=1/2$ (\textit{spinons}). The resulting fermionic system, in the saddle-point approximation, can be described by quadratic spinon Hamiltonians which fulfill either {\it all} or a set of symmetries of the initial spin Hamiltonian. The PSG approach enables one to classify these quadratic spinon Hamiltonians and thus map out an entire set of distinct spin liquid states which can be realized in the original spin system.

\begin{figure}
\includegraphics[width=1.0\linewidth]{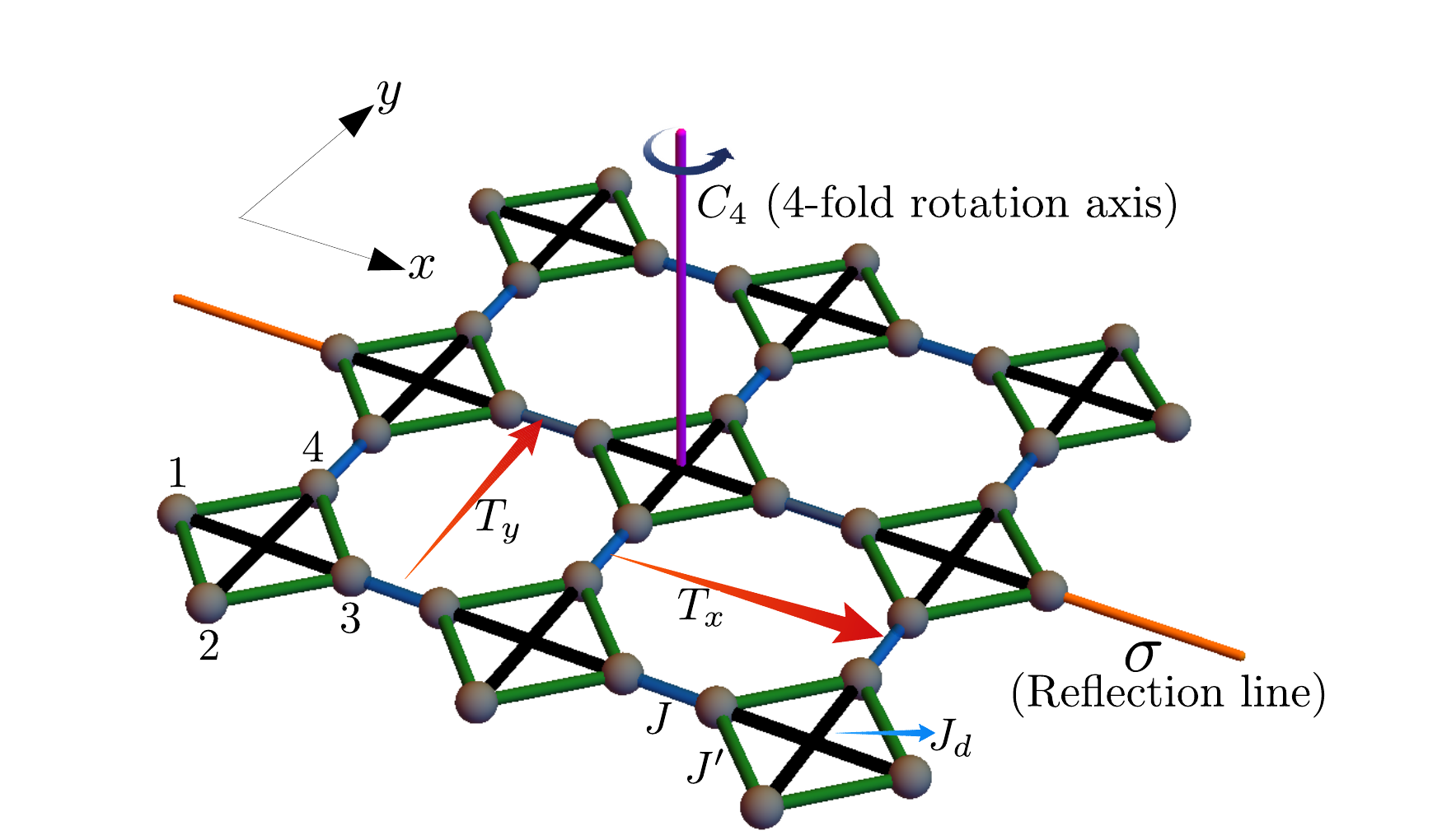}
\caption{The square-octagon lattice, formed by an underlying square Bravais lattice decorated with a four-site unit cell. The sublattice sites are labelled from $1$ to $4$ and form elementary square plaquettes inside the unit cells. The links between sites highlight the three exchange
couplings: (i) $J$ on bonds connecting sites of different unit cells (blue lines); (ii) $J'$ on the bonds forming the edges of the elementary square plaquettes  (green lines); (iii) $J_d$ connecting sites across the diagonals of the square plaquettes (black lines). The symmetries of the lattice are also shown (translations, rotations, reflections).
}
\label{fig:fig1}
\end{figure}

In this work, we perform a PSG analysis of the fermionic spin liquid states that can be realized in a Heisenberg model on square-octagon lattice (also called L4-L8, Fisher, CaVO lattice). The square-octagon network is obtained by decorating a square lattice with four sites in the unit cell, arranged as shown in Fig.~\ref{fig:fig1}. When restricted to Heisenberg interactions on the two symmetry inequivalent nearest-neighbor bonds, referred to as $J$ and $J'$ hereafter, the lattice structure is bipartite [see Fig.~\ref{fig:fig1}], and host to N\'eel~\cite{Farnell-2005,Farnell-2014,Farnell-2018} and valence bond crystal phases~\cite{Farnell-2009,Ueda-1996,Troyer-1996,Troyer-1997,Gelfand-1996,Weihong-1997,Albrecht-1996a,Meshkov-1996} with quantum phase transition phenomena~\cite{Zhang-2017}. However, the inclusion of diagonal couplings $J_{d}$ inside the elementary square plaquettes introduces frustration in the Hamiltonian and the resulting $J$-$J'$-$J_d$ model has been shown to host a rich variety of quantum paramagnetic phases~\cite{Maity-2020}. This motivates us to carry out a PSG classification of fermionic mean-field Ans\"atze with different low-energy gauge groups and respecting all symmetries of the spin Hamiltonian. Since, the corresponding classical Hamiltonian is host to coplanar magnetic orders in the $J_{1}$-$J_{2}$-$J_{d}$ parameter space, we restrict our analysis to nonchiral, i.e., symmetric spin liquids~\cite{Messio-2011}. The fermionic states thus obtained within a PSG analysis can be Gutzwiller projected and their properties, such as energy, static and dynamical structure factors computed within a variational Monte Carlo scheme~\cite{Capriotti-2001,Iqbal-2011b,Iqbal-2013,Iqbal_2014,Iqbal-2015,Iqbal_2016,Hu-2013,ferrari2018}. Furthermore, the quantum spin liquid Ans\"atze serve as parent states whose potential instabilities towards symmetry breaking dimerization patterns gives rise to valence-bond-crystal phases. Since, a previous work~\cite{Maity-2020} pointed to the presence of dimer orders in $J_1$-$J_2$-$J_d$ parameter space, our symmetric spin liquid Ans\"atze form the basis of variational Monte Carlo studies which have previously been employed on other lattices to investigate such potential instabilities~\cite{Iqbal-2011a,Iqbal_2012,Iqbal-2018,Iqbal-2021,Astrakhantsev-2021,Kiese-2023}. The comparison of these numerically computed properties with results from other numerical methods or neutron scattering data could potentially permit a microscopic characterization of the observed spin-liquid behavior for a given spin Hamiltonian or real material~\cite{Dodds-2013,Punk-2014,ferrari2022}. In this respect, the square-octagon lattice can be viewed as a cross-section of the three-dimensional Hollandite lattice which is realized in several materials such as KMn$_8$O$_{16}$, Ba-Mn-Ti oxides, Ba$_{1.2}$Mn$_8$O$_{16}$, and K$_{1.5}$(H$_3$O)$_x$Mn$_8$O$_{16}$~\cite{DeGuzman-1994,Suib-1994,Liu-2014,Ishiwata-2006,Sato-1999,Yamamoto-1974,Shen-2005,Luo-2009,Luo-2010,Crespo-2013a,Crespo-2013b,Strobel-1984,Mandal-2014}. The square-octagon network is topologically equivalent to the CaVO lattice family~\cite{Manuel-1998,Taniguchi-1995,Katoh-1995,Gelfand-1996,White-1996,Troyer-1996,Sano-1996,Albrecht-1996,Ueda-1996}, describing the periodic structure of the CaV$_{4}$O$_{9}$ compound, and more recently has been realized in the functional material ZnO~\cite{He-2012}. Furthermore, the potential synthesis of a two-dimensional carbon allotrope with a square-octagon network, dubbed octagraphene, has also been investigated~\cite{Kang-2019,Sheng-2012,Podlivaev-2013}.

 The article is organized as follows. In Sec.~\ref{model_method}, we discuss the spin Hamiltonian and describe the structure of the fermionic representation, and the mean-field approximation. The resulting symmetries and the projective symmetry group (PSG) framework is discussed in detail.  In Sec.~\ref{sec:mft_ansatz}, we carry out the PSG analysis for the square-octagon lattice and present the fully symmetric mean-field Ans\"atze classified according to their low-energy gauge groups of $SU(2)$, $U(1)$, and $\mathds{Z}_2$. For the frustrated regime of the Hamiltonian we self-consistently determine the mean-field parameters and discuss the spinon excitation spectrum for various Ans\"atze. Finally, we present an outlook in Sec.~\ref{sec:discussion}.

\section{Projective symmetry group framework}
\label{model_method}

In this section, we provide a general overview of the Abrikosov pseudofermion representation, the mean-field construction of quadratic spinon Hamiltonians, and their classification employing the PSG method. The starting point is provided by the Heisenberg Hamiltonian
\begin{equation}
\hat{\mathcal{H}}=\sum_{\langle i,j\rangle}J_{ij}\hat{\mathbf{S}}_{i} \cdot \hat{\mathbf{S}}_{j},
\label{eq:mod-ham}
\end{equation}
with spin-$1/2$ at each site $i$. The first step towards constructing a mean-field theory~\cite{Baskaran-1988} involves rewriting the spin operators $\hat{\mathbf{S}}_{i}$ at every site in terms of fermionic spinon operators ($\hat{f}_{i \sigma}$)~\cite{Abrikosov-1965}
\begin{equation}
\label{eq:abrikosov}
\hat{S}^{\alpha}_{i}=\frac{1}{2}\sum_{\sigma\sigma^{\prime}}\hat{f}^\dagger_{i\sigma}\tau^{\alpha}_{\sigma\sigma^\prime}\hat{f}_{i\sigma^\prime},
\end{equation}
where  $\sigma=\{\uparrow,\downarrow$\} and $\tau^\alpha$ ($\alpha=x,y,z$) are the Pauli matrices. As can be inferred from Eq.~\eqref{eq:abrikosov}, a spin operator is split into two fermionic operators, manifesting the fractional character of the spinons in this mathematical formalism. The fermionic representation is, however, associated with an artificial enlargement of the Hilbert space. Indeed, while the local Hilbert space of the spins contains two states ($\uparrow$ or $\downarrow$), the fermionic formalism introduces empty and doubly occupied sites, which are unphysical and must ultimately be projected out to obtain a {\it bona fide} fermionic wave function for the original spin Hamiltonian. However, it is this very property of Hilbert space enlargement which endows the pseudofermion representation with an additional {\it local} $SU(2)$ gauge symmetry~\cite{Affleck-1988,Dagotto-1988}\textemdash the backbone of the PSG method.

On substituting Eq.~\eqref{eq:abrikosov} in Eq.~\eqref{eq:mod-ham} one lands up with a quartic Hamiltonian
\begin{equation}
\hat{\mathcal{H}}=-\frac{1}{2}\sum_{i,j} \sum_{\sigma,\sigma} J_{ij} (\hat{f}^\dagger_{i\sigma}\hat{f}_{j\sigma}\hat{f}^\dagger_{j\sigma^\prime}\hat{f}_{i\sigma^\prime}+\frac{1}{2}\hat{f}^{\dagger}_{i\sigma}\hat{f}_{i\sigma}\hat{f}^{\dagger}_{j\sigma^{\prime}}\hat{f}_{j\sigma^{\prime}}).
\label{eq:fourfermion}
\end{equation}
A mean-field decoupling of the quartic Hamiltonian in the hopping and pairing channels is then performed by taking the ground state expectation value of the operators $\hat{f}^{\dagger}_{i\sigma}\hat{f}_{j\sigma^{\prime}}$ and $\hat{f}_{i\sigma}\hat{f}_{j\sigma^{\prime}}$ 
\begin{equation}
\chi_{ij}\delta_{\sigma\sigma^{\prime}}=2\langle\hat{f}^\dagger_{i\sigma}\hat{f}_{j\sigma^{\prime}}\rangle, \; \; \; \Delta_{ij}\epsilon_{\sigma\sigma^{\prime}}=-2\langle\hat{f}_{i\sigma}\hat{f}_{j\sigma^\prime}\rangle
\end{equation}
and replacing the operators in Eq.~\eqref{eq:fourfermion}, to obtain the quadratic Hamiltonian
\begin{eqnarray}
\hat{H}_{0}&& = \frac{3}{8}\sum_{i,j}J_{ij}\Bigg{[}\frac{1}{2}{\rm Tr}[u^\dagger_{ij} u_{ij}]-(\hat{\psi}^\dagger_iu_{ij}\hat{\psi}_j+{\rm h.c.})\bigg{]} \nonumber \\&&+\sum_{i,\mu}\hat{\psi}^\dagger_{i} a_\mu(i)\tau^\mu\hat{\psi}_{i}.
\label{eq:h0}
\end{eqnarray}
Here, we have introduced the doublet $\hat{\psi}_{i}^{\dagger}=(\hat{f}_{i,\uparrow}^{\dagger},\hat{f}_{i,\downarrow})$, and the site-dependent Lagrange multiplier terms $a_{\mu}(i)$ that enforce, on average, the one-fermion-per-site constraint,
\begin{equation}
\langle\hat{\psi}_{i}^{\dagger}\tau^{\mu}\hat{\psi}_{i}\rangle=0,   \;  \;  \; \mu=1,2,3 \; \; \; \forall~i.
\label{eq:constraint}
\end{equation} 
The link fields $u_{ij}$ ($=u_{ji}^\dagger$) are $2\times2$ matrices comprising of the mean-field hopping ($\chi_{ij}$) and pairing ($\Delta_{ij}$) amplitudes, and are referred to as the Ansatz of a quantum spin liquid state
\begin{eqnarray}
u_{ij} =
\begin{bmatrix}
\chi^\dagger_{ij} & \Delta_{ij}  \\
\Delta^\dagger_{ij} & -\chi_{ij}
\end{bmatrix}=\imath\lambda^0_{ij}\tau^0+\lambda^\alpha_{ij}\tau^\alpha
\label{eq:ansatz}
\end{eqnarray}
where $\lambda^\mu_{ij}$ are real parameters and $\tau^0$ is the $2 \times 2$ identity matrix. The power of the PSG approach lies in its ability to classify mean-field Hamiltonians of the type in Eq.~\eqref{eq:h0}, i.e., it enumerates distinct Ans\"atze $u_{ij}$ with a specified gauge structure [$\mathds{Z}_{2}$, $U(1)$, $SU(2)$] and desired lattice (and time-reversal) symmetries. Here, we classify Ans\"atze with all three aforementioned gauge groups, and identify the connections between them.

We now explain how the mean-field Ans\"atze respecting the complete set of symmetries of the lattice and having a $\mathds{Z}_{2}$ gauge group are classified. As mentioned before, the local $SU(2)$ gauge invariance of the fermionic representation is the key ingredient, as it implies the freedom to perform gauge transformations $\hat{\psi}_{i}\to W_{i}\hat{\psi}_{i}$ where $W_{i}$ are site-dependent generic $SU(2)$ matrices. In the fermionic Hilbert space, this transformation acts as a rotation {\it only} within the unphysical subspace of empty and doubly occupied sites, while the one-fermion per site subspace consisting of physical states remains untouched. This gauge transformation can equivalenty be formulated as an operation acting on the Ansatz instead of the spinor
\begin{equation}
    u_{ij}\to W_{i}^{\dagger}u_{ij}W_{j}.
    \label{eq:su2}
\end{equation}
It is worth mentioning that the symmetry group of a generic mean-field Hamiltonian, i.e., the low-energy gauge group, is independent of the $SU(2)$ symmetry of the fermionic representation [Eq.~\eqref{eq:abrikosov}] which is descriptive of the high-energy gauge structure. In particular, one can construct Ans\"atze with $SU(2)\times SU(2)$ symmetry which is thus larger than the $SU(2)$ gauge structure of the fermionic representation~\cite{Wen-2002}. The group $\mathcal{G}$ of symmetry operations which keep an Ansatz invariant
\begin{equation}
    u_{ij}=W_{i}^{\dagger}u_{ij}W_{j}, \; \; \; W_{i}\in\mathcal{G}
    \label{eq:IGG}
\end{equation}
is called the Invariant Gauge Group (IGG) of an Ansatz. Since, the global transformations $\mathcal{G}=\pm\mathds{1}_{2}$ leaves any Ansatz invariant, the IGG always contains $\mathds{Z}_{2}$ as a subgroup. Quantum spin liquids with a mean-field Ansatz with IGG $\mathds{Z}_{2}$ are referred to as $\mathds{Z}_{2}$ spin liquids, and similarly for $U(1)$ and $SU(2)$.  

The basic idea behind the PSG is that due to the presence of additional local $SU(2)$ gauge symmetry, any physical (lattice and time-reversal) symmetry operation can now be supplemented by gauge transformation
\begin{equation}
    u_{ij}\to W^{\dagger}_{\mathcal{S}(i)} u_{\mathcal{S}(i)\mathcal{S}(j)} W_{\mathcal{S}(j)},
    \label{eq:proj}
\end{equation}
where $\mathcal{S}$ is an element of the system’s symmetry group acting on the lattice sites. The observation that symmetries act projectively in the fermionic Hilbert space implies that an Ansatz $u_{ij}$ has the freedom to apparently break lattice and/or time-reversal symmetries as long as there exists a gauge transformation satisfying the generalized invariance condition
\begin{equation}
    G_{\mathcal{S}}^{\dagger}(\mathcal{S}(i))u_{\mathcal{S}(i)\mathcal{S}(j)}G_{\mathcal{S}}(\mathcal{S}(j))=u_{ij}.
    \label{eq:psg}
\end{equation}
Here, $G_{\mathcal{S}}(i)$ is a site-dependent gauge transformation which fulfills this condition. Hence, the different projective implementations $G_{\mathcal{S}}(i)$ fulfilling Eq.~\eqref{eq:psg} allow one to distinguish between different quantum spin liquid phases possessing the same physical (space-time) symmetries~\cite{Wen-2002}, and constitute the PSG. In other words, the PSG may be viewed as an extension of the system's symmetry group by the IGG. 
\begin{equation}
    {\rm PSG} = {\rm SG} \ltimes {\rm IGG}
\end{equation}
In this work, we classify all PSGs with the ${\rm C}_{4v}$ point group using Eq.~\eqref{eq:psg}. Subsequently, we construct Ans\"atze $u_{ij}$ and discuss their spinon band structures. The PSG method has been extensively employed to classify spin liquid Ans\"atze on a wide variety of two- and three-dimensional lattices~\cite{Wang-2006,Lu-2011a,Lu-2011b,Yang-2012,Lu-2016a,Lu-2016b,Huang-2017,Huang-2018,Liu-2019,Jin-2020,Sonnenschein-2020,Liu-2021,Chern-2021,Chern-2022}. 

\section{PSG classification for square-octagon lattice}
\label{sec:mft_ansatz}
The point group symmetry elements of the square-octagon lattice are those of $C_{4v}$. Its generators are given by four-fold rotations $C_{4}$ and reflections $\sigma$ (see Fig.~\ref{fig:fig1}). The action of these symmetry operations on a lattice site labelled $(x,y,\mu)$, where $(x,y)$ is the coordinate of the unit cell, and $\mu$ denotes the sublattice index, is given by
\begin{equation}
\label{eq:point}
\left.\begin{aligned}
&C_4(x,y,\mu)\rightarrow (-y,x,\mu+1),\\
&\sigma(x,y,1/3)\rightarrow (x,-y,1/3),\\
&\sigma(x,y,2/4)\rightarrow (x,-y,4/2).\\
\end{aligned}\right.
\end{equation}
Here, and in the remainder of the paper, the notation $\mu+1$ implies a modular arithmetic operation $\textrm{mod}(\mu-1,4)+1$, which permutes the sublattice sites $\mu=1,2,3,4$. The complete space group also includes translations 
\begin{equation}
\label{eq:trans}
\left.\begin{aligned}
&T_x(x,y,\mu)\rightarrow (x+1,y,\mu),\\
&T_y(x,y,\mu)\rightarrow (x,y+1,\mu).\\
\end{aligned}\right.
\end{equation}

In addition to these lattice symmetries, we also impose time-reversal symmetry since in this work we classify only fully symmetric Ans\"atze. The time-reversal operator $\mathcal{T}$ does not affect the lattice coordinates, and commutes with all symmetries, however, it has a nontrivial action on the spinon operators $\mathcal{T}(\hat{f}_{i\uparrow},\hat{f}_{i\downarrow})=(\hat{f}_{i\downarrow},-\hat{f}_{i\uparrow})$. Consequently, the gauge doublet transforms as $\mathcal{T}(\hat{\psi}_{i})=[(\imath\tau^{2}\hat{\psi}_{i})^{\dagger}]^{T}$. Upon performing a global gauge transformation $\hat{\psi}_{i}\to-\imath\tau^{2}\hat{\psi_{i}}$, the action of time reversal can be conveniently recast as $\mathcal{T}(\hat{\psi}_{i})=[(\hat{\psi}_{i})^{\dagger}]^{T}$. In this form, the action of $\mathcal{T}$ on the Ansatz takes the simplified form $\mathcal{T}(u_{ij})=-u_{ij}$, and similarly for the on-site terms $\mathcal{T}(a_{\mu}(i))=-a_{\mu}(i)$.

A \textit{bona fide} projective representation must obey the same algebraic relations as the space group itself, thus yielding a set of constraints on the representation. For example, the generators of the point group symmetries in Eq.~\eqref{eq:point} map back to the identity when applied twice (for $\sigma$) or four times (for $C_{4}$). Hence, the reflections (and time-reversal) should be represented by a cyclic group of order 2 while rotations form a cyclic group of order 4.

The algebraic relations between the generators of the symmetry group corresponding to the square-octagon lattice are
\begin{equation}
\label{algebraic}
\left.\begin{aligned}
&T_xT_y=T_yT_x,\\
&T_yC^{-1}_4T_xC_4=T^{-1}_xC^{-1}_4T_yC_4=\mathds{1},\\
&\sigma^{-1}T^{-1}_x\sigma T_x=\sigma^{-1}T^{-1}_y\sigma T_y=\mathds{1},\\
&C_4\sigma^{-1}C_4\sigma=\mathds{1},\\
&\mathcal{T}S=S\mathcal{T}; \; S\in\{T_x,T_y,C_4,\sigma\}, \\
&(C_4)^4=(\sigma)^2=(\mathcal{T})^2=\mathds{1},\\
\end{aligned}\right.
\end{equation}

In the above relations, we need to associate a gauge transformation with each of the symmetry generators. This will provide a set of gauge symmetry conditions [given by Eqs.~\eqref{c_translation},~\eqref{c_rotation},~\eqref{c_reflection},~\eqref{c_time} and~\eqref{c_rot_ref_time} in the Appendix~\ref{gen_condition}], or, in other words, a projective analogue of the above symmetry conditions in generic $SU(2)$ gauge space. 

\begin{figure}
\includegraphics[width=1.0\linewidth]{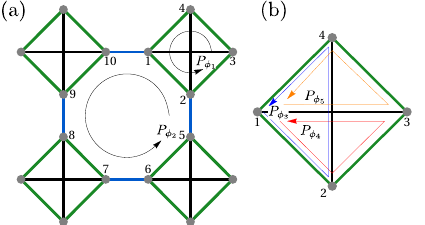}
\caption{ Schematic representation of loop operators $P_{\phi_1},P_{\phi_2},P_{\phi_3},P_{\phi_4},P_{\phi_5}$ around square, octagonal and three triangular plaquettes, respectively with an arbitrary base site `1'.}
\label{fig:loop_fig}
\end{figure}

In our construction of short-range mean-field states, we consider, in addition to the two nearest-neighbour amplitudes living on the 
$J$ and $J^\prime$ bonds, the $J_d$ bonds forming the diagonals inside the squares. It is helpful to characterize the gauge structure 
of the Ans\"atze $u_{ij}$ via the $SU(2)$ flux operator defined over suitable lattice loops~\cite{Affleck-1988,Lee-2006}. 
Considering $J$ and $J'$ bonds, there exist {\it only} two flux operators for a given base site. Adopting the sites labelling of Fig.~\ref{fig:loop_fig}, the two flux operators (for the base site marked as ``1'') can be written as
\begin{equation}
\label{eq:loop1}
\left.\begin{aligned}
&P_{\phi_1}=u_{1,2}u_{2,3}u_{3,4}u_{4,1},\\
&P_{\phi_2}=u_{1,10}u_{10,9}u_{9,8}u_{8,7}u_{7,6}u_{6,5}u_{5,2}u_{2,1}.\\
\end{aligned}\right.
\end{equation}
On the other hand, the inclusion of amplitudes on $J_d$ bonds requires the consideration of $SU(2)$ fluxes on additional loops
\begin{equation}
\label{eq:loop2}
\left.\begin{aligned}
&P_{\phi_3}=u_{1,2}u_{2,4}u_{4,1},\\
&P_{\phi_4}=u_{1,2}u_{2,3}u_{3,1},\\
&P_{\phi_5}=u_{1,3}u_{3,4}u_{4,1}\\
\end{aligned}\right.
\end{equation}

\begin{figure}[b]
\includegraphics[width=0.75\linewidth]{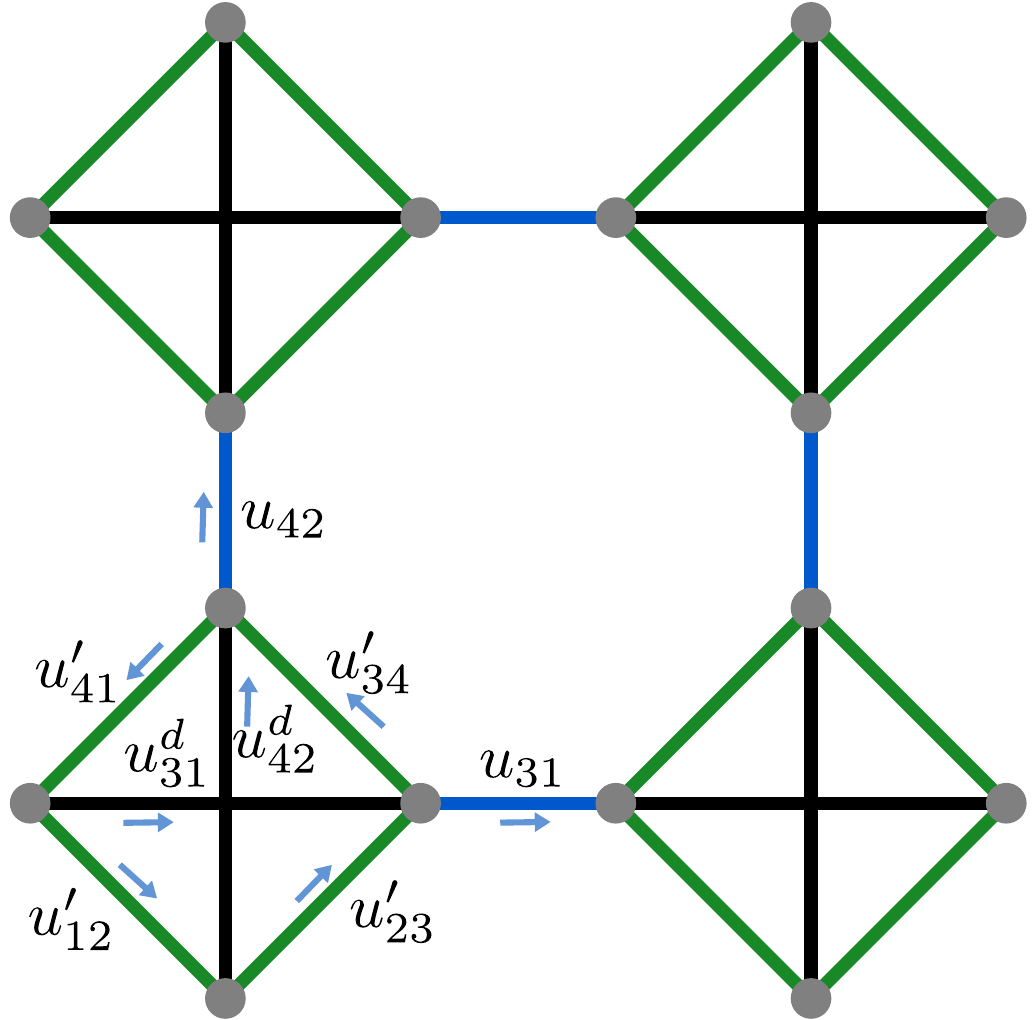}
\caption{Labelling of $u_{ij}$ link fields for the unit cell at $(x,y)=(0,0)$ (the arrows indicate the direction $i\rightarrow j$).}
\label{fig:links}
\end{figure}

The structure of the two loop operators given by Eq.~\eqref{eq:loop1} and \eqref{eq:loop2} determine the IGG of the Ansatz when considering only $J$, $J'$ and $J_d$ bonds. The generic form of $SU(2)$ flux operators is
\begin{equation}
\label{eq:su2_flux_even}
P_{\phi_i}=\tau^0\cos{\phi_i}+\imath(\hat{n}_i \cdot \hat{\tau})\sin{\phi_i}
\end{equation}
for even sided loops (i.e., $P_{\phi_1},P_{\phi_2}$)
and \begin{equation}
\label{eq:su2_flux_odd}
P_{\phi_i}=(\hat{n}_i \cdot \hat{\tau})\cos{\phi_i}+\imath\tau^0\sin{\phi_i}
\end{equation}
for odd sided loops (i.e., $P_{\phi_3},P_{\phi_4},P_{\phi_5}$), with $\hat{n}_i$ denoting a unit vector~\cite{Bieri-2016}.
There arise three possibilities. The first case occurs for those ans\"atze in which all SU(2) flux operators are trivial, i.e., $P_{\phi_i}\propto\tau^0\;\forall~i$. In this case, by choosing an appropriate gauge, the Ansatz can be written in terms of imaginary hopping only ($u_{ij}=\imath\chi^0_{ij}\tau^0$) and, thus, its IGG is manifestly $SU(2)$ (any global $SU(2)$ gauge transformation leaves the $u_{ij}$ matrices invariant). The fluxes through the even-sided plaquettes can only be $\phi_i=0$ or $\phi_i=\pi$. 
In the second case, the flux operators are non-trivial, i.e., they are directional in nature ($\phi_i\neq0,\pi$ for even-sided loops) and collinear ($[P_{\phi_i},P_{\phi_j}]=0$). In this case, we can choose a suitable gauge such that the Ansatz can be written in a form with only real and imaginary hopping terms ($u_{ij}=\imath\chi^0_{ij}\tau^0+\chi^3_{ij}\tau^3$). Thus, this collinear directional nature breaks the IGG $SU(2)$ down to $U(1)$. On the other hand when the flux operators are non-collinear ($[P_{\phi_i},P_{\phi_j}]\neq0$), in any gauge the Ansatz contains both hopping and pairing terms and thus the IGG lowers to $\mathds{Z}_{2}$.

We note that for $IGG\in SU(2)$ the Ansatz must vanish on the bonds responsible for forming odd-sided loops, such that the flux operator $P_{\phi_i}$ through those loops vanishes. Otherwise, one may end up with two consequences. When $\phi_i=\{0,\pi\}$, the flux operators through odd-sided loops are no longer trivial [Eq.~\eqref{eq:su2_flux_odd}] and thus $SU(2)$ IGG is broken. Also, when $\phi_i\neq0,\pi$, the time reversal symmetry is broken~\cite{Bieri-2016}. To respect time reversal symmetry, $\phi_i$ through odd sided loops can only take $0,\pi$, even for $IGG\in U(1),\mathds{Z}_2$. Furthermore, we also need to consider the commutation relations of the flux operators with the onsite terms ($a_\mu\tau^\mu$) when determining the IGG.\\

\subsection{$SU(2)$ Spin Liquids}
\label{sec:su2_sl}
The square-octagon lattice being bipartite, we begin by classifying mean-field Ans\"atze with $SU(2)$ IGG. For such mean-field Ansatz, all link fields $u_{ij}$ can be expressed in the canonical form $u_{ij}\propto \imath\tau^{0}$, which makes the $SU(2)$ gauge structure manifest~\cite{Wen-2002}. In this canonical representation, the elements of the IGG take the form of a global $SU(2)$ gauge transformation $\mathcal{G}=e^{\imath\theta \hat{n} \cdot \hat{\tau}}$ with site-independent $\theta$. However, given the $SU(2)$ gauge redundancy, one is always free to express any $SU(2)$ ansatz in noncanonical form, as we do in the following. Indeed, in order to make the connection with $U(1)$ and $\mathds{Z}_2$ ans\"atze more explicit, we choose a gauge form for $SU(2)$ spin liquids in which $u_{ij}$ is proportional to $\tau^3$ [see Appendix~\ref{app:gauge_dep_ansatz}].

The PSG solutions [see Appendix~\ref{su2_algebraic}] are given by :
\begin{equation}
\label{su2_solution}
\left.\begin{aligned}
&G_{T_x}(x,y,\mu)=\eta^y_yg_x\\
&G_{T_y}(x,y,\mu)=g_y\\
&G_{C_4}(x,y,\mu)=(\eta_{\sigma_x}\eta_{\sigma_y})^y\eta^{xy}_yg_{C_4}(\mu)\\
&G_\sigma(x,y,\mu)=\eta^x_{\sigma_x}\eta^y_{\sigma_y}g_\sigma(\mu)\\
&G_{\mathcal{T}}(x,y,\mu)=(-1)^{x+y}(-1)^{\mu}g_t
\end{aligned}\right.
\end{equation}
where the $g$'s are generic $SU(2)$ matrices. The gauge inequivalent choices of $g_{C_4}(\mu)$ and $g_{\sigma}(\mu)$ are given in Table~\ref{table:su2_psg}.

\begin{table}
	\caption{ All possible gauge inequivalent choices of 	$g_{C_4}(\mu)$ and $g_{\sigma}(\mu)$ in $SU(2)$ PSGs, and $\eta_c,\eta_\sigma=\pm1$.}
	\begin{ruledtabular}
		\begin{tabular}{cc}
			$g_{C_4}(\mu)$&$g_{\sigma}(\mu)$\\
			\hline
			$\{g_{C_4},g_{C_4},g_{C_4},\eta_c g_{C_4}\}$ & $\{g_{\sigma},\eta_\sigma g_{\sigma},\eta_c g_{\sigma},\eta_\sigma g_{\sigma}\}$\\
		\end{tabular}
	\end{ruledtabular}
	\label{table:su2_psg}
\end{table}

\begin{figure*}
\includegraphics[width=1.0\linewidth]{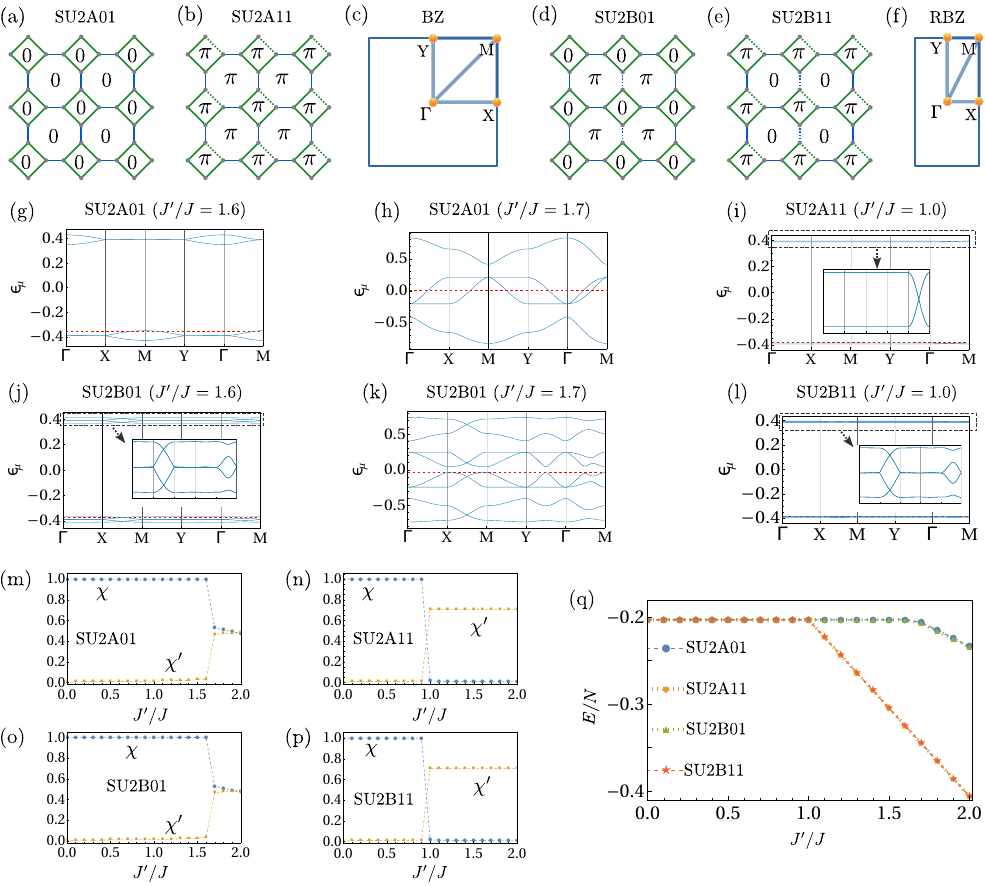}
  \caption{Sign structures of the hopping parameters in the four $SU(2)$ states (a) $SU2A01$, (b) $SU2A11$, (d) $SU2B01$ and (e) $SU2B11$. Solid and dotted lines represent positive and negative signs, respectively.   Definitions of high symmetry points for (c) single unit cell Ans\"atze in the first Brillouin zone (BZ) and (f) doubled unit cell Ans\"atze in the first reduced Brillouin zone (RBZ). Dispersions: (g) $SU2A01$ at $J'/J=1.6$, (h) $SU2A01$ at $J'/J=1.7$, (i) $SU2A11$ at $J'/J=1.0$,  (j) $SU2B01$ at $J'/J=1.6$, (k)   $SU2B01$ at $J'/J=1.7$, (l)  $SU2B11$ at $J'/J=1.0$. The red dashed line represents the Fermi level. The self-consistently determined mean-field parameters along the line $J_d=0$ for the $SU(2)$ Ans\"atze labelled by (m) $SU2A01$, (n) $SU2A11$, (o) $SU2B01$ and (p) $SU2B11$. (q) Ground state energies per site (in units of $J$) for these four Ans\"atze.}
\label{fig:fig4}
\end{figure*}

As there are five independent $\eta$ parameters in Eq.~\eqref{su2_solution}, we have $2^5=32$ $SU(2)$ PSGs. Upon restricting the mean-field Ansatz to $J$ and $J'$ bonds [see Fig.~\ref{fig:fig1}], only seven distinct PSGs can be realized. Among them, four states have nonvanishing amplitudes on {\it both} $J$ and $J'$ bonds while the remaining three have vanishing amplitudes on either $J$ or $J'$ bonds. We adopt the following convention to label the $SU(2)$ Ans\"atze:
\begin{equation}
\label{su2notation}
SU2(Class)\eta_{\sigma_x}\eta_{\sigma_y}
\end{equation}
where values $0$ and $1$ will be used to denote positive and negative sign of $\eta$ parameters, respectively. Here, $Class$ refers to the choice of sign of $\eta_y$. Based on this sign we group all Ans\"atze in two classes, `$A$' and `$B$' corresponding to $\eta_y=+1$ and $\eta_y=-1$, respectively. Henceforth, we adopt the same class notation for Ans\"atze with $U(1)$ and $\mathds{Z}_2$ IGG.

We now discuss the properties of the aforementioned $SU(2)$ states and the corresponding spinon dispersions (see Appendix~\ref{app:bands}), based on the results of a self-consistent mean-field treatment of the spin Hamiltonian, for different values of $J'/J$ (and $J_d=0$). It is worth noting that for $SU(2)$ spin liquid Ans\"atze no mean-field terms are allowed on the $J_d$ bonds. For this reason, the results of the self-consistent study of the spin Hamiltonian do not depend on the value of $J_d$, which can thus be set to zero.

\subsubsection{SU2A01 -- $(0,0)$ flux state}
In addition to the notation introduced previously, it is convenient to label the $SU(2)$ states by the fluxes through the squares and octagons. Since the $SU2A01$ state has zero flux through squares and octagons [see Fig.~\ref{fig:fig4}(a)], we label it as $(0,0)$. The Ansatz can be written as
\begin{equation}
\label{su1a_amp}
\left.\begin{aligned}
&u'_{12}=u'_{23}=u'_{34}=u'_{41}=\chi' \tau^3\\
&u_{31}=u_{42}=\chi \tau^3,\\
\end{aligned}\right.
\end{equation}
where the labelling convention of the $u$'s in the $(0,0$ unit cell is shown in Fig.~\ref{fig:links}.

In Fig.~\ref{fig:fig4}(g), we show the nature of the spinon excitation spectrum for the $(0,0)$ state at $J'/J=1.0$ along the high-symmetry path [see Fig.~\ref{fig:fig4}(c)], with the red dashed line denoting the Fermi level. We obtain four bands, each being doubly degenerate due to spin symmetry. The excitation spectrum is gapped. The band structure remains qualitatively similar (more squeezed) for the range $0 < J'/J\lesssim 1.6$.
For larger values of $J'/J$, the spectrum becomes gapless (with a spinon Fermi surface) as shown in Fig.~\ref{fig:fig4}(h) for $J'/J=1.7$.

\subsubsection{SU2A11 -- $(\pi,\pi)$ flux state} 
This state has a $\pi$-flux threading through both the squares and octagons [see Fig.~\ref{fig:fig4}(b)], and is hence dubbed $(\pi,\pi)$. The Ansatz can be written as
\begin{equation}
\label{su2a_amp}
\left.\begin{aligned}
&u'_{12}=u'_{23}=-u'_{34}=u'_{41}=\chi' \tau^3\\
&u_{31}=u_{42}=\chi \tau^3\\
\end{aligned}\right.
\end{equation}

The spectrum is gapped and consists of four nearly flat bands [see Fig.~\ref{fig:fig4}(i)]. A noticeable feature is the presence of Dirac nodal point at the midpoint of the segment $\overline{\Gamma M}$.\\

\subsubsection{SU2B01  -- $(0,\pi)$ flux state} 
This state is characterized by a $0$ flux threading the squares and $\pi$ flux threading the octagons, and hence we label it as $(0,\pi)$. To realize this flux pattern a doubling of the geometrical unit cell is required. Here, we adopt a gauge where the doubling occurs along the $x$-direction. Hence, the reduced Brillouin zone is defined by $-\pi/2\leqslant k_x\leqslant \pi/2$, $-\pi\leqslant k_y\leqslant \pi$, and the corresponding high symmetry points are shown in Fig.~\ref{fig:fig4}(f). The Ansatz takes the form

\begin{equation}
\label{su1b_amp}
\left.\begin{aligned}
&u'_{12}=u'_{23}=u'_{34}=u'_{41}=\chi' \tau^3\\
&u_{31}=u_{42}=\chi \tau^3\\
&u_{(x,y,4),(x,y+1,2)}=(-1)^xu_{42}=(-1)^x\chi \tau^3\\
\end{aligned}\right.
\end{equation}

The mean-field spectrum is gapped [see Fig.~\ref{fig:fig4}(j) for $J'/J=1.6$ and Fig.~\ref{fig:fig4}(k) for $J'/J=1.7$]. A noticeable characteristic is the presence of Dirac nodal points at the midpoint of the segment $\overline{X M}$. The bands are nearly flat within the range $0 < J'/J\lesssim 1.6$.\\

\subsubsection{SU2B11 -- $(\pi,0)$ flux state} 
This state has $\pi$ and $0$ fluxes threading through the squares and octagons, respectively, and hence, we label it as $(\pi,0)$. The Ansatz can be written as
\begin{equation}
\label{su2b_amp}
\left.\begin{aligned}
&u'_{12}=u'_{23}=-u'_{34}=u'_{41}=\chi' \tau^3\\
&u_{31}=u_{42}=\chi \tau^3\\
&u_{(x,y,4),(x,y+1,2)}=(-1)^xu_{42}=(-1)^x\chi \tau^3\\
\end{aligned}\right.
\end{equation}
This Ansatz features gapped excitations as can be seen from the Fig.~\ref{fig:fig4}(l) shown for $J'/J=1.0$. Similar to the $SU2B01$ state, this state has Dirac nodal points at the midpoint of the segment $\overline{X M}$. Nodal lines are also visible along the segments  $\overline{\Gamma X}$, $\overline{\Gamma Y}$ and $\overline{MY}$.

\subsubsection{Remaining $SU(2)$ states}

In addition to the above four $SU(2)$ states, there exist three more $SU(2)$ Ans\"atze which feature non-vanishing mean field amplitudes on either the $J$ bonds {\it or} the $J'$ bonds. One of them is composed of hoppings only on $J$ bonds, which we label as a $J$-VBS state~\cite{Maity-2020}. This is a trivial product states of fermionic singlets on the $J$ bonds. In the opposite case, where only the $J'$ bonds have non-vanishing amplitudes, two cases can be distinguished, (i) with $0$-flux through the squares, and (ii) with $\pi$-flux through the squares, both representing plaquette-RVB phases, henceforth labelled as PRVB$_1$ and PRVB$_2$, respectively. The mean-field parameters in these two cases are given by:

\begin{equation}
\label{su3ab_amp}
\left.\begin{aligned}
J-{\rm VBS}:\;&u'_{\mu,\mu'}=0,u_{31}=u_{42}=\chi \tau^3\\
{\rm PRVB}_1:\;&u'_{12}=u'_{23}=u'_{34}=u'_{41}=\chi' \tau^3,\\ 
&u_{31}=u_{42}=0\\
{\rm PRVB}_2:\;&u'_{12}=u'_{23}=-u'_{34}=u'_{41}=\chi' \tau^3,\\ 
&u_{31}=u_{42}=0\\
\end{aligned}\right.
\end{equation}

As expected, the spinon spectrum consists of trivial flat bands. However, it is worth noticing that the spectrum of the PRVB$_2$ state is gapped, while that of PRVB$_1$ is gapless (the Fermi energy cuts through degenerate flat bands). These properties directly descend from the spectrum of a half-filled square molecule (i.e., a $4$-sites chain) threaded by $\pi$ or $0$ flux, respectively.

\vspace{0.2cm}
In Figs.~\ref{fig:fig4}(m)-(p), we show the self-consistently determined mean-field parameters along the line $J_d=0$. It can be seen that for $0 < J'/J\lesssim 1.6$, the $SU2A01$ and $SU2B01$ Ans\"atze yield the $J-{\rm VBS}$ phase as the renormalized mean-field solution, while for $J'/J\gtrsim1.6$ they become actual spin liquid states. On the other hand, the $SU2A11$ and $SU2B11$ Ans\"atze are found to yield the $J-{\rm VBS}$ state for $0 < J'/J\lesssim 1.0$ and the plaquette PRVB$_2$ state for $J'/J\gtrsim 1.0$. In Fig.~\ref{fig:fig4}(q), we compare the ground state energies of the $SU2A01$, $SU2A11$, $SU2B01$ and $SU2B11$ Ans\"atze. The results indicate that all of these four Ans\"atze converge to a $J-{\rm VBS}$ state within the parameter range $0 < J'/J\lesssim 1.0$, whereas beyond this range the PRVB$_2$ state is found to be energetically preferable.

\begin{table*}
	\caption{The gauge inequivalent choices of $g_3(\theta^{\mu}_c)$ and $g_3(\theta^{\mu}_\sigma)$ matrices in the $U(1)$ PSGs. The $\eta$ parameters can take values $\pm 1$. $n_T=0$ corresponds to the cases with $G_{T_x}(x,y,\mu)=\eta^{y}_y\tau^0$ and $G_{T_y}(x,y,\mu)=\tau^0$, i.e., classes $U1A$ and $U1B$. $n_T=1$ corresponds to the cases with $G_{T_x}(x,y,\mu)=G_{T_y}(x,y,\mu)=\imath\tau^1$, i.e., class $U1C$. The parameters $\theta_2,\theta_3,\theta_\sigma,q$ are arbitrary $U(1)$ phases whose value is not fixed by the PSG constraints.}
	\begin{ruledtabular}
		\begin{tabular}{ccccccc}
		Ansatz No &	$n_{T}$&$n_{\sigma}$&$n_{C_4}$&$n_{\mathcal{T}}$&$g_3(\theta^\mu_c)$&$g_{3}(\theta^\mu_\sigma)$\\
			\hline
		$1$ &	0&0&0&0&$\{\tau^0,\tau^0,\tau^0,\eta_cg_3(-2q)\}$ & $\{\tau^0,g_3(q),\eta_c\tau^0,g_3(-q)\}$\\
			$2$ &	0&0&0&1&$\{\tau^0,\tau^0,\tau^0,\eta_c\tau^0\}$ & $\{\tau^0,\eta_\sigma\tau^0,\eta_c\tau^0,\eta_\sigma\tau^0\}$\\
				$3$&0&0&1&0/1&$\{\tau^0,\tau^0,\tau^0,\eta_c\tau^0\}$&$\{\tau^0,\eta_\sigma\tau^0,\eta_c\tau^0,\eta_\sigma\tau^0\}$\\
				$4$&0&1&0&0&$\{\tau^0,\tau^0,\tau^0,g_3(\theta_c)\}$ & $\{\tau^0,\eta_\sigma\tau^0,g_3(-\theta_c),\eta_\sigma\tau^0\}$\\
				$5$&0&1&0&1&$\{\tau^0,\tau^0,\tau^0,\eta_c\tau^0\}$ & $\{\tau^0,\eta_\sigma\tau^0,\eta_c\tau^0,\eta_\sigma\tau^0\}$\\
				$6$&0&1&1&0&$\{\tau^0,\tau^0,\tau^0,\eta_c\tau^0\}$ & $\{\tau^0,g_3(q),\eta_c\tau^0,g_3(q)\}$\\
				$7$ &0&1&1&1&$\{\tau^0,\tau^0,\tau^0,\tau^0\}$ & $\{\tau^0,\eta_\sigma\tau^0,\tau^0,\eta_\sigma\tau^0\}$\\
				$8$&1&0&0&0&$\{\tau^0,g_3(\theta_2),g_3(\theta_3),g_3(\theta_3)\}$ & $\{\tau^0,g_3(q),\eta_1\tau^0,g_3(-q)\}$\\
				$9$&1&0&0&1&$\{\tau^0,\eta_1\tau^0,\eta_c\tau^0,\eta_c\tau^0\}$ & $\{\tau^0,\eta_\sigma\tau^0,\eta_1\tau^0,\eta_\sigma\tau^0\}$\\
				$10$&1&0&1&0&$\{\tau^0,g_3(\theta_2),g_3(\theta_3),\eta_1g_3(\theta_3-\theta_2)\}$ & $\{\tau^0,g_3(q),\eta_1\tau^0,g_3(-q)\}$\\
				$11$&1&0&1&1&$\{\tau^0,\eta_2\tau^0,\eta_3\tau^0,\eta_1\eta_2\eta_3\tau^0\}$ & $\{\tau^0,\eta_\sigma\tau^0,\eta_1\tau^0,\eta_\sigma\tau^0\}$\\
			$12$&1&1&0&0&$\{\tau^0,\eta_c\tau^0,g_3(\theta_3),\eta_cg_3(\theta_3-\theta_\sigma)\}$ & $\{\tau^0,\eta_\sigma\tau^0,g_3(\theta_\sigma),\eta_\sigma\tau^0)\}$\\
			$13$&1&1&0&1&$\{\tau^0,\eta_2\tau^0,\eta_3\tau^0,\eta_1\eta_2\eta_3\tau^0\}$ & $\{\tau^0,\eta_\sigma\tau^0,\eta_1\tau^0,\eta_\sigma\tau^0\}$\\
			$14$&1&1&1&0&$\{\tau^0,g_3(\theta_2),g_3(\theta_3),g_3(\theta_2-\theta_3+\theta_\sigma)\}$ & $\{\tau^0,\eta_\sigma\tau^0,g_3(\theta_\sigma),\eta_\sigma\tau^0)\}$\\
			$15$&1&1&1&1&$\{\tau^0,\eta_2\tau^0,\eta_3\tau^0,\eta_2\eta_3\eta_c\tau^0\}$ & $\{\tau^0,\eta_\sigma\tau^0,\eta_c\tau^0,\eta_\sigma\tau^0\}$\\
			
		\end{tabular}
	\end{ruledtabular}
	\label{table:u1_psg}
\end{table*}

\subsection{$U(1)$ Spin Liquids}
\label{u1_sl}
By allowing for real hopping along with imaginary hopping terms, the IGG can be lowered from $SU(2)$ down to $U(1)$, i.e., $\mathcal{G}=e^{\imath\theta \tau^3}$. In this section, we classify the $U(1)$ Ans\"atze where the character of the IGG implies that in a projective construction the identity is defined up to a global gauge transformation $e^{\imath\theta \tau^3}$.  In general, the PSG trasformation $G_\mathcal{S}$ associated with a given symmetry operation $\mathcal{S}$ can be written in the conventional form $G_\mathcal{S}(x,y,\mu)=e^{\imath\phi(x,y,\mu)\tau^3}(\imath\tau^1)^{n_\mathcal{S}}$, which maintains the canonical form of the $U(1)$ Ans\"atze. Here, $n_\mathcal{S}=0,1$ and $\phi(x,y,\mu)$ is a local $U(1)$ phase (see Appendix~\ref{u1_algebraic} for details). We find that the projective implementation of $C_4$ symmetry imposes a constraint on the translational ``n'' parameters in that the solutions exist only for $n_{T_x}=n_{T_y}=n_T$. We have thus divided the PSG solutions into two parts depending on the value of $n_T$. 

The first case is associated with $n_T=0$ and the corresponding PSG solutions are given by:
\begin{equation}
\label{u1ab_solution_1}
\left.\begin{aligned}
G_{T_x}(x,y,\mu)=&\eta^y_yg_3(\theta_x)\\
G_{T_y}(x,y,\mu)=&g_3(\theta_y)\\
G_\sigma(x,y,\mu)=&\eta^{x\delta_{n_\sigma,0}+y\delta_{n_\sigma,1}}_{\sigma_{xy}}g_3(\theta^\mu_\sigma)(\imath\tau^1)^{n_\sigma}\\
G_{C_4}(x,y,\mu)=&\eta^y_{\sigma_{xy}}\eta^{xy}_yg_3(\theta^\mu_c)(\imath\tau^1)^{n_{C_4}}\\
G_{\mathcal{T}}(x,y,\mu)=&[(-1)^{x+y+\mu}\delta_{n_\mathcal{T},0}\\
&+\eta^{x+y}_tg_3(\theta^\mu_t)\delta_{n_\mathcal{T},1}](\imath\tau^1)^{n_\mathcal{T}}
\end{aligned}\right.
\end{equation}
where $g_3(\theta)=e^{\imath\theta\tau^3}$ is a generic $U(1)$ matrix. $\theta^\mu_c$, $\theta^\mu_\sigma$ and $\theta^\mu_t$ are $U(1)$ phases that do not depend on the unit cell coordinates $(x,y)$, but instead only on the sublattice index $\mu$, while $\theta_{x}$ and $\theta_y$ are instead global phases. Here, we adopt a similar class labelling scheme as used for the $SU(2)$ states, whereby, $\eta_y=+1$ and $\eta_y=-1$ are dubbed $U1A$ and $U1B$ classes, respectively. The gauge inequivalent choices of $g_{C_4,\sigma}$ for diferrent values of $\eta_{C_4,\sigma,\mathcal{T}}$ are given by the first seven rows in Table~\ref{table:u1_psg}. By counting the number of independent `$\eta$' parameters, we conclude that there are a total of 152 $U(1)$ PSGs corresponding to classes $U1A$ and $U1B$.

The second case is associated with $n_T=1$, and the corresponding PSGs are grouped in the $U1C$ class. The translation gauges in this class are given by $G_{T_x}(x,y,\mu)=g_3(\theta_x)\imath\tau^1$ and $G_{T_y}(x,y,\mu)=g_3(\theta_y)\imath\tau^1$, where $\theta_x$ and $\theta_y$ are two global $U(1)$ phases. The corresponding algebraic solutions are listed in Eqs.~\eqref{u1c_solution_1},~\eqref{u1c_solution_2},~\eqref{u1c_solution_3},~\eqref{u1c_solution_4} and~\eqref{u1c_solution_5} and subsequently discussed, making reference to Table~\ref{table:u1_psg} for the explicit form of the gauge transformations.

\begin{equation}
\label{u1c_solution_1}
\left.\begin{aligned}
G_\sigma(x,y,\mu)&=\eta^x_{\sigma_x}\eta^y_{\sigma_y}g_3(\theta^\mu_\sigma)\\
G_{C_4}(x,y,\mu)&=((-1)^m\eta_{\sigma_x}\eta_{\sigma_y})^yg_3((-1)^{x+y}\frac{m\pi}{4}+\theta^\mu_c)\\
G_{\mathcal{T}}(x,y,\mu)&=[(-1)^{x+y+\mu}\delta_{n_\mathcal{T},0}\\
&+\eta^{x+y}_t(-1)^{my}g_3(\theta^\mu_t)\delta_{n_\mathcal{T},1}](\imath\tau^1)^{n_\mathcal{T}}
\end{aligned}\right.
\end{equation}
In Eq.~\eqref{u1c_solution_1}, $m$ is an integer variable whose gauge inequivalent values are $0,1,2,3$. For $n_\mathcal{T}=0$, the sublattice-dependent $g_{3}(\theta^\mu_c)$ and $g_{3}(\theta^\mu_\sigma)$ matrices are given in the eighth row of Table~\ref{table:u1_psg} with the independent parameters as $\eta_1,\eta_{\sigma_x},\eta_{\sigma_y},m$. Hence, the number of PSG classes for $n_\mathcal{T}=n_{C_4}=n_{\sigma}=0, \; n_{T}=1$ are $2^3\times4=32$. Similarly, for $n_\mathcal{T}=1,\;n_{C_4}=n_{\sigma}=0, \; n_{T}=1$, the unit cell representations are given by the ninth row in Table~\ref{table:u1_psg} with 256 PSG classes corresponding to $\eta_1,\eta_c,\eta_{\sigma},\eta_{\sigma_x},\eta_{\sigma_y},\eta_t=\pm1$ and $m=0,1,2,3$.

\begin{equation}
\label{u1c_solution_2}
\left.\begin{aligned}
G_\sigma(x,y,\mu)&=\eta^x_{\sigma_x}\eta^y_{\sigma_y}g_3(\theta^\mu_\sigma)\\
G_{C_4}(x,y,\mu)&=(\eta_{\sigma_x}\eta_{\sigma_y})^yg_3((-1)^{x+y}\phi_c+\theta^\mu_c)\imath\tau^1\\
G_{\mathcal{T}}(x,y,\mu)&=(-1)^{x+y}(-1)^{\mu}g_3(\theta_t)
\end{aligned}\right.
\end{equation}
In Eq.~\eqref{u1c_solution_2}, the $g_{3}$-matrices are  given by the tenth row of Table~\ref{table:u1_psg}, and the parameter $\phi_c$ is an arbitrary $U(1)$ phase. The total number of PSGs are 8.

\begin{equation}
\label{u1c_solution_3}
\left.\begin{aligned}
G_\sigma(x,y,\mu)&=\eta^x_{\sigma_x}\eta^y_{\sigma_y}g_3(\theta^\mu_\sigma)\\
G_{C_4}(x,y,\mu)&=(\eta_{\sigma_x}\eta_{\sigma_y})^yg_3((-1)^{x+y}\frac{m\pi}{4}+\theta^\mu_c)\imath\tau^1,\;m\in\mathbb{Z}\\
G_{\mathcal{T}}(x,y,\mu)&=\eta^{x+y}_t(-1)^{my}g_3(\theta^\mu_t)\imath\tau^1
\end{aligned}\right.
\end{equation}
In Eq.~\eqref{u1c_solution_3}, the $g_{3}$-matrices are  given by the eleventh row of Table~\ref{table:u1_psg} and the number of PSGs are 512.

\begin{equation}
\label{u1c_solution_4}
\left.\begin{aligned}
G_\sigma(x,y,\mu)&=\eta^x_{\sigma_x}\eta^y_{\sigma_y}g_3(\theta^\mu_\sigma)\imath\tau^1\\
G_{C_4}(x,y,\mu)&=\eta^y_{\sigma_x}g_3((-1)^{x+y}\frac{m\pi}{4}+\theta^\mu_c),\;m\in\mathbb{Z}\\
G_{\mathcal{T}}(x,y,\mu)&=[(-1)^{x+y+\mu}\delta_{n_\mathcal{T},0}\\
&+\eta^{x+y}_t(-1)^{my}g_3(\theta^\mu_t)\delta_{n_\mathcal{T},1}](\imath\tau^1)^{n_\mathcal{T}}
\end{aligned}\right.
\end{equation}
In Eq.~\eqref{u1c_solution_4}, the $g_{3}$-matrices are  given by the twelfth and thirteenth rows of Table~\ref{table:u1_psg} for $n_\mathcal{T}=0$ and $n_{\mathcal{T}}=1$, respectively. The numbers of PSGs for these two cases are 64 and 512, respectively.

\begin{equation}
\label{u1c_solution_5}
\left.\begin{aligned}
G_\sigma(x,y,\mu)=&\eta^x_{\sigma_x}\eta^y_{\sigma_y}g_3(\theta^\mu_\sigma)\imath\tau^1\\
G_{C_4}(x,y,\mu)=&((-1)^m\eta_{\sigma_x}\eta_{\sigma_y})^yg_3(\theta^\mu_c)\imath\tau^1,\;m\in\mathbb{Z}\\
G_{\mathcal{T}}(x,y,\mu)=&[(-1)^{x+y+\mu}\delta_{n_\mathcal{T},0}\\
&+\eta^{x+y}_tg_3(\theta^\mu_t)\delta_{n_\mathcal{T},1}](\imath\tau^1)^{n_\mathcal{T}}
\end{aligned}\right.
\end{equation}
The  $g_3$-matrices for the above solutions [Eq.~\eqref{u1c_solution_5}] can be found in the fourteenth and fifteenth rows of the Table~\ref{table:u1_psg} for $n_\mathcal{T}=0$ and~$1$, respectively. The number of PSGs for these two cases are 16 and 256, respectively. 

\begin{center}
\begin{table}[]
\centering
    \caption{Symmetric $U(1)$ spin liquids perturbed around the $SU(2)$ states.}
     \label{tab:u1_su2}
    \begin{ruledtabular}
    \begin{tabular}{cc}
    Parent $SU(2)$ Ansatz &  Perturbed $U(1)$ Ans\"atze \\      
      \hline
 $SU2A01$ & $U1A0_+01_+$, $U1A0_+11_+$, $U1C0^-_+01_x0$,\\
 & $U1C1^{-x}_x100$, $U1A1_-00$, $U1C0^{-}_+00$,\\
 &$U1C0^-_+11_x0$, $U1C0^{-}_+100$, $U1C0^{-}_-100$\\
 $SU2A11$ & $U1A0_-01_+$, $U1A0_-11_+$, $U1C0^-_-01_x0$,  \\
 &$U1A1_-00$, $U1C0^{-}_+000$, $U1C0^{-}_+100$,\\
 &$U1C0^{-}_-100$, $U1C0^+_+11_x0$\\
 $SU2B01$ & $U1B0_+01_+$, $U1B0_+11_+$, $U1C0^-_+01_x1$,\\
 &$U1C0^-_+11_x1$, $U1C1^{x}_{x}101$, $U1B1_-00$, \\
 &$U1C1^{-}_{-}000$ \\
 $SU2B11$ & $U1B0_-01_+$, $U1B0_-11_+$, $U1C0^-_-01_x1$, \\
 &$U1C0^+_+11_x1$, $U1B1_-00$, $U1C1^{-}_{-}000$\\
        \end{tabular}
	\end{ruledtabular}
\end{table}
\end{center}

We thus land up with 1656 PSGs in the $U1C$ class which together with 152 PSGs in $U1A,B$ classes gives rise to a total of 1808 $U(1)$ PSGs. However, upon restricting the mean-field Ans\"atze up to diagonal $J_{d}$ bonds (inside squares), one realizes a small number, i.e., 24 distinct $U(1)$ Ans\"atze discussed below.

\begin{figure*}[t]
\includegraphics[width=1.0\linewidth]{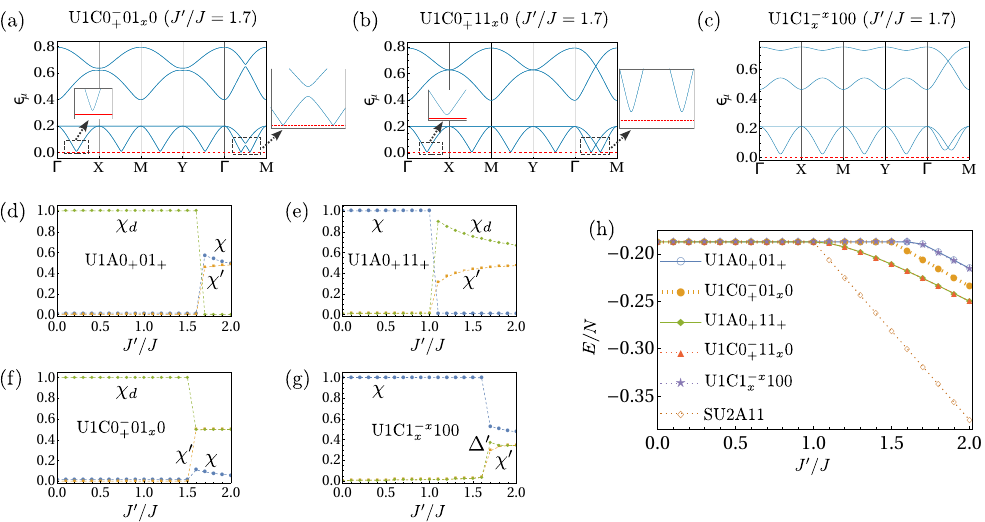}
  \caption{ Dispersion of $U(1)$ spin liquids around $SU2A01$. (a) $U1C0^-_+01_x0$, (b) $U1C0^-_+11_x0$ and (c) $U1C0^{-x}_x100$ at $J'/J=1.7$ with $J_d/J=0.25$. Self-consistently determined mean field parameters for (d)$U1A0_+01_+$ (e) $U1A0_+11_+$ and $U1C0^-_+11_x0$, (f) $U1C0^-_+01_x0$, (g) $U1C1^{-x}_x100$  at $J_d/J=1.0$. (h)  Ground state energies at $J_d/J=1.0$.}
\label{fig:fig5}
\end{figure*}

We now proceed towards systematically enumerating the $U(1)$ mean field Ans\"atze corresponding to the aforementioned PSGs. We divide these states into five groups depending upon how they are connected to the $SU(2)$ states $SU2A01$, $SU2A11$, $SU2B01$ and $SU2B11$ of Sec.~\ref{sec:su2_sl} by means of symmetric perturbations~\cite{Wen-2002}. These connections are summarized in Table~\ref{tab:u1_su2} and schematically represented by the chart in Fig.~\ref{fig:fig8}. We find five $U(1)$ states connected to the $SU2A01$ Ansatz, four $U(1)$ states connected to the $SU2A11$ Ansatz, five $U(1)$ states connected to $SU2B01$ Ansatz and four $U(1)$ states connected to the $SU2B11$ Ansatz. In addition, there is a group of six $U(1)$ states which are multiply connected to different parent $SU(2)$ states depending on the tuning of parameters. Among these 24 $U(1)$ Ans\"atze, there are 8 (4 connected to $SU2A01$ and $SU2B01$ each) which feature non-vanishing mean-field amplitude on $J_d$ bonds. It is thus worth investigating whether the inclusion of mean-field amplitudes on the $J_d$-bonds can potentially lower the energy, compared to the parent $SU(2)$ states, within a self-consistent treatment.

We henceforth adopt the following convention to label the different $U(1)$ states.

\noindent For $U1A/B$ when $n_\mathcal{T}=0$ :
\begin{equation}
\label{u1_notation}
U1(A/B)(n_\sigma)_{\eta_{\sigma_{xy}}}n_{C_4}0
\end{equation}
For $U1A/B$ when $n_\mathcal{T}=1$ :
\begin{equation}
\label{u2_notation}
U1(A/B)(n_\sigma)_{\eta_{\sigma_{xy}}}n_{C_4}1_{\eta_t}
\end{equation}
For $U1C$ when $n_\mathcal{T}=0$ :
\begin{equation}
\label{u3_notation}
U1C(n_\sigma)^{\eta_{\sigma_y}}_{\eta_{\sigma_x}}n_{C_4}0m
\end{equation}
and for $U1C$ when $n_\mathcal{T}=1$ :
\begin{equation}
\label{u4_notation}
U1C(n_\sigma)^{\eta_{\sigma_y}}_{\eta_{\sigma_x}}n_{C_4}1_{\eta_t}m
\end{equation}
We denote `+1' and `-1' values of $\eta$ parameters by `+' and `-', respectively. In the scenario when the $\eta$ parameter can take both signs, we use the index $x$.

\subsubsection{$U(1)$ mean field states around $SU2A01$ }
\label{sec:u00}
We discuss the descendent $U(1)$ Ans\"atze of the $SU2A01$ state. Here, in addition to the fluxes threading the square and octagonal plaquettes $(\phi_1,\phi_2)$, we can further characterize these states by the fluxes threading the triangular plaquettes $(\phi_3,\phi_4,\phi_5)$ [see Fig.~\ref{fig:loop_fig}] when $J_d\neq0$. We will henceforth label these Ans\"atze by the flux pattern $(\phi_1,\phi_2)-(\phi_3,\phi_4,\phi_5)$. For a self-consistent treatment of these Ans\"atze we consider two different values of $J_{d}$, namely, $J_d/J=0.25$ and $J_{d}/J=1$. For $J_d/J=0.25$, we obtain that qualitative properties of all $U(1)$ Ans\"atze are similar to those of the parent $SU2A01$ state.

\begin{equation}
\label{u11}
\left.\begin{aligned}
U1A0_+01_+:\;&u'_{12}=u'_{23}=u'_{34}=u'_{41}=\chi'\tau^3,\;a_3\neq0\\
&u_{31}=u_{42}=\chi\tau^3,\;u^d_{13}=u^d_{24}=\chi^d\tau^3\\
\end{aligned}\right.
\end{equation} 

Due to the presence of non-vanishing $J_d$ bonds, we find that $P_{\phi_{3,4,5}}\propto \tau^3$ which breaks the $SU(2)$ gauge structure down to $U(1)$. For this Ansatz, there is zero flux threading the square, octagonal and all triangular plaquettes, and we thus label it as $(0,0)-(0,0,0)$ state. The $J'/J$-dependency of the self-consistently determined mean-field parameters is shown in Fig.~\ref{fig:fig5}(d) for $J_d/J=1$ suggesting that the $J-VBS$ state is replaced by a $J_d-VBS$ variant~\cite{Maity-2020} within the range $0.1 < J'/J\lesssim 1.6$ leading to a gapped spectrum. Beyond this parameter regime, the properties are similar to the parent $SU(2)$ state. At $J'=0$ and $J_d/J=1$ both $J$-VBS and $J_d$-VBS orders have the same energy.

\begin{equation}
\label{u12}
\left.\begin{aligned}
U1A0_+11_+:\;&u'_{12}=u'_{23}=u'_{34}=u'_{41}=\chi'\tau^3\\
&u_{31}=u_{42}=\chi\tau^3,\;a_\mu=(-1)^\mu a_3\\
&u^d_{13}=-u^d_{24}=\chi^d\tau^3\\
\end{aligned}\right.
\end{equation}

This state has a $(0,0)-(\pi,0,0)$ flux structure. Figure~\ref{fig:fig5}(e) shows the mean field parameters at $J_d/J=1$. For $0.1 < J'/J\lesssim 1.1$, this state is in the $J$-VBS phase while for $J'/J\gtrsim 1.1$ it enters into a plaquette RVB phase with nonvanishing mean field parameters on $J$ and $J_d$ bonds. Hence, we refer to this state as $J'-J_d$-PRVB state, and it features nearly non-dispersive bands.

\begin{equation}
\label{u13}
\left.\begin{aligned}
U1C0^-_+01_x0:\;&u'_{12}=u'_{23}=u'_{34}=u'_{41}=\chi'\tau^3\\
&u^d_{13}=u^d_{24}=(-1)^{x+y}\chi^d\tau^3\\
&u_{31}=u_{42}=\chi\tau^3,\;a_\mu=(-1)^{x+y}a_3\\
\end{aligned}\right.
\end{equation} 

This Ansatz is defined within an eight-site unit cell and features fluxes $(\phi_3,\phi_4,\phi_5)$ in a staggered pattern of $(0,0,0)$ and $(\pi,\pi,\pi)$, and we thus label it as $(0,0)-(0,0,0)/(\pi,\pi,\pi)_{\rm stag}$ state. After a gauge transformation given by Eq.~\eqref{gauge_u13_pairing}, the Ansatz can be cast in the following translationally invariant form

\begin{equation}
\label{u13_pairing}
\left.\begin{aligned}
&u'_{12}=u'_{23}=u'_{34}=u'_{41}=\chi'\tau^3,\; a_\mu=(-1)^\mu a_1\\
&u_{31}=u_{42}=\chi\tau^3,\;u^d_{13}=-u^d_{24}=\Delta^d\tau^1\\
\end{aligned}\right.
\end{equation}  

At $J_d/J=0.25$, the excitation spectrum is gapped for $0 < J'/J\lesssim 1.6$ corresponding to a $J$-VBS state similar to its parent $SU(2)$ state. The spectrum at $J'/J=1.7$ is gapless at isolated $\mathbf{k}$-points on the segment $\overline{\Gamma M}$[see Fig.~\ref{fig:fig5}(a)]. At $J_d/J=1$, the $J$-VBS state is replaced by a $J_d$-VBS structure for $0.1 < J'/J\lesssim 1.6$ as can be seen from Fig.~\ref{fig:fig5}(f), while for $J'/J\gtrsim1.6$ the state tends to exhibit a $J'$-$J_d$-PRVB structure.

\begin{equation}
\label{u14}
\left.\begin{aligned}
U1C0^-_+11_x0:\;&u'_{12}=u'_{23}=u'_{34}=u'_{41}=\chi'\tau^3\\
&u^d_{13}=-u^d_{24}=(-1)^{x+y}\chi^d\tau^3\\
&u_{31}=u_{42}=\chi\tau^3,\;a_\mu=(-1)^{x+y+\mu}a_3\\
\end{aligned}\right.
\end{equation} 

For this Ansatz, the $(\phi_3,\phi_4,\phi_5)$ fluxes are staggered as $(\pi,0,0)$ and $(0,\pi,\pi)$. Hence, we refer to this state as $(0,0)-(\pi,0,0)/(0,\pi,\pi)_{\rm stag}$ state. Similar to the $U1C0^-_+01_x0$ Ansatz, after a gauge transformation [Eq.~\eqref{gauge_u13_pairing}], this state can be brought to the following translationally invariant form

\begin{equation}
\label{u14_pairing}
\left.\begin{aligned}
&u'_{12}=u'_{23}=u'_{34}=u'_{41}=\Delta'\tau^3,\;a_1\neq0\\
&u_{31}=u_{42}=\chi\tau^3,\;u^d_{13}=u^d_{24}=\Delta_d\tau^1.\\
\end{aligned}\right.
\end{equation} 

The spinon excitation spectrum in this state is gapped [see Fig.~\ref{fig:fig5}(b) for $J_d/J=0.25$ and $J'/J=1.7$]. A noticeable feature is the presence of two nodal points at the midpoint of the segment $\overline{\Gamma M}$. The properties at $J_d/J=1$ are qualitatively similar to those of $U1A0_+11_+$ [see Fig.~\ref{fig:fig5}(e)].

\begin{equation}
\label{u15}
\left.\begin{aligned}
U1C1^{-x}_x100:\;&u'_{12}=u'_{34}=\imath\chi'_0\tau^0+(-1)^{x+y}\chi'_3\tau^3\\
&u'_{23}=u'_{41}=-\imath\chi'_0\tau^0+(-1)^{x+y}\chi'_3\tau^3\\
&u_{31}=u_{42}=\chi\tau^3,\;u^d=0,\;a_\mu=0\\
\end{aligned}\right.
\end{equation} 

For this Ansatz, due to vanishing mean field parameter on $J_d$ bonds, the state is characterized only by the fluxes $(\phi_1,\phi_2)$. While the flux $\phi_1$ threading the square is zero, the $\phi_2$ flux in the octagonal plaquettes appears in a staggered pattern of $\phi$ and $-\phi$, where $\phi=4\tan^{-1}(\chi'_3/\chi'_0)$. Therefore, we refer to this state as $(0,\phi)/(0,-\phi)_{stag}$. Employing the gauge transformation in Eq.~\eqref{gauge_u15_pairing} the Ansatz takes the following translationally invariant form

\begin{equation}
\label{u15_pairing}
\left.\begin{aligned}
&u'_{12}=u'_{23}=u'_{34}=u'_{41}=\chi'\tau^3+\Delta'\tau^1\\
&u_{31}=u_{42}=\chi\tau^3,\;u^d=0,\;a_\mu=0\\
\end{aligned}\right.
\end{equation} 

For this state, the spectrum is gapped [shown in Fig.~\ref{fig:fig5}(c) for $J_d/J=0.25$ and $J'/J=1.7$], and there are nodal points at the midpoint of the segment $\overline{\Gamma M}$. The self-consistently determined properties are not noticeably different compared to its parent $SU(2)$ state [see Fig.~\ref{fig:fig5}(g)].

In Fig.~\ref{fig:fig5}(h), we compare the ground state energies of the aforementioned $U(1)$ states at $J_d/J=1$. We find that the inclusion of $J_d$ bonds (responsible for the $U(1)$ gauge structure) lowers the ground state energy of the $U1A0_+11_+$, $U1C0^-_+11_x0$ and $U1C0^-_+01_x0$ Ans\"atze compared to their parent $SU(2)$ states. It is interesting to note that while the above three descendant $U(1)$ Ans\"atze give rise to $J'$-$J_d$-PRVB order for $J'/J\geq 1.1$, the $SU2A11$ and $SU2B11$ states remain energetically preferable. For $J'/J\leq1$, the $J-VBS$ phase given by the Ans\"atze $U1A0_+11+$, $U1C0^-_+11_x0$, $U1C1^{-x}_x100$, and the $J_d$-VBS phase given by the Ans\"atze $U1A0_+01+$, $U1C0^-_+01_x0$ have the same energy at $J_d/J=1$. Thus, the $J_d/J=1$ line serves as a phase boundary between these two VBS phases.

\begin{figure*}
\includegraphics[width=1.0\linewidth]{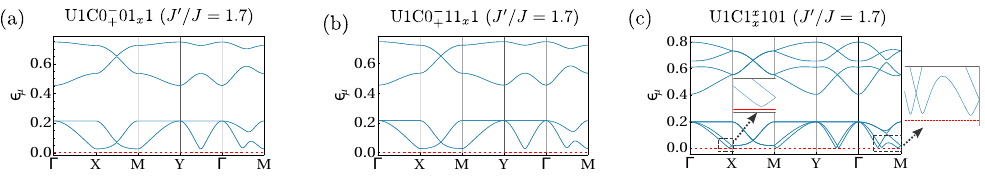}
  \caption{ Spinon spectrums of the Ans\"atze connected to the $SU2B01$ state (a) $U1C0^-_+01_x1$, (b) $U1C0^-_+11_x1$ and (c)  $U1C1^x_x101$ at $J'/J=1.7$, $J_d/J=0.25$.}
\label{fig:fig6}
\end{figure*}

\subsubsection{$U(1)$ mean field states around $SU2A11$}
\label{sec:u11}
In the following, we enlist the four $U(1)$ states which descend from the parent $SU2A11$ state characterized by the fluxes $(\phi_1,\phi_2)=(\pi,\pi)$. For all these Ans\"atze, the mean-field amplitudes vanish on $J_d$ bonds, and the $SU(2)$ IGG is broken down to $U(1)$ only due to the presence of on-site terms~\cite{Wen-2002}.

\begin{equation}
\label{u21}
\left.\begin{aligned}
U1A0_-01_+:\;&u'_{12}=u'_{23}=-u'_{34}=u'_{41}=\chi'\tau^3\\
&u_{31}=u_{42}=\chi\tau^3,\;a_3\neq0\\
\end{aligned}\right.
\end{equation}

\begin{equation}
\label{u22}
\left.\begin{aligned}
U1A0_-11_+:\;&u'_{12}=u'_{23}=-u'_{34}=u'_{41}=\chi'\tau^3\\
&u_{31}=u_{42}=\chi\tau^3,\;a_\mu=(-1)^\mu a_3\\
\end{aligned}\right.
\end{equation}

\begin{equation}
\label{u23}
\left.\begin{aligned}
U1C0^-_-01_x0:\;&u'_{12}=u'_{23}=-u'_{34}=u'_{41}=\chi'\tau^3\\
&u_{31}=u_{42}=\chi\tau^3,\;a_\mu=(-1)^{x+y}a_3\\
\end{aligned}\right.
\end{equation}

\begin{equation}
\label{u24}
\left.\begin{aligned}
U1C0^+_+11_x0:\;&u'_{12}=u'_{23}=-u'_{34}=u'_{41}=\chi'\tau^3\\
&u_{31}=u_{42}=\chi\tau^3,\;a_\mu=(-1)^{x+y+\mu}a_3\\
\end{aligned}\right.
\end{equation}
These states differ from their (gapped) parent $SU(2)$ state only by the presence of a chemical potential, which does not play a role in the self-consistent mean-field results, since it does not alter the ground state energy and the nature of the excitations. As a result, these Ans\"atze effectively behave in a similar manner as their parent $SU(2)$ state, i.e., $SU2A11$. Nevertheless, it is worth mentioning that the inclusion of third nearest-neighbor amplitudes (connecting sites 6 and 8 in Fig.~\ref{fig:loop_fig}) allows one to break the $SU(2)$ IGG down to $U(1)$ (without the need of a chemical potential). Further analysis of this third neighbour extension is beyond the scope of the current work.

\begin{figure*}[t]
\includegraphics[width=1.0\linewidth]{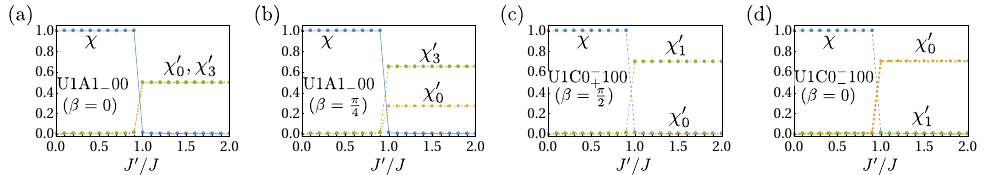}
  \caption{ Self-consistently determined mean field parameters for (a) $\beta=0$ (b) $\beta=\pi/4$ for $U1A1_-00$ and $U1B1_-00$. (c) $U1C0^-_+100$ at $\beta=\pi/2$ (d) $U1C1^-_-100$ at $\beta=0$.}
\label{fig:fig7}
\end{figure*}

\subsubsection{$U(1)$ mean field states around $SU2B01$}
\label{sec:u01}
We identify five $U(1)$ Ans\"atze which are connected to the $SU2B01$ state characterized by the fluxes $(\phi_1,\phi_2)=(0,\pi)$. The mean field parameters in these states are the same as those corresponding to the states listed in Sec.~\ref{sec:u00} with the only difference being that the sign of the mean-field amplitudes on the vertical $J$ bonds alternate along the $x$-direction,

\begin{equation}
\label{eq:doubling}
u_{(x,y,4),(x,y+1,2)}=(-1)^xu_{42}.
\end{equation}

Therefore, in these Ans\"atze, the $\phi_{1}$ and $(\phi_3,\phi_4,\phi_5)$ fluxes remain the same compared to the $U(1)$ Ans\"atze descending from the $SU2A01$ state except for the fact that there is now an additional $\pi$ flux threading the octagons. We label these states as $U1B0_+01_+$, $U1B0_+11_+$, $U1C0^-_+01_x1$, $U1C0^-_+11_x1$ and U1C1$_{x}^{x}101$, and whose mean-field ans\"atze are given by Eqs.~\eqref{u11}, \eqref{u12}, \eqref{u13}, \eqref{u14} and \eqref{u15} [combined with Eq.~\eqref{eq:doubling}], respectively.  \\
   The self consistently determined mean-field parameters and ground state energies at both $J_d/J=0.25$ and $J_d/J=1$ are the same as the $U(1)$ states descending from the $SU2A01$ counterpart [see Figs.~\ref{fig:fig5}(d)-(h)].
   
We now discuss the band structure at $J_d/J=0.25$ for the aforementioned states. The spectrum of $U1B0_+01_+$ and $U1B0_+11_+$ Ans\"atze is similar to their parent $SU(2)$ state $SU2B01$. The spectrum of the $U1C0^-_+01_x1$ is gapped [see Fig.~\ref{fig:fig6}(a), shown for $J'/J=1.7$]. The spectrum of $U1C0^-_+11_x1$ is shown in Fig.~\ref{fig:fig6}(b) which is also gapped and consists of two nodal points at the midpoint of the segment $\overline{XM}$. The spectrum of $U1C1^{x}_{x}101$ is also observed to be gapped when $J'/J\geq1.7$ [shown for $J'/J=1.7$ in Fig.~\ref{fig:fig6}(c)]. The noticeable features are the presence of nodal lines within the segment $\overline{XM}$ along with two nodal points at the midpoint of this segment. At $J_d/J=1$, all the above states have a gapped spectrum, and for $J'=0$ the $J$-VBS and $J_d$-VBS have the same energy.

\subsubsection{$U(1)$ mean field states around $SU2B11$}
\label{sec:u10}
In this section, we discuss the four $U(1)$ Ans\"atze which are connected to the $SU2B11$ state characterized by the fluxes $(\phi_1,\phi_2)=(\pi,0)$. The mean field parameters of these Ans\"atze are the same as those for states listed in Sec.~\ref{sec:u11} except the alternating sign structure of the vertical $J$ bonds given in Eq.~\eqref{eq:doubling}. These four states are labelled as $U1B0_-01_+$, $U1B0_-11_+$, $U1C0^-_-01_x1$ and $U1C0^+_+11_x1$, with mean field parameters given by Eqs.~\eqref{u21}, \eqref{u22}, \eqref{u23} and \eqref{u24} [combined with Eq.~\eqref{eq:doubling}, respectively. For all four aforementioned mean-field Ans\"atze, using the same argument as given in Sec.~\ref{sec:u11}, we may conclude that the spinon spectra remain gapped similar to the parent $SU(2)$ state, i.e., $SU2B11$ [see Fig.~\ref{fig:fig4}(l)], when self-consistent calculations are performed.

\subsubsection{$U(1)$ mean field states connected to multiple $SU(2)$ Ans\"atze}
Here, we present those mean field Ans\"atze which feature a nontrivial ($\neq~0~{\rm or}~\pi$) flux through the square or octagonal plaquettes and can thus be connected to multiple parent $SU(2)$ states, depending on the value of certain parameters. We find a total of six such mean-field Ans\"atze.

\begin{equation}
\label{u52}
\left.\begin{aligned}
U1A1_-00:\;&u'_{12}=u'_{34}=\imath\chi'_0\tau^0+\chi'_3\tau^3\\
&u'_{23}=u'_{41}=(\imath\chi'_0\tau^0+\chi'_3\tau^3)g_3(\beta)\\
&u_{31}=u_{42}=\chi\tau^3\\
\end{aligned}\right.
\end{equation}
Here, $\beta$ is a tuning parameter, depending on whose value the above state can be linked to the parent $SU(2)$ states, $SU2A01$ and $SU2A11$. The flux structure is $(\phi_1,\phi_2)=(\phi,-\phi)$ with $\phi=-4\tan^{-1}(\chi'_3/\chi'_0)+2\beta$. The connection with $SU2A01$ and $SU2A11$ ans\"atze can be readily seen if one chooses $2\beta=4\tan^{-1}(\chi'_3/\chi'_0)$ and $2\beta=4\tan^{-1}(\chi'_3/\chi'_0)+\pi$, respectively.

\begin{equation}
\label{u57}
\left.\begin{aligned}
U1B1_-00:\;&u'_{12}=u'_{34}=\imath\chi'_0\tau^0+\chi'_3\tau^3\\
&u'_{23}=u'_{41}=(\imath\chi'_0\tau^0+\chi'_3\tau^3)g_3(\beta)\\
&u_{31}=u_{42}=\chi\tau^3\\
&u_{(x,y,4),(x,y+1,2)}=(-1)^xu_{42}\\
\end{aligned}\right.
\end{equation}
Here, the mean-field parameters are the same as those of $U1A1_-00$. The only difference is in the sign pattern of $u_{42}$ which alternates along the $x$ direction. The flux structure of this state is $(\phi_1,\phi_2)=(\phi,-\phi+\pi)$ with $\phi=-4\tan^{-1}(\chi'_3/\chi'_0)+2\beta$.

\begin{equation}
\label{u53}
\left.\begin{aligned}
&U1C0^{-}_+100:\\
&u_{31}=u_{42}=\chi\tau^3,\;u'_{12}=u'_{34}=\imath\chi'_0\tau^0+(-1)^{x+y}\chi'_3\tau^3\\
&u'_{23}=u'_{41}=(\imath\chi'_0\tau^0+(-1)^{x+y}\chi'_3\tau^3)g_3((-)^{x+y}\beta)\\
\end{aligned}\right.
\end{equation}

\begin{equation}
\label{u54}
\left.\begin{aligned}
&U1C0^{-}_-100:\\
&u_{31}=u_{42}=\chi\tau^3,\;u'_{12}=-u'_{34}=\imath\chi'_0\tau^0+(-)^{x+y}\chi'_3\tau^3\\
&u'_{23}=u'_{41}=(\imath\chi'_0\tau^0+(-)^{x+y}\chi'_3\tau^3)g_3((-)^{x+y}\beta)\\
\end{aligned}\right.
\end{equation}
\begin{equation}
\label{u51}
\left.\begin{aligned}
U1C0^{-}_+000:\;&u_{31}=u_{42}=\chi\tau^3,\;u'_{12}=u'_{34}=\chi\tau^3\\
&u'_{23}=u'_{41}=\chi'_3\tau^3g_3((-1)^{x+y}\beta)\\
\end{aligned}\right.
\end{equation}
The above three Ans\"atze given in Eqs.~\eqref{u53},~\eqref{u54},~\eqref{u51} display a staggered pattern of  fluxes, namely $(\phi_1,\phi_2)=(\phi,-\phi)/(-\phi,\phi)_{\rm stag}$ with $\phi=4\tan^{-1}(\chi'_3/\chi'_0)+2\beta$, $\phi=4\tan^{-1}(\chi'_3/\chi'_0)+2\beta+\pi$ and $\phi=2\beta$, respectively. Upon performing a gauge transformation [Eq.~\eqref{gauge_beta_pairing}] these Ans\"atze take the translationally invariant form
\begin{equation}
\label{u53_pairing}
\left.\begin{aligned}
&u_{31}=u_{42}=\chi\tau^3,\;u'_{12}=u'_{34}=\imath\chi'_0\tau^0+\chi'_1\tau^1\\
&u'_{23}=u'_{41}=(\imath\chi'_0\tau^0+\chi'_1\tau^1)g_1(\beta)\\
\end{aligned}\right.
\end{equation}
\begin{equation}
\label{u54_pairing}
\left.\begin{aligned}
&u_{31}=u_{42}=\chi\tau^3,\;u'_{12}=-u'_{34}=\imath\chi'_0\tau^0+\chi'_1\tau^1\\
&u'_{23}=u'_{41}=(\imath\chi'_0\tau^0+\chi'_1\tau^1)g_1(\beta)\\
\end{aligned}\right.
\end{equation}
\begin{equation}
\label{u51_pairing}
\left.\begin{aligned}
&u_{31}=u_{42}=\chi\tau^3,\;u'_{12}=u'_{34}=\chi'_3\tau^3\\
&u'_{23}=u'_{41}=g_1(\beta)\chi'_3\tau^3\\
\end{aligned}\right.,
\end{equation}
respectively.

\begin{equation}
\label{u56}
\left.\begin{aligned}
U1C1^{-}_-000:\;&u_{31}=u_{42}=\chi\tau^3,\;u'_{12}=u'_{34}=\chi'\tau^3\\
&u'_{23}=u'_{41}=\chi'_3\tau^3g_3((-1)^{x+y}\beta)\\
&u_{(x,y,4),(x,y+1,2)}=(-1)^xu_{42}\\
\end{aligned}\right. 
\end{equation}
This Ansatz has the same mean-field parameters as those of $U1C0^-_+000$ [Eq.~\eqref{u51}] but with the difference that the vertical $J$ bonds alternate in sign along the $x$ direction.
The resulting flux structure is $(\phi_1,\phi_2)=(\phi,-\phi+\pi)/(-\phi,\phi+\pi)_{\rm stag}$ with $\phi=2\beta$. After the gauge transformation in Eq.~\eqref{gauge_beta_pairing} it can be recast as
\begin{equation}
\label{u56_pairing}
\left.\begin{aligned}
&u_{31}=u_{42}=\chi\tau^3,\;u'_{12}=u'_{34}=\chi'_3\tau^3\\
&u'_{23}=u'_{41}=g_1(\beta)\chi'_3\tau^3\\
&u_{(x,y,4),(x,y+1,2)}=(-1)^xu_{42}\\
\end{aligned}\right.
\end{equation}

\begin{table}
    \caption{Projective representation matrices $g_{c_4},\;g_\sigma\;{\rm and} \;g_T$ for the $\mathds{Z}_2$ PSG solutions. When two distinct solutions are possible, we write both choices for the matrices, separated by $/$ symbol.}
     \label{psg_rep}
    \begin{tabular}{cccc}
    \hline
    \hline
    PSG&  \textbf{$g_\tau(\mu)$} & \textbf{$g_{C_4}(\mu)$} & \textbf{$g_{\sigma}(\mu)$} \\      
      \hline
    1 &  $\imath\tau^2/\tau^0$ & $\tau^0$ & $\tau^0$ \\
    2 &  $\imath\tau^2/\tau^0$ & $\tau^0$ & $\imath\tau^1$ \\
    3 &  $\imath\tau^2$ & $\tau^0$ & $\imath\tau^2$ \\
    4 &  $\imath\tau^2/\tau^0$ & $\tau^0$ & $(-1)^{\mu+1}\tau^0$ \\
    5 &  $\imath\tau^2/\tau^0$ & $\tau^0$ & $(-1)^{\mu+1}\imath\tau^1$ \\
    6 &  $\imath\tau^2$ & $\tau^0$ & $(-1)^{\mu+1}\imath\tau^2$ \\
      7 &  $\imath\tau^2/\tau^0$ & $\tau^0,\tau^0,\tau^0,-\tau^0$ & $\imath\tau^1,\imath\tau^1,-\imath\tau^1,\imath\tau^1$ \\
    8 &  $\imath\tau^2/\tau^0$ & $\tau^0,\tau^0,\tau^0,-\tau^0$ & $\imath\tau^1,-\imath\tau^1,-\imath\tau^1,-\imath\tau^1$ \\
    9 &  $\imath\tau^2$ & $\tau^0,\tau^0,\tau^0,-\tau^0$ & $\imath\tau^2,-\imath\tau^2,-\imath\tau^2,-\imath\tau^2$ \\
    10 &  $\imath\tau^2$ & $\tau^0,\tau^0,\tau^0,-\tau^0$ & $-\imath\tau^2,-\imath\tau^2,\imath\tau^2,-\imath\tau^2$ \\
    11 &  $\imath\tau^2/\tau^0$ & $\tau^0,\tau^0,\tau^0,-\tau^0$ & $-\tau^0,-\tau^0,\tau^0,-\tau^0$ \\
    12 &  $\imath\tau^2/\tau^0$ & $\tau^0,\tau^0,\tau^0,-\tau^0$ & $\tau^0,-\tau^0,-\tau^0,-\tau^0$ \\
    13 &  $(-1)^{\mu+1}\imath\tau^2$ & $\tau^0$ & $\tau^0$ \\
    14 &  $(-1)^{\mu+1}\imath\tau^2$ & $\tau^0$ & $\imath\tau^1$ \\
    15 &  $(-1)^{\mu+1}\imath\tau^2$ & $\tau^0$ & $\imath\tau^2$ \\
    16 &  $(-1)^{\mu+1}\imath\tau^2$ & $\tau^0$ & $(-1)^{\mu+1}\tau^0$ \\
    17 &  $(-1)^{\mu+1}\imath\tau^2$ & $\tau^0$ & $(-1)^{\mu+1}\imath\tau^1$ \\
    18 &  $(-1)^{\mu+1}\imath\tau^2$ & $\tau^0$ & $(-1)^{\mu+1}\imath\tau^2$ \\
    19 &  $(-1)^{\mu+1}\imath\tau^2$ & $\tau^0,\tau^0,\tau^0,-\tau^0$ & $\imath\tau^1,\imath\tau^1,-\imath\tau^1,\imath\tau^1$ \\
    20 &  $(-1)^{\mu+1}\imath\tau^2$ & $\tau^0,\tau^0,\tau^0,-\tau^0$ & $\imath\tau^1,-\imath\tau^1,-\imath\tau^1,-\imath\tau^1$ \\
    21 &  $(-1)^{\mu+1}\imath\tau^2$ & $\tau^0,\tau^0,\tau^0,-\tau^0$ & $\imath\tau^2,-\imath\tau^2,-\imath\tau^2,-\imath\tau^2$ \\
    22 &  $(-1)^{\mu+1}\imath\tau^2$ & $\tau^0,\tau^0,\tau^0,-\tau^0$ & $-\imath\tau^2,-\imath\tau^2,\imath\tau^2,-\imath\tau^2$ \\
    23 &  $(-1)^{\mu+1}\imath\tau^2$ & $\tau^0,\tau^0,\tau^0,-\tau^0$ & $-\tau^0,-\tau^0,\tau^0,-\tau^0$ \\
    24 &  $(-1)^{\mu+1}\imath\tau^2$ & $\tau^0,\tau^0,\tau^0,-\tau^0$ & $\tau^0,-\tau^0,-\tau^0,-\tau^0$ \\
    25 &  $\tau^0$ & $\tau^0,\imath\tau^3,\tau^0,\imath\tau^3$ & $\imath\tau^1,-\imath\tau^2,\imath\tau^1,-\imath\tau^2$ \\
    26 &  $\tau^0$ & $\tau^0,\imath\tau^3,\tau^0,\imath\tau^3$ & $\imath\tau^1,\imath\tau^2,\imath\tau^1,\imath\tau^2$ \\
    27 &  $\tau^0$ & $\tau^0,\imath\tau^3,\tau^0,\imath\tau^3$ & $\imath\tau^3,-\tau^0,-\imath\tau^3,\tau^0$ \\
    28 &  $\tau^0$ & $\tau^0,\imath\tau^3,\tau^0,\imath\tau^3$ & $-\imath\tau^3,\tau^0,\imath\tau^3,-\tau^0$ \\
    29 &  $\tau^0$ & $\tau^0,\imath\tau^3,\tau^0,\imath\tau^3$ & $\tau^0,\imath\tau^3,-\tau^0,-\imath\tau^3$ \\
    30 &  $\tau^0$ & $\tau^0,\imath\tau^3,\tau^0,\imath\tau^3$ & $-\tau^0,-\imath\tau^3,\tau^0,\imath\tau^3$ \\
    \hline
    \hline
        \end{tabular}
\end{table}

We performed a self-consistent analysis for different choices of $\beta$ to study which values yield the lowest energy. Interestingly, we find that the ground state energies of the $U1A1_-00$ and $U1B1_-00$ states are independent of the $\beta$ parameter and equal to those of the $SU2A11$ and $SU2B11$ Ans\"atze. The self-consistently determined mean field parameters [shown in Fig.~\ref{fig:fig7}(a) and Fig.~\ref{fig:fig7}(b) for $\beta=0$ and $\beta=\pi/4$, respectively] are such that $2\beta=-4\tan^{-1}(\chi'_3/\chi'_0)+\pi$ is satisfied. Thus, in the ground state, the $U1A1_-00$ and $U1B1_-00$ Ans\"atze effectively reduce to the $SU2A11$ and $SU2B11$ states, respectively, which realize the $J$-VBS phase for $0 < J'/J\lesssim 1.0$, while a transition to the $P$RVB phase occurs at $J'/J\approx1.1$. A similar scenario occurs for the states $U1C0^-_+100$ and $U1C0^-_-100$ but at $(\beta=\pi/2,\phi\approx\pi/2)$ and $(\beta=0,\phi\approx0)$, respectively. The corresponding mean-field parameters are shown in Fig.~\ref{fig:fig7}(c) and Fig.~\ref{fig:fig7}(d), respectively. The remaining two states $U1C0^-_+000$ and $U1C0^-_-000$, at $\beta=\pi$, reduce to $SU2A11$ and $SU2B11$ states, respectively. 

\subsection{$\mathds{Z}_2$ Spin liquids}
\label{z2_sl}

Finally allowing for additional pairing terms along with the hopping terms, the Ans\"atze with the lowest invariant gauge group $\mathds{Z}_2$, i.e., $\mathcal{G}=\pm\tau^0$ can be obtained. In this section, we list the $\mathds{Z}_2$ spin-liquid Ans\"atze. The $\mathds{Z}_2$ character of the IGG implies that in the projective construction the identity is defined up to a global sign factor. The algebraic solutions of $\mathds{Z}_2$ PSG are given by [see Appendix~\ref{z2_algebraic} for details]
\begin{equation}
\label{gauge}
\left.\begin{aligned}
&G_{T_x}(x,y,\mu)=\eta^y_y\tau^0,\;G_{T_y}(x,y,\mu)=\tau^0,\\
&G_{c_4}(x,y,\mu)=(\eta_{\sigma_x},\eta_{\sigma_y})^y\eta^{xy}_{y}g_{c_4}(\mu),\\
&G_{\sigma}(x,y,\mu)=\eta^x_{\sigma_x}\eta^y_{\sigma_y}g_{\sigma}(\mu),\\
&G_{\mathcal{T}}(x,y,\mu)=\eta^{x+y}_{\mathcal{T}}g_{T}(\mu).\\
\end{aligned}\right.
\end{equation}
We note that in the above each of the parameters $\eta_y,\eta_{\sigma_x},\eta_{\sigma_y},\eta_\mathcal{T}$ can take the values $\pm1$, and there are 38 inequivalent sets of $g_S(\mu)$ [$g_{c_4},\;g_\sigma\; {\rm and} \;g_\mathcal{T}$] matrices which satisfy Eq.~\eqref{gauge} and are given in Table~\ref{psg_rep}. Also, from  Eq.~\eqref{gauge} we note that there are four $\eta$ parameters each of which can take values $\pm1$ and thus there are $2^4\times38=608$ algebraic PSGs on the square-octagon lattice. However, it is important to note that those PSGs for which time-reversal acts trivially, i.e., $\eta_{\mathcal{T}}=1$ and $g_{\mathcal{T}}=\mathds{1}$, the mean-field Ansatz must vanish since $\mathcal{T}(u_{ij})=-u_{ij}$. This leads to  $38\times2^4 - 14\times2^4=384$ non-vanishing Ans\"atze which need to be considered.  

We can first split our Ans\"atze based on the value of the $\eta_{y}$ parameter. The case $\eta_{y}=1$ corresponds to translationally invariant Ans\"atze (henceforth referred to as Class A), while those corresponding to  $\eta_{y}=-1$ require a doubling of the unit-cell at the mean-field level (henceforth referred to as Class B). We shall label the  mean-field states with the notation: 
\begin{equation}
\label{notation}
Z\;(\#PSG)\;(Class)\;\eta_\mathcal{T}\eta_{\sigma_x}\eta_{\sigma_y}
\end{equation}
Here Z stands for IGG $\mathds{Z}_2$. In place of $\eta_\mathcal{T},\eta_{\sigma_x},\eta_{\sigma_y}$ we shall use `$0/1$' for $\eta_S=\pm1$ respectively. If any Ansatz corresponds to both signs of a given $\eta_S$ parameter, we shall write `x' in its place. In the following, we list all the $\mathds{Z}_2$ mean-field states which have non-vanishing $u_{ij}$ matrix on both $J$ and $J'$-bonds. 

\begin{figure*}[t]
\includegraphics[width=0.85\linewidth]{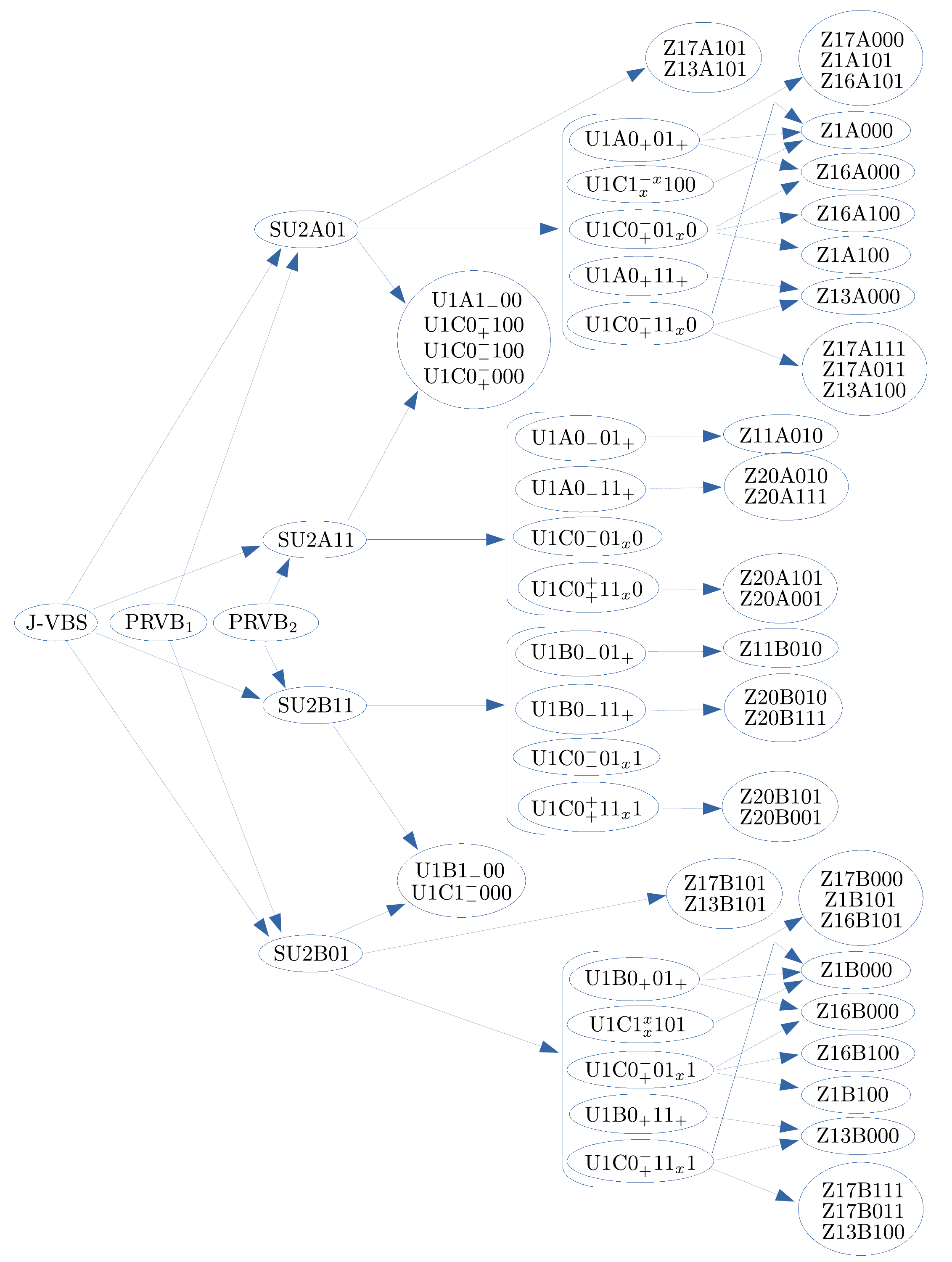}
  \caption{The hierarchy of mean-field Ans\"atze with different gauge structures}
\label{fig:fig8}
\end{figure*}

In the following, we split the discussion of $\mathds{Z}_2$ states in different subsections, depending on the parent $SU(2)$ Ansatz. Many of these states are also descendant
of some of the $U(1)$ Ans\"atze, as summarized in Table~\ref{tab:z2_u1} and schematically shown in the chart of Fig.~\ref{fig:fig8}.
We enlist all $\mathds{Z}_2$ Ans\"atze along with the nature of their spinon excitation spectra based on the self-consistently determined mean-field parameters. More specifically, we focus on those Ans\"atze whose parent states are gapless to verify whether the addition of paring terms opens a gap in the spinon band structure. For example, the parent $SU(2)$ state and two of its $U(1)$ descendants, $U1A0_+01_+$ and $U1A0_+11_+$, display an extended Fermi surface in the spinon dispersion. An important question is whether breaking the IGG down to $\mathds{Z}_{2}$ leads to a removal of the Fermi surface, either by a complete opening of a gap, or by reducing it to being gapless at isolated {\bf k} points. In our analysis, we set $J_d=0.25$ and consider different values of $J'/J$.

\subsubsection{$\mathds{Z}_2$ mean field states around $(0,0)$ flux state }
\label{sec:z00}
We find a total of thirteen $\mathds{Z}_2$ Ans\"atze which appear in the vicinity of the $SU2A01$ state. Most of them can be connected to some $U(1)$ Ans\"atze descending from the $SU2A01$ state. Additionally, we find two direct $\mathds{Z}_2$ descendants of the parent $SU(2)$ state. 

Let us first consider the following two direct $\mathds{Z}_2$ descendants of the parent $SU(2)$ state:
\begin{equation}
\label{z11}
\left.\begin{aligned}
Z17A101:\;&u'_{12}=u'_{34}=\chi'\tau^3+(-)^{x+y}\Delta'\tau^1\\
&u'_{23}=u'_{41}=\chi'\tau^3-(-)^{x+y}\Delta'\tau^1,\\
&u^d_{13}=-u^d_{24}=(-)^{x+y}\Delta_{d}\tau^1\\
&u_{31}=u_{42}=\chi\tau^3,\;a_\mu=(-)^{x+y+\mu}a_1\\
\end{aligned}\right.
\end{equation}

\begin{equation}
\label{z12}
\left.\begin{aligned}
Z13A101:\;&u'_{12}=u'_{34}=u'_{23}=u'_{41}=\chi'\tau^3,\;u^d_{13}=\\
&u^d_{24}=(-)^{x+y}(\Delta_{d,1}\tau^1+\Delta_{d,2}\tau^2)\\
&u_{31}=u_{42}=\chi\tau^3,\;a_\mu=(-)^{x+y}a_1\\
\end{aligned}\right.
\end{equation}

The ansatz $Z17A101$ has a non-trivial flux $(\phi_1,\phi_2)=(\phi,-\phi)$ threading the square and octagon plaquette. The self-consistently determined mean-field parameters are such that $\phi$ becomes $\pi$ and thus it turns into an effective the $(\pi,\pi)$ flux state, which behaves as dimerized phases  ($PRVB_2$ phase when $J'>J$ and $J$-VBS when $J'\leq J$). On the other hand for $Z13A101$, we find a $J$-VBS ground state for $J'/J<1.7$, and a gapless spin liquid otherwise. However, contrary to the parent $SU(2)$ state, which displays an extended Fermi surface,  the spinon dispersion of the $Z13A101$ state for $J'/J\geq1.7$ is gapless only at isolated $\mathbf{k}$-points.

The following five $\mathds{Z}_2$ states descend from the $U1A0_+01_+$ Ansatz
\begin{equation}
\label{z13}
\left.\begin{aligned}
Z17A000:\;&u'_{12}=u'_{34}=\chi'\tau^3+\Delta'\tau^1\\
&u'_{23}=u'_{41}=\chi'\tau^3-\Delta'\tau^1,\;a_3\neq0\\
&u_{31}=u_{42}=\chi\tau^3,\;u^d_{13}=u^d_{24}=\chi_d\tau^3\\
\end{aligned}\right.
\end{equation}

\begin{equation}
\label{z14}
\left.\begin{aligned}
Z1A101:\;&u'_{12}=u'_{34}=u'_{23}=u'_{41}=\chi'\tau^3,\;u^d_{13}=\\
&u^d_{24}=(\chi_{d}\tau^3+(-)^{x+y}\Delta_{d}\tau^1)\\
&u_{31}=u_{42}=\chi\tau^3,\;a_\mu=(a_3,(-)^{x+y}a_1)\\
\end{aligned}\right.
\end{equation}

\begin{equation}
\label{z15}
\left.\begin{aligned}
Z16A101:\;&u'_{12}=u'_{34}=u'_{23}=u'_{41}=\chi'\tau^3,\\
&u_{31}=u_{42}=\chi\tau^3,\;u^d_{13}=u^d_{24}=\chi_d\tau^3\\
&a_\mu= (a_3,(-1)^{x+y+\mu} a_1)\\
\end{aligned}\right.
\end{equation}

\begin{equation}
\label{z16}
\left.\begin{aligned}
Z1A000:\;&u'_{12}=u'_{23}=u'_{34}=u'_{41}=\chi'\tau^3+\Delta'\tau^1\\
&u_{31}=u_{42}=\chi\tau^3,\;u^d_{13}=\chi_d\tau^3+\Delta_d\tau^1\\
&u^d_{24}=\chi_d\tau^3+\Delta_d\tau^1,\;a_\mu=(a_3,a_1)\\
\end{aligned}\right.
\end{equation}

\begin{equation}
\label{z17}
\left.\begin{aligned}
Z16A000:\;&u'_{12}=u'_{23}=u'_{34}=u'_{41}=\chi'\tau^3\\
&u_{31}=u_{42}=\chi\tau^3,\;u^d_{13}=\chi_d\tau^3+\Delta_d\tau^1\\
&u^d_{24}=\chi_d\tau^3-\Delta_d\tau^1,\;a_\mu= (a_3,(-1)^\mu a_1)\\
\end{aligned}\right.
\end{equation}
Similar to the $Z17A101$ state, the presence of a nontrivial $\phi_1$ in $Z17A000$ allows the mean-field approach to minimize the energy by transforming this Ansatz into (the dimerized phases of) the $SU2A11$ state. On the other hand,  
$Z1A101$, $Z1A000$, $Z16A101$ and $Z16A000$ yield a $J-VBS$ state when $J'/J<1.7$. For $J'/J\geq1.7$, $Z1A101$ and $Z1A000$ give gapped spin liquid states, while $Z16A101$ and $Z16A000$ are spin liquids which are gapless at two isolated $\mathbf{k}$-points on the $\overline{\Gamma M}$ segment. Thus, in this case also, breaking the IGG from $U(1)$ down to $\mathds{Z}_2$ removes the Fermi surface.

Now let us discuss the descendants of  $U1C0^-_+01_x0$ where the extended Fermi surface of the parent $SU(2)$ state disappears due to the inclusion of amplitudes on $J_{d}$ bonds, except at isolated $\mathbf{k}$-points on the $\overline{\Gamma M}$ segment. $U1C0^-_+01_x0$ has three $\mathds{Z}_2$ descendants in total. Among them, two are listed in the following and the other one ($Z16A000$) is a common descendant of $U1A0_+01+$ (previously discussed).
\begin{equation}
\label{z18}
\left.\begin{aligned}
Z16A100:\;&u'_{12}=u'_{23}=u'_{34}=u'_{41}=\chi'\tau^3\\
&u_{31}=u_{42}=\chi\tau^3,\;a_\mu=(-1)^\mu a_1\\
&u^d_{13}=-u^d_{24}=\Delta_{d,1}\tau^1+\Delta_{d,2}\tau^2\\
\end{aligned}\right.
\end{equation}

\begin{equation}
\label{z19}
\left.\begin{aligned}
Z1A100:\;&u'_{12}=u'_{23}=u'_{34}=u'_{41}=\chi'\tau^3\\
&u_{31}=u_{42}=\chi\tau^3,\;a_\mu=(-1)^\mu (a_3,a_1)\\
&u^d_{13}=-u^d_{24}=\Delta_{d,1}\tau^1+\Delta_{d,2}\tau^2\\
\end{aligned}\right.
\end{equation}
$Z16A100$ remains gapless while a gap opens for $Z1A100$.

There is only one descendant of $U1A0_+11_+$, given by
\begin{equation}
\label{z110}
\left.\begin{aligned}
Z13A000:\;&u'_{12}=u'_{23}=u'_{34}=u'_{41}=\chi'\tau^3\\
&u_{31}=u_{42}=\chi\tau^3,\;u^d_{13}=\chi_d\tau^3+\Delta_d\tau^1\\
&u^d_{24}=-\chi_d\tau^3+\Delta_d\tau^1,\;a_\mu=((-1)^\mu a_3,a_1)\\
\end{aligned}\right.
\end{equation}
 In this case, the Fermi surface of the parent $U(1)$ state completely disappears. 
 
Finally, the following three Ans\"atze, along with $Z13A000$ (previously discussed as the only descendant of $U1A0_+11_+$), appear in the vicinity of the $U1C0^-_+11_x0$ state:
\begin{equation}
\label{z111}
\left.\begin{aligned}
Z13A100:\;&u'_{12}=u'_{23}=u'_{34}=u'_{41}=\chi'\tau^3\\
&u_{31}=u_{42}=\chi\tau^3,\;a_1\neq0\\
&u^d_{13}=u^d_{24}=\Delta_{d,1}\tau^1+\Delta_{d,2}\tau^2\\
\end{aligned}\right.
\end{equation}

\begin{equation}
\label{z112}
\left.\begin{aligned}
Z17A111:\;&u_{31}=u_{42}=\chi\tau^3,\;u'_{12}=u'_{23}=\\
&u'_{34}=u'_{41}=\imath\chi'_0\tau^0+\chi'_3\tau^3\\
&u^d_{13}=u^d_{24}=\Delta_d\tau^1,\;a_1\neq0\\
\end{aligned}\right.
\end{equation}

\begin{equation}
\label{z113}
\left.\begin{aligned}
Z17A011:\;&a_1\neq0,\;u'_{12}=u'_{34}=\chi'\tau^3+\Delta'\tau^1\\
&u'_{23}=u'_{41}=\chi'\tau^3-\Delta'\tau^1,\\
&u_{31}=u_{42}=\chi\tau^3,\;u^d_{13}=u^d_{24}=\Delta_d\tau^1\\
\end{aligned}\right.
\end{equation}
$Z13A100$ reduces to the $J$-VBS state for $J'/J<1.7$, and to a gapped spin liquid state otherwise, similar to its parent $U(1)$ state. Self-consistent analysis turns the non-trivial flux threading the square plaquettes in $Z17A111$ and $Z17A011$ ans\"atze into a trivial ($\pi$) flux, and reduces these states to the dimerized $J$-VBS and $PRVB_2$ states obtained from $SU2A11$.

\subsubsection{$\mathds{Z}_2$ mean field states around $(\pi,\pi)$ flux state }
\label{sec:z11}
In this section, we enlist five $\mathds{Z}_2$ Ans\"atze which appear in the vicinity of the $SU2A11$ state. None of these is a direct descendant of the parent $SU(2)$ state. All of them can however be connected to $U(1)$ descendants of the parent $SU(2)$ state. They all reduce to a $J-VBS$ phase for $J'<J$ and to the $PRVB_2$ state when $J'>J$.
\begin{equation}
\label{z21}
\left.\begin{aligned}
Z20A111:\;&u'_{12}=-u'_{34}=\chi'\tau^3+(-)^{x+y}\Delta'\tau^1,\\
&u'_{23}=u'_{41}=\chi'\tau^3-(-)^{x+y}\Delta'\tau^1
\\&u_{31}=u_{42}=\chi\tau^3,\;a_\mu=(-)^{\mu}a_3\\
\end{aligned}\right.
\end{equation}

\begin{equation}
\label{z22}
\left.\begin{aligned}
Z20A101:\;&u'_{12}=-u'_{34}=u'_{23}=u'_{41}=\chi'\tau^3+(-)^{x+y}\Delta'\tau^1,\\
&u_{31}=u_{42}=\chi\tau^3,\;a_\mu=(-)^{x+y+\mu}a_3\\
\end{aligned}\right.
\end{equation}

\begin{equation}
\label{z23}
\left.\begin{aligned}
Z11A010:\;&u'_{12}=-u'_{34}=u'_{23}=u'_{41}=\chi'_3\tau^3+\Delta'_1\tau^1,\\
&u_{31}=u_{42}=\chi\tau^3,\;a_\mu=(a_3,a_1)\\
\end{aligned}\right.
\end{equation}

\begin{equation}
\label{z24}
\left.\begin{aligned}
Z20A001:\;&u'_{12}=u'_{23}=\chi'_3\tau^3+\Delta'_1\tau^1,\;-u'_{34}=\\&u'_{41}=\chi'_3\tau^3-\Delta'_1\tau^1,\;u_{31}=u_{42}=\chi\tau^3,\\
&a_\mu=(-)^{x+y+\mu}a_3\\
\end{aligned}\right.
\end{equation}

\begin{equation}
\label{z25}
\left.\begin{aligned}
Z20A010:\;&u'_{12}=u'_{23}=\chi'_3\tau^3+\Delta'_1\tau^1,\\
&u'_{34}=u'_{41}=\chi'_3\tau^3-\Delta'_1\tau^1\\
&u_{31}=u_{42}=\chi\tau^3,\;a_\mu=(-)^{\mu}a_3
\end{aligned}\right.
\end{equation}

\subsubsection{$\mathds{Z}_2$ mean field states around $(0,\pi)$ flux state }
\label{sec:z01}
The 13 $\mathds{Z}_2$ mean-field states in the vicinity of the $(0,\pi)$ flux state have the same form as the 13 states around the $(0,0)$ flux state (Sec.~\ref{sec:z00}), with the only difference being the signs of $u_{ij}$ matrices on the vertical $J$-bonds, which alternate along the $x$ direction [as given by Eq.~\eqref{eq:doubling}]. We can thus simply list
the $\mathds{Z}_2$ mean-field states around the $(0,\pi)$ flux state giving, within parenthesis, the equations of the corresponding $\mathds{Z}_2$ Ans\"atze around the $(0,0)$ flux state: $Z17B101$ [Eq.~\eqref{z11}], $Z13B101$ [Eq.~\eqref{z12}], $Z17B000$ [Eq.~\eqref{z13}], $Z1B101$ [Eq.~\eqref{z14}], $Z16B101$ [Eq.~\eqref{z15}], $Z1B000$ [Eq.~\eqref{z16}], $Z16B000$ [Eq.~\eqref{z17}], $Z16B100$ [Eq.~\eqref{z18}], $Z1B100$ [Eq.~\eqref{z19}], $Z13B000$ [Eq.~\eqref{z110}], $Z13B100$ [Eq.~\eqref{z111}],  $Z17B111$ [Eq.~\eqref{z112}] and $Z17B011$ [Eq.~\eqref{z113}]. As the parent state already features a gapped spinon dispersion, we skip further discussion on the nature of the excitations in its $\mathds{Z}_2$ descendants.

\subsubsection{$\mathds{Z}_2$ mean field states around $(\pi,0)$ flux state }
\label{sec:z10}
The Ans\"atze of the five mean-field states in the vicinity of the $(\pi,0)$ flux state have same form as those of the $\mathds{Z}_2$ states around the $(\pi,\pi)$ flux state, except for the signs of the $u_{ij}$ matrices over the vertical $J$-bonds, which alternate along the $x$ direction [as given by Eq.~\eqref{eq:doubling}]. Analogous to the previous subsection, we list the $\mathds{Z}_2$ states around the $(\pi,0)$ flux state, indicating, within parenthesis, the equations of the corresponding $\mathds{Z}_2$ Ans\"atze around the $(\pi,\pi)$ flux state: $Z20B111$ [Eq.~\eqref{z21}], $Z20B101$ [Eq.~\eqref{z22}], $Z11B010$ [Eq.~\eqref{z23}], $Z20B001$ [Eq.~\eqref{z24}] and $Z20B010$ [Eq.~\eqref{z25}]. The self-consistent mean-field analysis gives analogous results as those obtained by starting from the parent state.

\begin{center}
\begin{table}[]
\centering
    \caption{Symmetric $\mathds{Z}_2$ spin liquids perturbed around the $U(1)$ states. There are four $\mathds{Z}_2$ Ans\;atze labelled by $Z17A101$, $Z13A101$, $Z17A101$ and $Z13A101$ which are the direct descendants of $SU(2)$ Ans\"atze. Among them the former two Ans\"atze can be connected to $SU2A01$ while the later two appear in the vicinity of $SU2B01$.}
     \label{tab:z2_u1}
	\begin{ruledtabular}
    \begin{tabular}{cc}
    Parent $U(1)$ Ansatz &  Perturbed $\mathds{Z}_2$ Ans\"atze \\      
      \hline
 $U1A0_+01_+$ & $Z1A000$, $Z16A000$, $Z17A000$ \\
   &  $Z1A101$, $Z16A101$ \\
 $U1A0_+11_+$ & $Z1A100$, $Z13A000$ \\
 $U1C0^-_+01_x0$ & $Z16A100$, $Z16A000$, $Z1A100$\\
 $U1C0^-_+11_x0$ & $Z13A100$, $Z17A111$, $Z1A000$,\\
 & $Z13A000$, $Z17A011$\\
 $U1C^{-x}_x100$ & $Z1A000$\\
 
  $U1B0_+01_+$ & $Z1B000$, $Z16B000$, $Z17B000$ \\
   &  $Z1B101$, $Z16B101$ \\
 $U1B0_+11_+$ & $Z1B100$, $Z13B000$ \\
 $U1C0^-_+01_x1$ & $Z16B100$, $Z16B000$, $Z1B100$\\
 $U1C0^-_+11_x1$ & $Z13B100$, $Z17B111$, $Z1B000$,\\
 & $Z13B000$, $Z17B011$\\
 $U1C^{x}_x101$ & $Z1B000$\\
 
  $U1A0_-01_+$ &  $Z11A010$\\
 $U1A0_-11_+$ & $Z22A111$, $Z20A010$\\
 $U1C0^-_-01_x0$ & $-$\\
$U1C0^+_+11_x0$ &  $Z20A101$, $Z20A001$\\
  $U1B0_-01_+$ &  $Z11B010$\\
 $U1B0_-11_+$ & $Z22B111$, $Z20B010$\\
 $U1C0^-_-01_x1$ & $-$\\
$U1C0^+_+11_x1$ &  $Z20B101$, $Z20B001$\\
        \end{tabular}
	\end{ruledtabular}
\end{table}
\end{center}

\section{Conclusions}
\label{sec:discussion}
In this work, we have undertaken a systematic classification of symmetric quantum spin liquids with different gauge structures. Employing the Abrikosov fermion representation of spin-$1/2$ and the projective symmetry group approach we obtain the parton (spinon) mean-field Ans\"atze with $SU(2)$, $U(1)$ and $\mathds{Z}_2$ gauge structures. Although we obtain a large number, $32$ $SU(2)$, $1808$ $U(1)$, and $384$ $\mathds{Z}_{2}$ algebraic PSGs, upon restricting to short-range ($J$, $J'$, and $J_d$ bonds) mean-field amplitudes these PSGs collapse to only $4$ $SU(2)$, $24$ $U(1)$, and $36$ $\mathds{Z}_{2}$ distinct Ans\"atze. The existence of such a large number of $U(1)$ PSGs compared to that of $\mathds{Z}_2$ indicates that there must be a number of $U(1)$ solutions which appear with only $U(1)$ gauge structure and no symmetric perturbation is possible which can lower the IGG down to $\mathds{Z}_2$. Furthermore, we constructed mean-field spinon Hamiltonians and self-consistently determined the mean-field parameters at few representative points of interest in the parameter space of the $J-J'-J_d$ model. Consequently, we determined ground-state energies and studied the nature of the spinon dispersion in this zeroth-order mean-field approach. The states were found to have gapless or gapped excitations depending on the choices of parameter values. 

Our work sets the stage for different avenues of future investigations. For example, some nodal structures in the spinon spectrum exist for all choices of parameters. This raises the question as to whether these nodal band structures are protected by any space group symmetries which are acting projectively, i.e., whether the projective implementation of symmetries governs the protections~\cite{Liu-2021}. In these cases, such a nodal structure will survive even if we add amplitudes on further neighbour bonds (beyond $J_d$) consistent with the PSG. Furthermore, it can be examined whether there is any topological protection, i.e., whether they correspond to topological semimetals. As we have considered only the fully symmetric case which includes time-reversal symmetry, the calculation of the $\mathds{Z}_2$ invariant can be utilized to examine the latter possibility. Another question is what happens to the properties of the mean-field spin-liquid states once gauge fluctuations are taken into account, and their competition with magnetically ordered states upon Gutzwiller projection via variational Monte Carlo simulations~\cite{Capriotti-2001}. The corresponding classification for chiral spin liquids~\cite{Bieri-2016} on the square-octagon lattice is another interesting endeavor for future investigations.

\section*{Acknowledgements}
We thank Ganapathy Baskaran and Rajesh Narayanan for encouraging us to look at this problem. F. F. acknowledges financial support from the Deutsche Forschungsgemeinschaft (DFG, German Research Foundation) for funding through TRR 288 -- 422213477 (project A05). R.\,T. and F.\,F. thank IIT Madras for a visiting position under the IoE program which facilitated the completion of this research work. Y.I. acknowledges financial support by the Science and Engineering Research Board (SERB), DST, India through the Startup Research Grant No.~SRG/2019/000056, MATRICS Grant No.~MTR/2019/001042, and the Indo-French Centre for the Promotion of Advanced Research (CEFIPRA) Project No. 64T3-1. This research was supported in part by the National Science Foundation under Grant No.~NSF~PHY-1748958, ICTP through the Associates Programme and from the Simons Foundation through grant number 284558FY19, IIT Madras through the Institute of Eminence (IoE) program for establishing the QuCenDiEM CoE (Project No. SP22231244CPETWOQCDHOC), the International Centre for Theoretical Sciences (ICTS), Bengaluru, India during a visit for participating in the program “Frustrated Metals and Insulators” (Code: ICTS/frumi2022/9). Y.I. thanks IIT Madras for IoE travel grant to Germany which facilitated progress on the research work. Y.I. acknowledges the use of the computing resources at HPCE, IIT Madras. SM acknowledges SAMKHYA (High-Performance
Computing Facility provided by the Institute of Physics, Bhubaneswar) at initial phases of the work being carried out.

\appendix
\section{General Conditions on Projective Symmetry Group}
\label{gen_condition}
The symmetry generators satisfy the algebraic relations listed in Eq.~\eqref{algebraic}, which dictate the constraints on the projective symmetry groups. Together with a given IGG($\mathcal{G}$) this enables one to find allowed projective extensions of the symmetry group. Using Eq.~\eqref{algebraic}, the conditions on the projective matrices are
\begin{equation}
\label{c1}
\left.\begin{aligned}
&G_{T_x}T_xG_{T_y}T_yT^{-1}_xG^{-1}_{T_x}T^{-1}_yG^{-1}_{T_y}\in \mathcal{G}\\
\end{aligned}\right.
\end{equation}
\begin{equation}
\label{c2}
\left.\begin{aligned}
&G_{T_y}T_yC^{-1}_4G^{-1}_{C_4}G_{T_x}T_xG_{C_4}C_4\in \mathcal{G}\\
&T^{-1}_xG^{-1}_{T_x}C^{-1}_4G^{-1}_{C_4}G_{T_y}T_yG_{C_4}C_4\in \mathcal{G}\\
\end{aligned}\right.
\end{equation}

\begin{equation}
\label{c3}
\left.\begin{aligned}
&\sigma^{-1}G^{-1}_{\sigma}T^{-1}_xG^{-1}_{T_x}G_\sigma\sigma G_{T_x}T_x \in \mathcal{G}\\
&\sigma^{-1}G^{-1}_{\sigma} G_{T_y}T_yG_\sigma\sigma G_{T_y}T_y \in \mathcal{G}\\
\end{aligned}\right.
\end{equation}

\begin{equation}
\label{c4}
\left.\begin{aligned}
&T^{-1}_xG^{-1}_{T_x}\mathcal{T}^{-1}G^{-1}_{\mathcal{T}}G_{T_x}T_xG_{\mathcal{T}}\mathcal{T}\in \mathcal{G}\\
&T^{-1}_yG^{-1}_{T_y}\mathcal{T}^{-1}G^{-1}_{\mathcal{T}}G_{T_y}T_yG_{\mathcal{T}}\mathcal{T}\in \mathcal{G}\\
\end{aligned}\right.
\end{equation}

\begin{equation}
\label{c5}
\left.\begin{aligned}
&G_{C_4}C_4\sigma^{-1}G^{-1}_{\sigma}G_{C_4}C_4G_{\sigma}\sigma\in \mathcal{G}\\
&C^{-1}_4G^{-1}_{C_4}\mathcal{T}^{-1}G^{-1}_{\mathcal{T}}G_{C_4}C_4G_{\mathcal{T}}\mathcal{T}\in \mathcal{G}\\
&\sigma^{-1}G^{-1}_{\sigma}\mathcal{T}^{-1}G^{-1}_{\mathcal{T}}G_{\sigma}\sigma G_{\mathcal{T}}\mathcal{T}\in \mathcal{G}\\
\end{aligned}\right.
\end{equation}

\begin{equation}
\label{c6}
\left.\begin{aligned}
&G_{C_4}C_4G_{C_4}C_4G_{C_4}C_4G_{C_4}C_4\in \mathcal{G}\\
&G_{\sigma}\sigma G_{\sigma}\sigma\in \mathcal{G}\\
&G_{\mathcal{T}}\mathcal{T} G_{\mathcal{T}}\mathcal{T}\in \mathcal{G}\\
\end{aligned}\right.
\end{equation}
which imply 
\begin{equation}
\label{c_translation}
\left.\begin{aligned}
&G_{T_x}(i)G_{T_y}(T^{-1}_x(i))G^{-1}_{T_x}(T^{-1}_y(i))G^{-1}_{T_y}(i)\in \mathcal{G}\\
\end{aligned}\right.
\end{equation}
\begin{equation}
\label{c_rotation}
\left.\begin{aligned}
&G_{T_y}(C^{-1}_4(i)G^{-1}_{C_4}(T_x(i))G_{T_x}(T_x(i))G_{C_4}(i)\in \mathcal{G}\\
&G^{-1}_{T_x}(T_xC^{-1}_4(i))G^{-1}_{C_4}(T_y(i))G_{T_y}(T_y(i))G_{C_4}(i)\in \mathcal{G}\\
\end{aligned}\right.
\end{equation}

\begin{equation}
\label{c_reflection}
\left.\begin{aligned}
&G^{-1}_\sigma(T^{-1}_x\sigma(i))G^{-1}_{T_x}(\sigma(i))G_\sigma(\sigma(i)) G_{T_x}(i) \in \mathcal{G}\\
&G^{-1}_\sigma(\sigma T^{-1}_y(i)) G_{T_y}(T_y\sigma(i))G_\sigma(\sigma(i)) G_{T_y}(i) \in \mathcal{G}\\
\end{aligned}\right.
\end{equation}

\begin{equation}
\label{c_time}
\left.\begin{aligned}
&G^{-1}_{T_x}(\mathcal{T}^{-1}T_x(i))G^{-1}_\mathcal{T}(T_x(i))G_{T_x}(T_x(i))G_\mathcal{T}\mathcal{T}(i)\in \mathcal{G}\\
&G^{-1}_{T_y}(\mathcal{T}^{-1}T_y(i))G^{-1}_\mathcal{T}(T_y(i))G_{T_y}(T_y(i))G_\mathcal{T}\mathcal{T}(i)\in \mathcal{G}\\
\end{aligned}\right.
\end{equation}

\begin{equation}
\label{c_rot_ref_time}
\left.\begin{aligned}
&G_{C_4}(\sigma^{-1}(i)G^{-1}_\sigma(C_4(i))G_{C_4}(C_4(i))G_\sigma(i)\in \mathcal{G}\\
&G^{-1}_{C_4}(\mathcal{T}^{-1}C_4(i))G^{-1}_\mathcal{T}(C_4(i))G_{C_4}(C_4(i))G_\mathcal{T}\mathcal{T}(i)\in \mathcal{G}\\
&G^{-1}_\sigma(\mathcal{T}^{-1}\sigma(i))G^{-1}_\mathcal{T}(\sigma(i))G_\sigma(T_x(i))G_\mathcal{T}\mathcal{T}(i)\in \mathcal{G}\\
\end{aligned}\right.
\end{equation}

\begin{equation}
\label{cyclic}
\left.\begin{aligned}
&G_{C_4}(C^{-1}_4(i))G_{C_4}(C^2_4(i))G_{C_4}(C_4(i))G_{C_4}(i)\in \mathcal{G}\\
&G_{\sigma}(\sigma(i)) G_{\sigma}(i)\in \mathcal{G}\\
&G_{\mathcal{T}}(\mathcal{T}(i)) G_{\mathcal{T}}(i)\in \mathcal{G}\\
\end{aligned}\right.
\end{equation}
In the ensuing sections, we show how the above conditions completely determine the algebraic PSGs corresponding to different IGGs.

\section{Derivation of $SU(2)$ Algebraic Relations}
\label{su2_algebraic}

Henceforth, we will express the Ans\"atze of $SU(2)$ spin liquids, in the following canonical gauge~\cite{Wen-2002} wherein the $SU(2)$ gauge structure is manifest
\begin{equation} 
\label{su2_canonical}
u_{ij}=\imath\chi_{ij}\tau^0
\end{equation}
This is achieved by writing the projective matrices as
\begin{equation} 
G(x,y,\mu)=\eta(x,y)g(\mu)
\end{equation}
where $g(\mu)$ is a generic $SU(2)$ matrix and $\eta(x,y)$ can take values $\{+1,-1\}$ depending on site $i$.
In this way the projective extension of translational symmetry operators takes form
\begin{equation}
\label{su2_trans_1}
\left.\begin{aligned}
&G_{T_x}(x,y,\mu)=\eta_y(x,y,\mu)g_x\\
&G_{T_y}(x,y,\mu)=\eta_x(x,y,\mu)g_y\\
\end{aligned}\right.
\end{equation}
where $g_x,g_y\in SU(2)$. We can always perform a local gauge transformation which does not affect the form of the Ans\"atze in the canonical gauge
\begin{equation} 
W_i=\eta(i)\tau^0
\end{equation}
This enables one to fix the sign of $\eta_x$ to be positive on the entire lattice and that of $\eta_y$ to be positive along the line $y=0$, 
i.e., $\eta_y(x,y,\mu)=+1, \eta_x(x,0,\mu)=+1$. Thus, the simplified form of $G_{T_{x,y}}$ can be written as
\begin{equation}
\label{su2_trans_2}
\left.\begin{aligned}
&G_{T_x}(x,y,\mu)=\eta_y(x,y,\mu)g_x\\
&G_{T_y}(x,y,\mu)=g_y\\
\end{aligned}\right.
\end{equation}

Using Eq.~\eqref{c_translation} we have
\begin{equation} 
\eta_y(x,y+1,\mu)\eta_y(x,y,\mu)g^{-1}_yg^{-1}_xg_yg_x \in SU(2),
\end{equation}
implying the projective gauge corresponding to translational symmetries is given by
\begin{equation}
\label{su2_translation}
\left.\begin{aligned}
&G_{T_x}(x,y,\mu)=\eta^y_yg_x\\
&G_{T_y}(x,y,\mu)=g_y\\
\end{aligned}\right.
\end{equation}
Hence, there can be only two possible PSGs ($\eta=\pm1$) associated with translational symmetric Ans\"atze.

We shall now proceed towards obtaining the gauge transformation matrices corresponding to the point group symmetry generators. Eq.~\eqref{c_rotation} and 
Eq.~\eqref{c_reflection}, after simplification, take the following form
\begin{equation}
\left.\begin{aligned}
&\eta^y_yG^{-1}_{C_4}(x+1,y,\mu)G_{C_4}(x,y,\mu)\in SU(2)\\
&\eta^x_yG^{-1}_{C_4}(x,y+1,\mu)G_{C_4}(x,y,\mu)\in SU(2)\\
\end{aligned}\right.
\end{equation}

\begin{equation}
\left.\begin{aligned}
&G^{-1}_\sigma(x-1,y,\mu)G_\sigma(x,y,\mu) \in SU(2)\\
&G^{-1}_\sigma(x,y+1,\mu) G_\sigma(x,y,\mu)  \in SU(2)\\
\end{aligned}\right.
\end{equation}

These relations suggest the following solutions for $G_\sigma$ and $G_{C_4}$ 
\begin{equation}
\label{point1}
\left.\begin{aligned}
&G_{C_4}(x,y,\mu)=\eta^x_{cx}\eta^y_{cy}\eta^{xy}_yg_{C_4}(\mu)\\
&G_\sigma(x,y,\mu)=\eta^x_{\sigma_x}\eta^y_{\sigma_y}g_\sigma(\mu)\\
\end{aligned}\right.
\end{equation}
where $\eta_{cx,cy,\sigma_x,\sigma_y}$ take the values $\pm1$ and $g_{C_4,\sigma}\in SU(2)$. We now observe that under a local gauge transformation $W_{x,y,\mu}=\eta^x\tau^0$, the $G_{T_x,T_y,\sigma}$ are affected at most by a global sign factor which can, however, be absorbed by a redefinition of the unit cell representation matrices. Under such a transformation, the $G_{C_4}$ transform as
\begin{equation}
\left.\begin{aligned}
&\tilde{G}_{C_4}(x,y,\mu)=(\eta_{cx}\eta)^x(\eta_{cy}\eta)^y\eta^{xy}_yg_{C_4}(\mu)\\
\end{aligned}\right.
\end{equation} 
Since, we can always choose $\eta=\eta_{cx}$, upon introducing $\eta_{cxy}=\eta_{cx}\eta_{cy}$, the solutions~\eqref{point1} can be recast as
\begin{equation}
\label{point2}
\left.\begin{aligned}
&G_{C_4}(x,y,\mu)=\eta^y_{cxy}\eta^{xy}_yg_{C_4}(\mu)\\
&G_\sigma(x,y,\mu)=\eta^x_{\sigma_x}\eta^y_{\sigma_y}g_\sigma(\mu)\\
\end{aligned}\right.
\end{equation}
In a similar manner, Eq.~\eqref{c_time} invokes us to write $G_\mathcal{T}$ in the following form
\begin{equation}
\label{time1}
\left.\begin{aligned}
&G_{\mathcal{T}}(x,y,\mu)=\eta^x_{tx}\eta^y_{ty}g_{\mathcal{T}}(\mu)\\
\end{aligned}\right.
\end{equation}
where $\eta_{tx,ty}=\pm1$ and $g_{\mathcal{T}}\in SU(2)$.
Now, using the symmetry relation between $G_{C_4}$ and $G_\sigma$ in Eq.~\eqref{c_rot_ref_time} yields
\begin{equation}
\left.\begin{aligned}
&\eta^{x+y}_{cxy}\eta^{x+y}_{\sigma_x}\eta^{x+y}_{\sigma_y}g_{C_4}(\bar{\mu})g^{-1}_\sigma(\mu+1)g_{C_4}(\mu+1)g_\sigma(\mu)\in SU(2)\\
\end{aligned}\right.
\end{equation}
giving the following constraint among the $\eta's$
\begin{equation}
\label{eta_con_1}
\left.\begin{aligned}
&\eta_{cxy}\eta_{\sigma_x}\eta_{\sigma_y}=1\\
\end{aligned}\right.
\end{equation}
From the symmetry relation between $G_{C_4}$ and $G_\mathcal{T}$, and $G_{\sigma}$ and $G_\mathcal{T}$ in Eq.\eqref{c_rot_ref_time}, we find
\begin{equation}
\left.\begin{aligned}
&\eta^{x+y}_{tx}\eta^{x+y}_{ty} g^{-1}_{C_4}(\mu)g^{-1}_\mathcal{T}(\mu)g_{C_4}(\mu)g_\mathcal{T}(\mu-1)\in SU(2)\\
& g^{-1}_{\sigma}(\mu)g^{-1}_\mathcal{T}(\mu)g_{\sigma}(\mu)g_\mathcal{T}(\bar{\mu})\in SU(2)\\
\end{aligned}\right.
\end{equation}
The first one of the above two relations requires
\begin{equation}
\label{eta_con_2}
\left.\begin{aligned}
&\eta_{tx}\eta_{ty}=1\\
\end{aligned}\right.
\end{equation}
We can set $\eta_{tx}=\eta_{ty}=\eta_{\mathcal{T}}$, and Eq.~\eqref{time1} can thus be rewritten as
\begin{equation}
\label{time2}
\left.\begin{aligned}
&G_{\mathcal{T}}(x,y,\mu)=\eta^{x+y}_{\mathcal{T}}g_{\mathcal{T}}(\mu)\\
\end{aligned}\right.
\end{equation}
Using the cyclic relation associated with $C_4,\sigma,\mathcal{T}$ in Eq.~\eqref{cyclic} one finds
\begin{equation}
\label{su2_cyclic_relation}
\left.\begin{aligned}
&g_{C_4}(1)g_{C_4}(2)g_{C_4}(3)g_{C_4}(4)\in SU(2)\\
&g_{\sigma}(\mu)g_\sigma(\bar{\mu})\in SU(2)\\
&[g_{\mathcal{T}}(\mu)]^2\in SU(2)\\
\end{aligned}\right.
\end{equation}
which does not impose any further constraints on the $\eta$ parameters. Furthermore, it is clear that for $\eta_{\mathcal{T}}=+1$, the mean field 
amplitudes vanish. To obtain a non-vanishing Ansatz of the form~\eqref{su2_canonical}, the following conditions need to be imposed on $G_{\mathcal{T}}$
\begin{equation}
\label{time3}
\left.\begin{aligned}
&\eta_{\mathcal{T}}=-1\\
&g_{\mathcal{T}}(\mu)=\{g_t,-g_t,g_t,-g_t\}\\
\end{aligned}\right.
\end{equation}
Therefore, combining Eqs.~\eqref{point2},~\eqref{eta_con_1},~\eqref{time2},~\eqref{time3}, the solutions of the $G$ matrices with IGG $SU(2)$ can be written as 
\begin{equation}
\label{su2_sol}
\left.\begin{aligned}
&G_{T_x}(x,y,\mu)=\eta^y_yg_x,\;G_{T_y}(x,y,\mu)=g_y\\
&G_{C_4}(x,y,\mu)=(\eta_{\sigma_x}\eta_{\sigma_y})^y\eta^{xy}_yg_{C_4}(\mu)\\
&G_\sigma(x,y,\mu)=\eta^x_{\sigma_x}\eta^y_{\sigma_y}g_\sigma(\mu)\\
&G_{\mathcal{T}}(x,y,\mu)=(-1)^{x+y}(-1)^{\mu}g_t
\end{aligned}\right.
\end{equation}
To maintain the explicit $SU(2)$ canonical form given by Eq.~\eqref{su2_canonical}, the unit cell representation matrices must have 
the following forms
\begin{equation} 
\left.\begin{aligned}
&g_{C_4}(\mu)=\eta_c(\mu)g_c\\
&g_\sigma(\mu)=\eta_\sigma(\mu)g_\sigma\\
\end{aligned}\right.
\end{equation}
where $g_c,g_\sigma\in SU(2)$ and $\eta_c(\mu),\eta_\sigma(\mu)=\pm1$. A sublattice dependent gauge transformation of the form $W(x,y,\mu)=\eta(\mu)\tau^0$ enables one to fix some of these signs as
\begin{equation}
    \eta_c(1)=\eta_c(2)=\eta_c(3)=+1,\; \eta_c(4)=\eta_c
\end{equation}
Also, given the freedom of choosing a global sign one can set $\eta_\sigma(1)=+1$. Furthermore, using Eq.~\eqref{su2_cyclic_relation}, additional fixing of signs can be achieved as $\eta_\sigma(2)=\eta_\sigma(4)=\eta_\sigma$ and $\eta_\sigma(3)=\eta_c$. We summarize the gauge inequivalent choices of $(g_{C_4}(\mu),g_\sigma(\mu))$ in Table~\ref{table:su2_psg}.

\section{Derivation of $U(1)$ Algebraic Relations}
\label{u1_algebraic}
In this section, we obtain the algebraic solutions of PSG with IGG $U(1)$. The canonical gauge will be such that all the elements of IGG point in a particular direction in a vector space spanned by Pauli matrices, for example, the $\tau^3$ direction. Therefore, the IGG must take the following form:
\begin{equation}
\label{u1_igg}
\mathcal{G}=\{e^{\imath\theta\tau^3}|\;\;0\leq\theta<2\pi\}
\end{equation}
with the Ans\"atze expressed as
\begin{equation}
\label{u1_canonical}
u_{ij}=\imath\chi^0_{ij}\tau^0+\chi^3_{ij}\tau^3
\end{equation}

This form of the Ans\"atze suggests that the loop operators must be directed along the $\tau^3$ direction. Now, the translational symmetric Ans\"atze demand that the loop operators connected by translations can vary, at most, by a sign~\cite{Wen-2002}, fixing the form of the loop operator to
\begin{equation}
P_{C_i}=(\tau^1)^{n_i}P_{C_0}(\tau^1)^{n_i},\;\;n_i=0,1
\end{equation} 
where $P_{C_i}$ is the loop operator corresponding to the loop $C_i$ with the base point $i$ when $n_i=0$ and $P_{C_i}=P_{C_0}$. The corresponding gauge will be referred to as the uniform gauge.

Considering the generic case, i.e., for both $n_i=0,1$, we require a gauge transformation having a structure which preserves the canonical gauge form~\eqref{u1_canonical} of the Ans\"atze. This implies that the gauge structure must have either of the following forms in order to have non-vanishing Ans\"atze
\begin{equation}
\label{u1_gauge1}
G_\mathcal{S}(i)=g_3(\theta(i))(\imath\tau^1)^{n_\mathcal{S}}
\end{equation}
where,
\begin{equation}
g_3(\theta)=e^{\imath\theta\tau^3}
\end{equation}
and $n_\mathcal{S}=0,1$ with $\mathcal{S}\in \{T_x,T_y,C_4,\sigma,\mathcal{T}\}$.

Consequently, the gauge transformations corresponding to translation can have the following four sets of choices
\begin{equation}
\label{xy1}
G_{T_x}(i)=g_3(\theta_x(i)),\;\;\;G_{T_y}(i)=g_3(\theta_y(i))
\end{equation}
\begin{equation}
\label{xy2}
G_{T_x}(i)=g_3(\theta_x(i))\imath\tau^1,\;\;\;G_{T_y}(i)=g_3(\theta_y(i))\imath\tau^1
\end{equation}
\begin{equation}
\label{xy3}
G_{T_x}(i)=g_3(\theta_x(i))\imath\tau^1,\;\;\;G_{T_y}(i)=g_3(\theta_y(i))
\end{equation}
\begin{equation}
\label{xy4}
G_{T_x}(i)=g_3(\theta_x(i)),\;\;\;G_{T_y}(i)=g_3(\theta_y(i))\imath\tau^1
\end{equation}
where the last two sets are equivalent due to the $C_4$ symmetry of the lattice. Hence we need to consider a total of three sets of translational gauges.

Let us first proceed with the choice~\eqref{xy1}. In this case, it is easy to check that the translational invariance of the loop operators imply
\begin{equation}
\left.\begin{aligned}
&P_{C_{i-\hat{x}}}=G^{-1}_{T_x}(i)P_{C_i}G_{T_x}(i)=P_{C_i}\\
&P_{C_{i-\hat{y}}}=G^{-1}_{T_y}(i)P_{C_i}G_{T_y}(i)=P_{C_i}\\
\end{aligned}\right.
\end{equation}
thus corresponding to a uniform gauge. Furthermore, we notice that a local gauge transformation of the form $W_i=g_3(\phi_i)$ preserves the canonical form~\eqref{u1_canonical} of the Ans\"atz, and the resulting PSG stays in its canonical gauge. This gauge freedom also allows us to fix $G_{T_x},G_{T_y}$ to
\begin{equation}
\label{u1_freedom}
\theta_y(x,y,\mu)=\theta_y,\;\;\theta_x(x,0,\mu)=\theta_x
\end{equation} 
With the help of this simplification, the condition~\eqref{c_translation} gives
\begin{equation}
G^{-1}_{T_x}(x,y+1,\mu)G_{T_x}(x,y,\mu)=g_3(\phi)
\end{equation}
whose solution can be written as
\begin{equation}
\label{u1_tran_1}
\left.\begin{aligned}
&G_{T_x}(x,y,\mu)=g_3(y\phi+\theta_x)\\
&G_{T_y}(x,y,\mu)=g_3(\theta_y)\\
\end{aligned}\right.
\end{equation}

Now, let us consider the choice~\eqref{xy2}. The translational invariance of the loop operators implies
\begin{equation}
\left.\begin{aligned}
&P_{C_{i-\hat{x}}}=G^{-1}_{T_x}(i)P_{C_i}G_{T_x}(i)=\tau^1P_{C_i}\tau^1\\
&P_{C_{i-\hat{y}}}=G^{-1}_{T_y}(i)P_{C_i}G_{T_y}(i)=\tau^1P_{C_i}\tau^1\\
\end{aligned}\right.
\end{equation}
This means the loops connected by both $T_x$ and $T_y$ change their signs. With the help of ~\eqref{u1_freedom} the condition~\eqref{c_translation} simplifies to
\begin{equation}
\left.\begin{aligned}
&\tau^1G^{-1}_{T_x}(x,y+1,\mu)\tau^1G_{T_x}(x,y,\mu)=g_3(\phi)\\
&\implies \theta_x(x,y+1,\mu)+\theta_x(x,y,\mu)=\phi\\
\end{aligned}\right.
\end{equation}
This leads to the following solution
\begin{equation}
G_{T_x}(x,y,\mu)=g_3((-1)^y\theta(i))\imath\tau^1
\end{equation}
where $\theta(i)$ satisfies
\begin{equation}
\theta(x,y,\mu)-\theta(x,y-1,\mu)=(-1)^y\phi
\end{equation}
The solution which satisfies the above equation is given by
\begin{equation}
\theta(x,y,\mu)=\phi_x+(-1)^y\theta_x
\end{equation}
Hence,
\begin{equation}
G_{T_x}(x,y,\mu)=g_3((-1)^y\phi_x+\theta_x)\imath\tau^1
\end{equation}
Following a gauge transformation, $W_i=g_3((-1)^y\phi_x/2)$, the final solution can be written as
\begin{equation}
\label{u1_tran_2}
\left.\begin{aligned}
&G_{T_x}(x,y,\mu)=g_3(\theta_x)\imath\tau^1\\
&G_{T_y}(x,y,\mu)=g_3(\theta_y)\imath\tau^1\\
\end{aligned}\right.
\end{equation}
Let us now proceed with the case~\eqref{xy3}, where translational invariance implies
\begin{equation}
\left.\begin{aligned}
&P_{C_{i-\hat{x}}}=G^{-1}_{T_x}(i)P_{C_i}G_{T_x}(i)=\tau^1P_{C_i}\tau^1\\
&P_{C_{i-\hat{y}}}=G^{-1}_{T_y}(i)P_{C_i}G_{T_y}(i)=P_{C_i}\\
\end{aligned}\right.
\end{equation}
This implies that the loops connected by $T_x$ differ in their signs while the signs do not differ for loops connected by $T_y$. However, such a choice is not compatible with the $C_4$ symmetry as can be readily seen from Eq.~\eqref{c_rotation}, and thus this case need not be pursued.

We now study how a generalized translational symmetric Ans\"atze looks like in the cases given by Eqs.~\eqref{u1_tran_1} and \eqref{u1_tran_2}. For the uniform gauge [Eq.~\eqref{u1_tran_1}], the Ansatz on the bond connecting sites $i(x_i,y_i,\mu)$ and $j(x_j,y_j,\nu)$ has the form
\begin{equation}
\label{tran_ansatz_old}
u_{ij}=\imath\lambda_{ij}g_3(-x_iy_{ij}\phi+\xi_{ij}),\;\;y_{ij}=y_j-y_i
\end{equation}
In order to realize such an Ans\"atze on a lattice, $\phi$ must have the form $\phi=(p/q)\pi$ where $p$ and $q$ are integers. We divide this case into three classes $U1A$, $U1B$ and $U1(p,q)$ similar to the convention adopted in Ref.~\cite{Wen-2002}. $U1A$ corresponds to $\phi=0$, $U1B$ corresponds to $p=q$, i.e., $\phi=\pi$ and $U1(p,q)$ represents generic choices of $\phi$, i.e., $p\neq0,\;p\neq q$. The complete classificaton of $U1(p,q)$ type PSGs falls beyond the scope of this work, however, for any generic rational fraction $p/q\in(0,1)$, there will exist at least one mean-field Ansatz with $(p/q)\pi$ flux threading the plaquettes~\cite{Wen-2002}. We denote the class with the translational gauge given by Eq.~\eqref{u1_tran_2} by $U1C$.

We now find the gauge transformations corresponding to the symmetry operations $\sigma,C_4,\mathcal{T}$ for the classes $U1A$ and $U1B$ with the translational gauges given by
\begin{equation}
\label{u1ab_tran_1}
\left.\begin{aligned}
&G_{T_x}(x,y,\mu)=\eta^y_yg_3(\theta_x)\\
&G_{T_y}(x,y,\mu)=g_3(\theta_y)\\
\end{aligned}\right.
\end{equation}
$n^y_y=\pm1$ correspond to $U1A$ and $U1B$, respectively.

The relation~\eqref{c_time} gives
\begin{equation}
\left.\begin{aligned}
&G^{-1}_\mathcal{T}(x+1,y,\mu)G_\mathcal{T}(x,y,\mu)\in U(1)\\
&G^{-1}_\mathcal{T}(x,y+1,\mu)G_\mathcal{T}(x,y,\mu)\in U(1)\\
\end{aligned}\right.
\end{equation}
This implies the solutions
\begin{equation}
G_\mathcal{T}(x,y,\mu)=g_3(x\phi^t_x+y\phi^t_y+\theta^\mu_t),\;g_3(x\phi^t_x+y\phi^t_y+\theta^\mu_t)\imath\tau^1
\end{equation}
which after the imposition of the third condition of Eq.~\eqref{cyclic}, becomes
\begin{equation}
\label{time_u1a_1}
G_\mathcal{T}(x,y,\mu)=\eta^x_{tx}\eta^y_{ty}g_3(\theta^\mu_t),\;g_3(x\phi^t_x+y\phi^t_y+\theta^\mu_t)\imath\tau^1
\end{equation} 
Similarly, conditions~\eqref{c_reflection} and \eqref{c_rotation} become
\begin{equation}
\left.\begin{aligned}
&G^{-1}_\sigma(x+1,y,\mu)G_\sigma(x,y,\mu)\in U(1)\\
&G^{-1}_\sigma(x,y+1,\mu)G_\sigma(x,y,\mu)\in U(1)\\
\end{aligned}\right.
\end{equation}
\begin{equation}
\left.\begin{aligned}
&\eta^y_yG^{-1}_{C_4}(x+1,y,\mu)G_{C_4}(x,y,\mu)\in U(1)\\
&\eta^x_yG^{-1}_{C_4}(x,y+1,\mu)G_{C_4}(x,y,\mu)\in U(1)\\
\end{aligned}\right.
\end{equation}

These lead to the following solutions
\begin{equation}
\left.\begin{aligned}
&G_\sigma(x,y,\mu)=g_3(x\phi^x_\sigma+y\phi^y_\sigma+\theta^\mu_\sigma),\;g_3(x\phi^x_\sigma+y\phi^y_\sigma+\theta^\mu_\sigma)\imath\tau^1\\
&G_{C_4}(x,y,\mu)=\eta^{xy}_yg_3(x\phi^x_c+y\phi^y_c+\theta^\mu_c),\\
&\;\;\;\;\;\;\;\;\;\;\;\;\;\;\;\;\;\;\;\eta^{xy}_yg_3(x\phi^x_c+y\phi^y_c+\theta^\mu_c)\imath\tau^1\\
\end{aligned}\right.
\end{equation}
Without affecting the form of the translational gauges, one can perform a gauge transformation of the form $W(x,y,\mu)=g_3(x\zeta_x+y\zeta_y)$ which transforms this into
\begin{equation}
\left.\begin{aligned}
&G_\sigma(x,y,\mu)=g_3(x\phi_\sigma+\theta^\mu_\sigma),\;g_3(y\phi_\sigma+\theta^\mu_\sigma)\imath\tau^1\\
&G_{C_4}(x,y,\mu)=\eta^{xy}_yg_3(y\phi_c+\theta^\mu_c),\eta^{xy}_yg_3(y\phi_c+\theta^\mu_c)\imath\tau^1\\
\end{aligned}\right.
\end{equation}
The above solutions, with the help of the first two conditions of Eq.~\eqref{cyclic}, take the forms
\begin{equation}
\left.\begin{aligned}
&G_\sigma(x,y,\mu)=\eta^x_{\sigma x}g_3(\theta^\mu_\sigma),\;\eta^y_{\sigma y}g_3(\theta^\mu_\sigma)\imath\tau^1\\
&G_{C_4}(x,y,\mu)=\eta^{xy}_yg_3(y\phi_c+\theta^\mu_c),\eta^{xy}_yg_3(y\phi_c+\theta^\mu_c)\imath\tau^1\\
\end{aligned}\right.
\end{equation}
Now, the interrelation~\eqref{c_rot_ref_time} between $C_4$ and $\sigma$ fixes $\phi_c$. Also, the interrelation between $C_4$ and $\mathcal{T}$ fixes $\eta_{tx}=\eta_{ty}=\eta_t$, and the solution thus becomes
\begin{equation}
\label{time_u1ab_1}
G_\mathcal{T}(x,y,\mu)=\eta^{x+y}_{t}g_3(\theta^\mu_t),\;\eta^{x+y}_{t}g_3(\theta^\mu_t)\imath\tau^1
\end{equation} 
To obtain a nonvanishing time-reversal symmetric Ans\"atze on $J$ and $J'$ bonds we need to fix $n_t=-1,\;\theta^\mu_t=(-1)^\mu\theta_t$ for the first case i.e., $n_\mathcal{T}=0$. Therefore, the PSGs for classes $U1A$ and $U1B$ are given by 
\begin{equation}
\label{u1ab000(0,1)}
\left.\begin{aligned}
&U1(A/B)000(0/1):\\
&G_{T_x}(x,y,\mu)=\eta^y_yg_3(\theta_x),\;G_{T_y}(x,y,\mu)=g_3(\theta_y)\\
&G_\sigma(x,y,\mu)=\eta^x_{\sigma_x}g_3(\theta^\mu_\sigma)\\
&G_{C_4}(x,y,\mu)=\eta^y_{\sigma_x}\eta^{xy}_yg_3(\theta^\mu_c)\\
&G_{\mathcal{T}}(x,y,\mu)=(-1)^{x+y}(-1)^{\mu}g_3(\theta_t),\;\eta^{x+y}_tg_3(\theta^\mu_t)\imath\tau^1\\
\end{aligned}\right.
\end{equation}
\begin{equation}
\label{u1ab001(0,1)}
\left.\begin{aligned}
&U1(A/B)001(0/1):\\
&G_{T_x}(x,y,\mu)=\eta^y_yg_3(\theta_x),\;G_{T_y}(x,y,\mu)=g_3(\theta_y)\\
&G_\sigma(x,y,\mu)=\eta^x_{\sigma_x}g_3(\theta^\mu_\sigma)\\
&G_{C_4}(x,y,\mu)=\eta^y_{\sigma_x}\eta^{xy}_yg_3(\theta^\mu_c)\imath\tau^1\\
&G_{\mathcal{T}}(x,y,\mu)=(-1)^{x+y}(-1)^{\mu}g_3(\theta_t),\;\eta^{x+y}_tg_3(\theta^\mu_t)\imath\tau^1\\
\end{aligned}\right.
\end{equation}
\begin{equation}
\label{u1ab010(0,1)}
\left.\begin{aligned}
&U1(A/B)010(0/1):\\
&G_{T_x}(x,y,\mu)=\eta^y_yg_3(\theta_x),\;G_{T_y}(x,y,\mu)=g_3(\theta_y)\\
&G_\sigma(x,y,\mu)=\eta^y_{\sigma_y}g_3(\theta^\mu_\sigma)\imath\tau^1\\
&G_{C_4}(x,y,\mu)=\eta^y_{\sigma_y}\eta^{xy}_yg_3(\theta^\mu_c)\\
&G_{\mathcal{T}}(x,y,\mu)=(-1)^{x+y}(-1)^{\mu}g_3(\theta_t),\;\eta^{x+y}_tg_3(\theta^\mu_t)\imath\tau^1\\
\end{aligned}\right.
\end{equation}

\begin{equation}
\label{u1ab011(0,1)}
\left.\begin{aligned}
&U1(A/B)011(0/1):\\
&G_{T_x}(x,y,\mu)=\eta^y_yg_3(\theta_x),\;G_{T_y}(x,y,\mu)=g_3(\theta_y)\\
&G_\sigma(x,y,\mu)=\eta^y_{\sigma_y}g_3(\theta^\mu_\sigma)\imath\tau^1\\
&G_{C_4}(x,y,\mu)=\eta^y_{\sigma_y}\eta^{xy}_yg_3(\theta^\mu_c)\imath\tau^1\\
&G_{\mathcal{T}}(x,y,\mu)=(-1)^{x+y}(-1)^{\mu}g_3(\theta_t),\;\eta^{x+y}_tg_3(\theta^\mu_t)\imath\tau^1\\
\end{aligned}\right.
\end{equation}

Similarly, for class $U1C$, i.e., for the translational gauges given by $G_{T_x}(x,y,\mu)=g_3(\theta_x)\imath\tau^1,\;G_{T_y}(x,y,\mu)=g_3(\theta_y)\imath\tau^1$, the algebraic PSGs can be obtained. Employing these translational gauges, condition~\eqref{c_time} reads as
\begin{equation}
\left.\begin{aligned}
&\tau^1G^{-1}_\mathcal{T}(x+1,y,\mu)\tau^1G_\mathcal{T}(x,y,\mu)\in U(1)\\
&\tau^1G^{-1}_\mathcal{T}(x,y+1,\mu)\tau^1G_\mathcal{T}(x,y,\mu)\in U(1).\\
\end{aligned}\right.
\end{equation}
After substituting the general solution $G_\mathcal{T}(x,y,\mu)=g_3(\theta_\mathcal{T}(x,y,z))(\imath\tau^1)^{n_\mathcal{T}}$ we get
\begin{equation}
\left.\begin{aligned}
&(-)^{n_\mathcal{T}}[\theta_\mathcal{T}(x+1,y,\mu)+\theta_\mathcal{T}(x,y,\mu)]= \phi_\mathcal{T}\\
&(-)^{n_\mathcal{T}}[\theta_\mathcal{T}(x,y+1,\mu)+\theta_\mathcal{T}(x,y,\mu)]= \phi_\mathcal{T}\\
\end{aligned}\right.
\end{equation}
where $n_\mathcal{T}=0,1$. From the above equation we find two possible solution of $\theta_\mathcal{T}(x,y,\mu)$,
\begin{equation}
\left.\begin{aligned}
\theta_\mathcal{T}(x,y,\mu)= &(-)^{x+y}\phi_\mathcal{T}+\theta^\mu_\mathcal{T},\\
&(-)^{x+y}\phi_\mathcal{T}+x\pi+\theta^\mu_\mathcal{T}\\
\end{aligned}\right.
\end{equation}
By imposing the third condition of Eq.~\eqref{cyclic}, we obtain the following solution for $G_\mathcal{T}$.
\begin{equation}
\label{ap_time_u1c}
\left.\begin{aligned}
G_\mathcal{T}(x,y,\mu)= &\eta^x_{xt}\eta^y_{yt}g_3(\theta^\mu_\mathcal{T}),\\
&g_3((-)^{x+y}\phi_\mathcal{T}+\theta^\mu_\mathcal{T})\imath\tau^1,\\
&g_3((-)^{x+y}\phi_\mathcal{T}+x\pi+\theta^\mu_\mathcal{T})\imath\tau^1\\
\end{aligned}\right.
\end{equation}
where $\eta_{xt},\eta_{yt}=\pm1$.  Now conditions~\eqref{c_reflection} and \eqref{c_rotation} take the following forms,
\begin{equation}
\left.\begin{aligned}
&G^{-1}_\sigma(x-1,y,\mu)\tau^1G_\sigma(x,y,\mu)\tau^1\in U(1)\\
&G^{-1}_\sigma(x,y,\mu)\tau^1G_\sigma(x,y-1,\mu)\tau^1\in U(1).\\
\end{aligned}\right.
\end{equation}
and
\begin{equation}
\left.\begin{aligned}
&\tau^1G^{-1}_{C_4}(x+1,y,\mu)\tau^1G_{C_4}(x,y,\mu)\in U(1)\\
&\tau^1G^{-1}_{C_4}(x,y+1,\mu)\tau^1G_{C_4}(x,y,\mu)\in U(1).\\
\end{aligned}\right.
\end{equation}
 The solutions have the general forms $G_\sigma(x,y,\mu)=g_3(\theta_\sigma(x,y,z))(\imath\tau^1)^{n_\sigma}$ and $G_{C_4}(x,y,\mu)=g_3(\theta_{C_4}(x,y,z))(\imath\tau^1)^{n_{C_4}}$ where $n_{\sigma},n_{C_4}=0,1$. Similar to time reversal phase, in these two cases also the solutions can be written as :
\begin{equation}
\left.\begin{aligned}
\theta_\sigma(x,y,\mu)= &(-)^{x+y}\phi_\sigma+\theta^\mu_\sigma,\\
&(-)^{x+y}\phi_\sigma+x\pi+\theta^\mu_\sigma\\
\end{aligned}\right.
\end{equation}
and
\begin{equation}
\left.\begin{aligned}
\theta_{C_4}(x,y,\mu)= &(-)^{x+y}\phi_{c}+\theta^\mu_{c},\\
&(-)^{x+y}\phi_{c}+x\pi+\theta^\mu_{c}.\\
\end{aligned}\right.
\end{equation}
Now applying the cyclic condition~\eqref{cyclic} for $\sigma$ we obtain
\begin{equation}
\label{ap_ref_u1c}
\left.\begin{aligned}
G_\sigma(x,y,\mu)= &\eta^x_{\sigma_x}\eta^y_{\sigma_y}g_3(\theta^\mu_\sigma),\\
&g_3((-)^{x+y}\phi_\sigma+\theta^\mu_\sigma)\imath\tau^1,\\
&g_3((-)^{x+y}\phi_\sigma+x\pi+\theta^\mu_\sigma)\imath\tau^1\\
\end{aligned}\right.
\end{equation}
where $\eta_{\sigma_x},\eta_{\sigma_y}=\pm1$. Similarly, the cyclic condition for $C_4$ [Eq.~\eqref{cyclic}] imposes no constraint on $\phi_c$ for $n_{C_4}=1$, but for $n_{C_4}=0$ $\phi_{c}=\frac{m\pi}{4}$ where $m$ is any integer. Thus, for $C_4$ the solution can be written as 
\begin{equation}
\label{ap_rot_u1c}
\left.\begin{aligned}
G_{C_4}(x,y,\mu)= &\eta^x_{cx}\eta^y_{cy}g_3((-)^{x+y}\frac{m\pi}{4}+\theta^\mu_{c}),\\
&g_3((-)^{x+y}\phi_{c}+\theta^\mu_{c})\imath\tau^1,\\
&g_3((-)^{x+y}\phi_{c}+\theta^\mu_{c})\imath\tau^1.\\
\end{aligned}\right.
\end{equation}
In the obtained solutions all the parameters are not independent. The fixing of gauges can be carried out further using the conditions given by Eq.~\eqref{c_rot_ref_time} and then following some appropriate local gauge transformations the complete PSG solutions corresponding to IGG $U(1)$ are given in the Eqs.~\eqref{u1c_solution_1},\eqref{u1c_solution_2},\eqref{u1c_solution_3},\eqref{u1c_solution_4} and \eqref{u1c_solution_5}. Furthermore, the choices of the $U(1)$ phases $\theta^\mu_c$ and $\theta^\mu_\sigma$ are not independent and arbitrary. They can be fixed upon imposing the symmetry conditions given by Eqs.~\eqref{c_rot_ref_time} and \eqref{cyclic}. The choices are summarized in Table~\ref{table:u1_psg}.

\section{Derivation of $\mathds{Z}_2$ Algebraic Relations}
\label{z2_algebraic}
In this section, we derive the algebraic solutions for PSG considering IGG $\mathds{Z}_2$. In this case, the Ans\"atze has the generalized form [Eq.~\eqref{eq:ansatz}] consisting of real hopping and real pairing terms.

Let us first find the gauge transformation associated with the translational symmetry operators $T_x$, $T_y$. By exploiting local $SU(2)$ gauge symmetry $G_S(i)\rightarrow\tilde{G}_S(i)=W^\dagger_iG_S(i)W_{S^{-1}(i)},\;\;W_i\in SU(2)$, we can always fix the translational gauges as following
\begin{equation}
\label{psg_gauge}
G_{T_x}(x,0,\mu)=G_{T_y}(x,y,\mu)=\tau^0.
\end{equation}
Now, we need to fix $G_{T_x}(x,y,\mu)$ for $y\neq0$ sites. This can be done by substituting~\eqref{psg_gauge} in the condition~\eqref{c_translation} leading to
\begin{equation}
\label{tx_trans}
\left.\begin{aligned}
G_{T_x}(x,y,\mu)&G_{T_y}(x-1,y,\mu)\\
&G^{-1}_{T_x}(x,y-1,\mu)G^{-1}_{T_y}(x,y,\mu)=\eta_y\tau^0\\
\implies G_{T_x}(x,y,\mu)&=\eta_yG_{T_x}(x,y-1,\mu)\\
\end{aligned}\right.
\end{equation}
The solution of the above equation is given by $G_{T_x}(x,y,\mu)=\eta^y_yG_{T_x}(x,0,\mu)=\eta^y_y\tau^0$.
Now, we shall use the above result to find the projective solutions of point group symmetry generators $G_{C_4},G_\sigma,G_\mathcal{T}$ using the conditions given by Eqs.~\eqref{c_rotation},~\eqref{c_reflection},~\eqref{c_time},~\eqref{cyclic} and~\eqref{c_rot_ref_time}.

The condition given by Eq.~\eqref{c_reflection} relating the $C_4$ and $T_{x/y}$ results in
\begin{equation}
\left.\begin{aligned}
&G_{T_y}(x,y,\mu)G^{-1}_{C_4}(-y+1,x,\mu+1)\\
&G_{T_x}(-y+1,x,\mu+1)G_{C_4}(-y,x,\mu+1)=\eta_{cx}\tau^0\\
&G^{-1}_{T_x}(x+1,y,\mu)G^{-1}_{C_4}(-y,x+1,\mu+1)\\
&G_{T_y}(-y+1,x+1,\mu+1)G_{C_4}(-y,x,\mu+1)=\eta_{cy}\tau^0.\\
\end{aligned}\right.
\end{equation}
which on substitution of $G_{T{x/y}}$ gives
\begin{equation}
\left.\begin{aligned}
&G_{C_4}(x,y,\mu)=\eta_{cx}\eta^y_yG_{C_4}(x-1,y,\mu)\\
&G_{C_4}(x,y,\mu)=\eta_{cy}\eta^x_yG_{C_4}(x,y-1,\mu).\\
\end{aligned}\right.
\end{equation}
The above two equations can be combined into an algebraic solution for $G_{C_4}$
\begin{equation}
G_{C_4}(x,y,\mu)=\eta^x_{cx}\eta^y_{cy}\eta^{xy}_yg_{C_4}(\mu)
\end{equation}
where $g_{C_4}(\mu)=G_{C_4}(0,0,\mu)\in SU(2)$.
Similarly, for $G_\sigma$ from condition~\eqref{c_reflection},
\begin{equation}
\left.\begin{aligned}
&G^{-1}_{\sigma}(x,-y,1/3)G^{-1}_{T_x}(x+1,-y,1/3)\\
&G_\sigma(x+1,-y,1/3) G_{T_x}(x+1,y,1/3) =\eta_{\sigma_x}\tau^0\\
&G^{-1}_{\sigma}(x,-y,1/3)G_{T_y}(x,-y,1/3)\\
&G_\sigma(x,-y-1,1/3) G_{T_y}(x,y+1,1/3)=\eta_{\sigma_y}\tau^0\\
&and\\
&G^{-1}_{\sigma}(x,-y,4/2)G^{-1}_{T_x}(x+1,-y,4/2)\\
&G_\sigma(x+1,-y,4/2) G_{T_x}(x+1,y,2/4)=\eta_{\sigma_x}\tau^0\\
&G^{-1}_{\sigma}(x,-y,4/2)G_{T_y}(x,-y,4/2)\\
&G_\sigma(x,-y-1,4/2) G_{T_y}(x,y+1,2/4)=\eta_{\sigma_y}\tau^0\\
&\implies G_\sigma(x,y,\mu)=\eta_{\sigma_x}G_\sigma(x-1,y,\mu)\\
&\;\;\;\;\&\;\;\; G_\sigma(x,y,\mu)=\eta_{\sigma_y}G_\sigma(x,y-1,\mu)
\end{aligned}\right.
\end{equation}
which results in the solution of $G_\sigma$, 
\begin{equation}
G_\sigma(x,y,\mu)=\eta^x_{\sigma_x}\eta^y_{\sigma_y}g_\sigma(\mu)
\end{equation}

Similarly, the solutions can be obtained for the remaining point group generators $\sigma,\mathcal{T}$
\begin{equation}
\left.\begin{aligned}
G^{-1}_{T_x}&(x+1,y,\mu)G^{-1}_{\mathcal{T}}(x+1,y,\mu)\\
&G_{T_x}(x+1,y,\mu)G_{\mathcal{T}}(x,y,\mu)=\eta_{tx}\tau^0\\
G^{-1}_{T_y}&(x,y+1,\mu)G^{-1}_{\mathcal{T}}(x,y+1,\mu)\\
&G_{T_y}(x,y+1,\mu)G_{\mathcal{T}}(x,y,\mu)=\eta_{ty}\tau^0\\
&\implies G_\mathcal{T}(x,y,\mu)=\eta_{tx}G_\mathcal{T}(x-1,y,\mu)\\
&\;\;\;\;{\rm and}\;\;\; G_\mathcal{T}(x,y,\mu)=\eta_{ty}G_\mathcal{T}(x,y-1,\mu)
\end{aligned}\right.
\end{equation}
which get combined to give
\begin{equation}
\label{st}
G_{\mathcal{T}}(x,y,\mu)=\eta^{x}_{tx}\eta^{y}_{ty}g_{\mathcal{T}}(\mu)
\end{equation}
Now, to find the unit cell PSG representation matrices $g_{C_4}, g_{\sigma},g_\mathcal{T}$ we need to find the relations among them using the algebraic relations among the point group generators given in Eq.~\eqref{c_rot_ref_time}. The expression relating $C_4$ and $\sigma$ gives
\begin{equation}
\left.\begin{aligned}
&g_{C_4}(\mu+1)g_\sigma(\mu)g_{C_4}(\bar{\mu})=\pm g_\sigma(\mu+1),\\
&\eta_{cx}\eta_{cy}=\eta_{\sigma_x}\eta_{\sigma_y},\\
\end{aligned}\right.
\end{equation}
Similarly, for $S=C_4,\sigma$, $\mathcal{T}S=S\mathcal{T}$ leads to following relations
\begin{equation}
\left.\begin{aligned}
&g_{C_4}(\mu)g_\mathcal{T}(\mu-1)=\pm g_\mathcal{T}(\mu)g_{C_4}(\mu),\\
&g_{\sigma}(\mu)g_\mathcal{T}(\bar{\mu})=\pm g_\mathcal{T}(\mu)g_{\sigma}(\mu),\\
&\eta_{tx}\eta_{ty}=1,\;\implies\eta_{tx}=\eta_{ty}=\eta_\mathcal{T}(say)
\end{aligned}\right.
\end{equation}
Another three relations can also be constructed using the cyclic nature (condition~\eqref{cyclic}) of the point group operations:
\begin{equation}
\prod^4_{\mu=1} g_{c_4}(\mu)=\pm\tau^0,\;\;g_{\sigma}(\bar{\mu})g_{\sigma}(\mu)=\pm\tau^0,\;\;[g_{T}(\mu)]^2=\pm\tau^0
\end{equation}

All these algebraic solutions are combined in Eq.~\eqref{gauge} after applying a local gauge transformation $W(x,y,\mu)=\eta^x\tau^0$ where $\eta=\pm1$. This transformation allows further fixing of $\eta_{cx}$ and $\eta_{cy}$ and we thus obtained the particular form of $G_{C_4}$ given in Eq.~\eqref{gauge}. 

\section{Construction of short-range mean-field Ans\"atze}
\label{algebraic_short_ranged}
Once we are equipped with all the solutions for algebraic PSG solutions, using the condition~\eqref{eq:psg} we can obtain the mean-field Ans\"atze which can be realized on the square-octagon lattice. We introduce a concise notation for the links ($u_{ij}$) on all bonds in a reference unit cell such as $(x,y)=(0,0)$ as follows.
\begin{equation}
\label{eq:bond_def}
\left.\begin{aligned}
    \text{On $J$-bonds: }&
    u_{(0,0,3),(1,0,1)}=u_{31},\;u_{(0,0,4),(0,1,2)}=u_{42}.\\ 
    \text{On $J'$-bonds: }&
    u_{(0,0,1),(0,0,2)}=u'_{12},\;u_{(0,0,2),(0,0,3)}=u'_{23},\\
    &u_{(0,0,3),(0,0,4)}=u'_{34},\;u_{(0,0,4),(0,0,1)}=u'_{41}.\\ 
    \text{On $J_d$-bonds: }&
    u_{(0,0,1),(0,0,3)}=u^d_{13},\;u_{(0,0,2),(0,0,4)}=u^d_{24}.\\
    \end{aligned}\right.
\end{equation}
All other links can be obtained from the above-mentioned links by the application of translations. The reference bonds given by Eq.~\eqref{eq:bond_def} cannot be chosen arbitrarily. There always exists a nontrivial element of the point group $\mathcal{O}$ which imposes a condition that may leave a link invariant or exchange the pair of sites associated with the link, i.e., $\mathcal{O}(u_{ij})\rightarrow u_{ij}$ or $\mathcal{O}(u_{ij})\rightarrow u_{ji}=u^\dagger_{ij}$. Also, there are many symmetry elements which connect different links. In the following, we enlist such symmetry conditions. 
\begin{equation}
\left.\begin{aligned}
&\text{Conditions for the Ans\"atze on $J$ bonds}\\
&T_xC^2_4\;:\;u_{31}\rightarrow u^\dagger_{31},\\
&T_yC^2_4\;:\;u_{42}\rightarrow u^\dagger_{42},\\
&C_4\;:\;u_{31}\rightarrow u_{42}\\
&\sigma\;:\;u_{31}\rightarrow u_{31},\\
&T_y\sigma\;:\;u_{42}\rightarrow u^\dagger_{42}.\\
\end{aligned}\right.
\end{equation} 
\begin{equation}
\left.\begin{aligned}
&\text{Conditions for the Ans\"atze on $J'$ bonds}\\
&\sigma C^3_4\;:\;u'_{12}\rightarrow (u'_{12})^\dagger;\;u'_{34}\rightarrow (u'_{34})^\dagger,\\
&\sigma C_4\;:\;u'_{23}\rightarrow (u'_{23})^\dagger;\;u'_{41}\rightarrow (u'_{41})^\dagger,\\
&C_4\;:\;u'_{12}\rightarrow u'_{23},\;u'_{23}\rightarrow u'_{34},\;u'_{34}\rightarrow u'_{41}.
\end{aligned}\right.
\end{equation} 
\begin{equation}
\left.\begin{aligned}
&\text{Conditions for the Ans\"atze on $J_d$ bonds}\\
&C^2_4\;:\;u^d_{13}\rightarrow (u^d_{13})^\dagger;\;u^d_{24}\rightarrow (u^d_{24})^\dagger\\
&\sigma\;:\;u^d_{13}\rightarrow u^d_{13};\;u^d_{24}\rightarrow (u^d_{24})^\dagger,\\
&C_4\;:\;u^d_{13}\rightarrow u^d_{24};\;u^d_{24}\rightarrow (u^d_{13})^\dagger
\end{aligned}\right.
\end{equation}
Incorporating projective construction~\eqref{eq:psg}, the above relations can be written as follows
\begin{widetext}
 \begin{equation}
\label{eq:j_condition}
\left.\begin{aligned}
&\text{Conditions for the Ans\"atze on $J$ bonds: }\\
&G^\dagger_{C_4}(0,0,4)G^\dagger_{C_4}(0,0,1)G^\dagger_{T_x}(1,0,1)u^\dagger_{31}G_{T_x}(0,0,3)G_{C_4}(-1,0,3)G_{C_4}(0,1,2)=u_{31}\\
&G^\dagger_{C_4}(0,0,1)G^\dagger_{C_4}(0,0,2)G^\dagger_{T_y}(0,1,2)u^\dagger_{42}G_{T_y}(0,0,4)G_{C_4}(0,-1,4)G_{C_4}(-1,0,3)=u_{42}\\
&G^\dagger_{C_4}(0,0,4)u_{42}G_{C_4}(0,1,2)=u_{31},\;G^\dagger_{\sigma}(0,0,3)u_{31}G_{\sigma}(1,0,1)=u_{31}\\
&G^\dagger_{\sigma}(0,0,2)G^\dagger_{T_y}(0,0,4)u^\dagger_{42}G_{T_y}(0,1,2)G_\sigma(0,-1,4)=u_{42}
\end{aligned}\right.
\end{equation}   
\begin{equation}
\label{eq:jprime_condition}
\left.\begin{aligned}
&\text{Conditions for the Ans\"atze on $J'$ bonds}\\
&G^\dagger_{C_4}(0,0,2)G^\dagger_{C_4}(0,0,3)G^\dagger_{C_4}(0,0,4)G^\dagger_{\sigma}(0,0,2)(u'_{12})^\dagger G_{\sigma}(0,0,1)G_{C_4}(0,0,1)G_{C_4}(0,0,4)G_{C_4}(0,0,3)=u'_{12}\\
&G^\dagger_{C_4}(0,0,4)G^\dagger_{C_4}(0,0,1)G^\dagger_{C_4}(0,0,2)G^\dagger_{\sigma}(0,0,4)(u'_{12})^\dagger G_{\sigma}(0,0,3)G_{C_4}(0,0,3)G_{C_4}(0,0,2)G_{C_4}(0,0,1)=u'_{34}\\
&G^\dagger_{C_4}(0,0,3)G^\dagger_{\sigma}(0,0,3)(u'_{23})^\dagger G_{\sigma}(0,0,2)G_{C_4}(0,0,4)=u'_{23}\\
&G^\dagger_{C_4}(0,0,1)G^\dagger_{\sigma}(0,0,1)(u'_{41})^\dagger G_{\sigma}(0,0,4)G_{C_4}(0,0,2)=u'_{41}\\
&G^\dagger_{C_4}(0,0,2)u'_{23} G_{C_4}(0,0,3)=u'_{12},\;G^\dagger_{C_4}(0,0,3)u'_{34} G_{C_4}(0,0,4)=u'_{23},\;G^\dagger_{C_4}(0,0,4)u'_{41} G_{C_4}(0,0,1)=u'_{34}.\\
\end{aligned}\right.
\end{equation} 
\begin{equation}
\label{eq:jd_condition}
\left.\begin{aligned}
&\text{Conditions for the Ans\"atze on $J_d$ bonds}\\
&G^\dagger_{C_4}(0,0,2)G^\dagger_{C_4}(0,0,3)(u^d_{13})^\dagger G_{C_4}(0,0,1)G_{C_4}(0,0,4)=u^d_{13}\\
&G^\dagger_{C_4}(0,0,3)G^\dagger_{C_4}(0,0,4)(u^d_{24})^\dagger G_{C_4}(0,0,1)G_{C_4}(0,0,2)=u^d_{24}\\
&G^\dagger_{C_4}(0,0,2)u^d_{24}G_{C_4}(0,0,4)=u^d_{13},\;G^\dagger_{\sigma}(0,0,1)u^d_{13}G_{\sigma}(0,0,3)=u^d_{13},\;G^\dagger_{\sigma}(0,0,4)(u^d_{24})^\dagger G_{\sigma}(0,0,2)=u^d_{24}\\
\end{aligned}\right.
\end{equation}
\end{widetext}
In addition to these, the condition for time-reversal reads as the follows
\begin{equation}
\label{eq:sym_con_time_reversal}
    G^\dagger_{\mathcal{T}}(x,y,\mu)u_{(x,y,\mu),(x,y,\nu)}G_{\mathcal{T}}(x,y,\nu)=u_{(x,y,\mu),(x,y,\nu)}
\end{equation}

Let us now construct the Ans\"atze with IGG $SU(2)$ which can be realized on the square-octagon lattice. Since these Ans\"atze have the canonical form given by Eq.~\eqref{su2_canonical}, in order to define the link variables on the reference bonds we need to define in general $8$ mean-field parameters given by $\chi_{31}$, $\chi_{42}$, $\chi'_{12}$, $\chi'_{23}$, $\chi'_{34}$, $\chi'_{41}$, $\chi^d_{13}$ and $u^d_{24}$. These parameters are not independent and must obey the aforementioned symmetry conditions. They can be fixed once the solutions~\eqref{su2_sol} are substituted in the Eqs.~\eqref{eq:j_condition},~\eqref{eq:jprime_condition},~\eqref{eq:jd_condition},~\eqref{eq:sym_con_time_reversal}. 

From Eqs.~\eqref{eq:j_condition},~\eqref{eq:jprime_condition} we obtain the following three conditions given by Eqs.~\eqref{eq:su2_j_condition},~\eqref{eq:su2_jprime_condition}, respectively.
\begin{equation}
\label{eq:su2_j_condition}
\left.\begin{aligned}
&\eta_c\eta_{\sigma_x}\eta_{\sigma_y}=-1\\
&\chi_{31}=\eta_c\eta_{\sigma_x}\eta_{\sigma_y}u_{42}\\
&\eta_c\eta_{\sigma_x}=1,\;\eta_{\sigma_y}=-1\\
\end{aligned}\right.
\end{equation} 
\begin{equation}
\label{eq:su2_jprime_condition}
\left.\begin{aligned}
&\eta_\sigma=-1\\
&u'_{12}=\chi'_{23}=\eta_c u'_{34}=u'_{12}.\\
\end{aligned}\right.
\end{equation} 
Now imposing Eq.~\eqref{eq:sym_con_time_reversal} on $J_d$-bond links, we have
\begin{equation}
    \chi^d_{13}=-\chi^d_{13}\implies\chi^d_{13}=0.
\end{equation}

These conditions allow one to fix the link parameters on the reference bonds and upon applying translation operations on these reference links give us $u_{ij}$ on all other bonds. Notice that $\eta_y$ does not appear in the above constraints. This indicates that the mean-field parameters are the same for both signs of $\eta_y$. However, they can differ by the signs of the parameter when one translates throughout the lattice. The projective implication of translations gives
\begin{equation}
\left.\begin{aligned}
   G^\dagger_{T_{x}}(x,y,\mu)u_{(x,y,\mu),(x',y',\nu)}&G_{T_{x}}(x',y',\nu)\\
   &=u_{(x+1,y,\mu),(x'+1,y',\nu)}\\
   G^\dagger_{T_{y}}(x,y,\mu)u_{(x,y,\mu),(x',y',\nu)}& G_{T_{y}}(x',y',\nu)\\
   &=u_{(x,y+1,\mu),(x',y'+1,\nu)}\\
    \end{aligned}\right.
\end{equation}
Application of this on the reference bonds. i.e., $(x,y)=(x',y')=0$ affects only the vertical $J$-bonds as following.
\begin{equation}
    u_{(1,0,4),(1,1,2)}=\eta_yu_{42}.
\end{equation}
Thus, for $\eta_y=-1$, the vertical $J$-bond alters its sign when $T_x$ operates on it, yielding, $u_{(x,y,4),(x,y+1,2)}=\eta^x_yu_{42}$ while all other parameters remain the same. In the following, we enumerate all such Ans\"atze.

Let us first consider the cases where there are non-vanishing amplitudes on both $J$ and $J'$ bonds. For that Eq.~\eqref{eq:su2_j_condition} and Eq.~\eqref{eq:su2_jprime_condition} give $\eta_{\sigma_y}=\eta_\sigma=-1$ and $\eta_{\sigma_x}=\eta_c$. So there are four such Ans\"atze among which two correspond to $\eta_y=+1$ and the other two correspond to $\eta_y=-1$. For $\eta_{\sigma_x}=\eta_c=1$, we have
\begin{equation}
\left.\begin{aligned}
&u'_{12}=u'_{23}=u'_{34}=u'_{41}=\imath\chi' \tau^0\\
&u_{31}=-u_{42}=\imath\chi \tau^0\\
\end{aligned}\right.    
\end{equation}
We label this case by $SU2A01$ and $SU2B01$ for $\eta_y=+1$ and $\eta_y=-1$, respectively (labelling scheme is given by the Eq.~\eqref{su2notation}). Now, for $\eta_{\sigma_x}=\eta_c=-1$, we have
\begin{equation}
\left.\begin{aligned}
&u'_{12}=u'_{23}=-u'_{34}=u'_{41}=\imath\chi' \tau^0\\
&u_{31}=-u_{42}=\imath\chi \tau^0\\
\end{aligned}\right.    
\end{equation}
We label this case by $SU2A11$ and $SU2B11$ for $\eta_y=+1$ and $\eta_y=-1$, respectively. Now, let us consider the case with $\eta_\sigma=+1$, $\eta_{\sigma_y}=-1$ and $\eta_{\sigma_x}=\eta_c=1$. From Eq.~\eqref{eq:su2_jprime_condition}, $u'_{12}=u'_{23}=u'_{34}=u'_{41}=0$, we have nonvanishing amplitudes only on $J'$-bonds. This corresponds to a dimerized state which we label as $J$-VBS and the Ansatz is given by
\begin{equation}
\left.\begin{aligned}
&u'_{12}=u'_{23}=-u'_{34}=u'_{41}=0\\
&u_{31}=u_{42}=\imath\chi \tau^0.\\
\end{aligned}\right.    
\end{equation}
Now, let us consider the case with $\eta_{\sigma_y}=+1$, $\eta_\sigma=-1$ and $\eta_{\sigma_x}=\eta_c=1$. Here, $\eta_{\sigma_y}=+1$ immediately sets the mean-field parameters on $J$-bonds. The Ansatz is given by
\begin{equation}
\left.\begin{aligned}
&u'_{12}=u'_{23}=\eta_c u'_{34}=u'_{41}=\imath\chi' \tau^0\\
&u_{31}=u_{42}=0.\\
\end{aligned}\right.    
\end{equation} 
We label the cases with $\eta_c=+1$ and $\eta_c=-1$ by PRVB$_1$ and PRVB$_2$, respectively. By a suitable gauge transformation given by Eq.~\eqref{gauge_su2_real_hopping}, all these cases take the form given in Sec.~\ref{sec:su2_sl}. Similarly, all Ans\"atze with IGG $SU(2)$ and $U(1)$ can be obtained using the conditions given in Eqs.~\eqref{eq:j_condition},~\eqref{eq:jprime_condition},~\eqref{eq:jd_condition},~\eqref{eq:sym_con_time_reversal}.  

\section{Fermionic Hamiltonians and spinon dispersions}
\label{app:bands}

When the mean-field Hamiltonian consists of only hopping terms, as in the case of $SU(2)$ spin liquids (within a suitable gauge), it can be expressed in Fourier space as follows
\begin{equation}
\label{eq:hop_ham}
    \hat{\mathbf{H}}_{\rm mf}(\mathbf{k})=\hat{\mathbf{\Psi}}^\dagger_\mathbf{k}\mathbb{D}_\mathbf{k}\hat{\mathbf{\Psi}}_\mathbf{k}
\end{equation}
where we have used the vector $\hat{\mathbf{\Psi}}_\mathbf{k}=(\hat{f}_{\mathbf{k},s,\uparrow},\hat{f}_{\mathbf{k},s,\downarrow})^T$, with $s$ denoting the sublattice index. Since only singlet hopping terms are considered, the $\mathbb{D}_\mathbf{k}$ matrix is block diagonal, with the $\uparrow$ and the $\downarrow$ sectors being equal, i.e. $\mathbb{D}_{\mathbf{k}}^{\uparrow\uparrow}=\mathbb{D}_{\mathbf{k}}^{\downarrow\downarrow}$. Thus, the resulting spinon bands are doubly degenerate and the one-fermion-per-site constraint in the ground state is fulfilled by simply filling the lower half of the energy eigenstates. In the main text, when plotting the spinon bands of the fermionic Hamiltonians with only hopping terms, we show the eigenvalues of one of the blocks of the $\mathbb{D}_{\mathbf{k}}$ matrix.

In the presence of pairing terms, the matrix expression of Eq.~\eqref{eq:hop_ham} is no longer suitable to describe the mean-field Hamiltonian. Therefore, one can resort to Nambu vectors of the form $\hat{\mathbf{\Psi}}_\mathbf{k}=(\hat{f}_{\mathbf{k},s,\uparrow},\hat{f}^\dagger_{\mathbf{-k},s,\downarrow})^T$, within which $\mathbb{D}_\mathbf{k}$ takes the form of a Bogoliubov-de Gennes Hamiltonian, resulting in eigenvalues which come in positive/negative pairs, i.e. ($\omega_{\mathbf{k},s,\uparrow},-\omega_{-\mathbf{k},s,\downarrow}$). After a suitable Bogoliubov transformation, the Hamiltonian can be written in a diagonal form in terms of the Bogoliubov quasiparticle operators $\hat{d}_{\mathbf{k},s,\uparrow}$ and $\hat{d}_{\mathbf{k},s,\downarrow}$
\begin{equation}
\label{eq:pair_ham}
\left.\begin{aligned}
    \hat{\mathbf{H}}_{\rm mf}=&\sum_{\mathbf{k}}(\omega_{\mathbf{k},s,\uparrow}\hat{d}^\dagger_{\mathbf{k},s,\uparrow}\hat{d}_{\mathbf{k},s,\uparrow}-\omega_{-\mathbf{k},s,\downarrow}\hat{d}_{\mathbf{-k},s,\downarrow}\hat{d}^\dagger_{\mathbf{-k},s,\downarrow})\\
    =&\sum_{\mathbf{k}}(\omega_{\mathbf{k},s,\uparrow}\hat{d}^\dagger_{\mathbf{k},s,\uparrow}\hat{d}_{\mathbf{k},s,\uparrow}+\omega_{-\mathbf{k},s,\downarrow}\hat{d}^\dagger_{\mathbf{-\mathbf{k}},s,\downarrow}\hat{d}_{\mathbf{-\mathbf{k}},s,\downarrow})\\
    &-\sum_{\mathbf{k}}\omega_{-\mathbf{k},s,\downarrow}\\
    \end{aligned}\right.
\end{equation}
In the main text, when displaying the spinon bands for the fermionic Hamiltonians with hopping and pairing terms, we will plot only the positive quasiparticle energies for each momentum $\mathbf{k}$, similar to the convention adopted in Ref.~\cite{Chern-2021}.

\section{The different gauge dependent representations of the Ans\"atze}
\label{app:gauge_dep_ansatz}
Due to the presence of $SU(2)$ gauge redundancy in the mean-field Hamiltonian of Eq.~\eqref{eq:h0}, a given Ansatz may not be in the canonical form of the corresponding IGG. For example, one notices that all $SU(2)$ Ans\"atze in Sec.~\ref{sec:su2_sl} feature only real hopping terms though in the canonical form of $SU(2)$ Ans\"atze, only imaginary hopping terms are allowed. They can be brought back to their canonical form after affecting the following gauge transformation
\begin{equation}
\label{gauge_su2_real_hopping}
\left.\begin{aligned}
    W(x,y,\mu)&=g_3((x+y)\pi/2) W(\mu)\\ W(\mu)&=\{\tau^0,\imath\tau^3,-\tau^0,-\imath\tau^3\}.\\
    \end{aligned}\right.
\end{equation}
Choosing to express an Ansatz in a suitable gauge has the advantage of inferring its IGG manifestly and in determining whether it is symmetrically connected to other Ans\"atze.

Let us consider the Ansatz $U1C0^-_+01_x0$ given by Eq.~\eqref{u13}. To realize it one needs to consider an eight-site unit cell. However, $SU(2)$ gauge redundancy can be exploited to write it in a form of Eq.~\eqref{u13_pairing} which can be realized within a four site unit cell. The corresponding gauge transformation is given by
\begin{equation}
\label{gauge_u13_pairing}
   W(x,y,\mu)=g_2(-(-1)^{x+y+\mu}\pi/4)
\end{equation}
where, $g_2(\phi)=e^{\imath\phi\tau^2}$. The same gauge transformation can be utilized to transform Eq.~\eqref{u14} into the gauge invariant form Eq.~\eqref{u14_pairing}.
Similarly, the gauge transformations to write Eq.~\eqref{u15} in the form given by Eq.~\eqref{u15_pairing} is given by
\begin{equation}
\label{gauge_u15_pairing}
\left.\begin{aligned}
    W(x,y,\mu)&=g_1((x+y)\pi/4) W(\mu)\\ W(\mu)&=\{\tau^0,\imath\tau^1,\tau^0,\imath\tau^1\},\\
    \end{aligned}\right.
\end{equation}
where, $g_1(\phi)=e^{\imath\phi\tau^1}$. Finally, the gauge transformation which transforms Eqs.~\eqref{u53},~\eqref{u54},~\eqref{u51} and ~\eqref{u56} into the simple form given by Eqs.~\eqref{u53_pairing},~\eqref{u54_pairing},~\eqref{u51_pairing} and~\eqref{u56_pairing} is
\begin{equation}
\label{gauge_beta_pairing}
\left.\begin{aligned}
    W(x,y,\mu)&=g_1((x+y)\pi/4)\\
    \end{aligned}\right.
\end{equation}


%

\end{document}